\documentclass{emulateapj}

\usepackage{graphicx}
\usepackage{amssymb}
\usepackage{longtable}
\usepackage{epsf}
\usepackage{apjfonts}
\usepackage{mathptmx}

\slugcomment{Accepted for publication in the Astrophysical Journal}

\shorttitle{Rest-frame optical LFs at $2 \leq z \leq 3.5$}
\shortauthors{Marchesini et al.}

\begin{document}

\title{The rest-frame optical luminosity functions of galaxies at 
$2 \leq z \leq 3.5$ \altaffilmark{1}}

\author{
D.~Marchesini\altaffilmark{2},
P.~van Dokkum\altaffilmark{2},
R.~Quadri\altaffilmark{2}, 
G.~Rudnick\altaffilmark{3},
M.~Franx\altaffilmark{4},
P.~Lira\altaffilmark{5},
S.~Wuyts\altaffilmark{4},
E.~Gawiser\altaffilmark{2,5,6},
D.~Christlein\altaffilmark{2,5},
S.~Toft\altaffilmark{2,7}
}

\altaffiltext{1}{Based on observations with the NASA/ESA 
\textit{Hubble Space Telescope}, obtained at the Space Telescope 
Science Institute, which is operated by AURA, Inc., under NASA contract 
NAS5-26555. Also based on observations collected at the European Southern 
Observatories on Paranal, Chile as part of the ESO program 164.O-0612.}

\altaffiltext{2}{Department of Astronomy, Yale University, New Haven, 
CT}
\email{danilom@astro.yale.edu}
\altaffiltext{3}{Goldberg Fellow, National Optical Astronomical Observatory, 
Tucson, AZ}
\altaffiltext{4}{Leiden Observatory, Leiden University, Leiden, Netherlands}
\altaffiltext{5}{Departamento de Astronom\'\i{}a, Universidad de Chile, 
Santiago, Chile}
\altaffiltext{6}{National Science Foundation Astronomy and Astrophysics
Postdoctoral Fellow}
\altaffiltext{7}{European Southern Observatory, Garching bei M\"{u}nchen, 
Germany}

\begin{abstract}
We present the rest-frame optical ($B$, $V$, and $R$ band) luminosity 
functions (LFs) of galaxies at $2 \leq z \leq 3.5$, measured from 
a $K$-selected sample constructed from the deep NIR MUSYC, the ultradeep 
FIRES, and the GOODS-CDFS. This sample is unique for its combination of 
area and range of luminosities. The faint-end slopes of the LFs at $z>2$ 
are consistent with those at $z\sim0$. The characteristic magnitudes are 
significantly brighter than the local values (e.g., $\sim1.2$~mag in the 
$R$ band), while the measured values for $\Phi^{\star}$ are typically 
$\sim5$ times smaller. The $B$-band luminosity density at $z \sim2.3$ is 
similar to the local value, and in the $R$ band it is $\sim2$ times smaller 
than the local value. We present the LF of Distant Red Galaxies (DRGs), 
which we compare to that of non-DRGs. While DRGs and non-DRGs are 
characterized by similar LFs at the bright end, the faint-end slope of 
the non-DRG LF is much steeper than that of DRGs. The contribution of DRGs 
to the global densities down to the faintest probed luminosities is 
14\%-25\% in number and 22\%-33\% in luminosity. From the derived 
rest-frame $U-V$ colors and stellar population synthesis models, we 
estimate the mass-to-light ratios ($M/L$) of the different subsamples. The 
$M/L$ ratios of DRGs are $\sim5$ times higher (in the $R$ and $V$ bands) 
than those of non-DRGs. The global stellar mass density at 
$2 \leq z \leq 3.5$ appears to be dominated by DRGs, whose contribution 
is of order $\sim$60\%-80\% of the global value. Qualitatively similar 
results are obtained when the population is split by rest-frame $U-V$ 
color instead of observed $J-K$ color. 
\end{abstract}



\section{INTRODUCTION}\label{sec-in}

Understanding the formation mechanisms and evolution with cosmic time 
of galaxies is one of the major goals of observational cosmology. In 
the current picture of structure formation, dark matter halos build up 
in a hierarchical fashion controlled by the nature of the dark matter, 
the power spectrum of density fluctuations, and the parameters of the 
cosmological model. The assembly of the stellar content of galaxies is 
governed by much more complicated physics, such as the mechanisms of 
star formation, gaseous dissipation, the feedback of stellar and central 
supermassive black hole energetic output on the baryonic material of 
the galaxies, and mergers.

The mean space density of galaxies per unit luminosity, or luminosity 
function (LF), is one of the most fundamental of all cosmological 
observables, and it is one of the most basic descriptors of a galaxy 
population. The shape of the LF retains the imprint of galaxy formation 
and evolution processes; the evolution of the LF as a function of cosmic 
time, galaxy type and environment provides insights into the physical 
processes that govern the assembly and the following evolution of galaxies. 
Therefore, the LF represents one of the fundamental observational tools 
to constrain the free parameters of theoretical models.

The local ($z\sim0$) LF has been very well determined from several 
wide-area, multi-wave band surveys with follow-up spectroscopy 
(\citealt{norberg02}; \citealt{blanton01, blanton03}; 
\citealt{kochanek01}; \citealt{cole01}). At intermediate redshifts 
($z\sim0.75$), spectroscopic surveys found a steepening of the 
faint-end LF with increasing redshift in the global LF, mainly due to 
the contribution by later type galaxies (\citealt{lilly95}; 
\citealt{ellis96}). From the COMBO-17 survey, \citet{wolf03} measured 
the rest-frame optical LF up to $z\sim1.2$, finding that early-type 
galaxies show a decrease of a factor of $\sim10$ in the characteristic 
density $\Phi^{\star}$ of the LF. The latest type galaxies show a 
brightening of $\sim1$~mag in $M^{\star}$ (the characteristic magnitude) 
and an increase of $\sim1.6$ in $\Phi^{\star}$ in their highest redshift 
bin in the blue band. Further progress in the measurement of the LF at 
$z<2$ was obtained with the VIMOS VLT Deep Survey (VVDS; 
\citealt{lefevre04}) and the DEEP-2 Galaxy Redshift Survey \citep{davis03}. 
From the VVDS data, \citet{ilbert05} measured the rest-frame optical LF 
from $z=0.05$ to $z=2$. From the same data set, \citet{zucca06} performed 
a similar analysis for different spectral galaxy types, finding a 
significant steepening of the LF going from early to late types. Their 
results indicate a strong type-dependent evolution of the LF, and identify 
the latest spectral types as responsible for most of the evolution of the 
UV-optical LF out to $z=1.5$. 

Contrary to low-redshift studies, the selection of high-redshift ($z>2$) 
galaxies still largely relies on their colors. One of the most efficient 
ways to select high-redshift galaxies is the Lyman drop-out technique, 
which enabled Steidel and collaborators to build large samples of 
$z\sim3$ star-forming galaxies (\citealt{steidel96}, 1999). Extensive 
studies of these optically (rest-frame ultraviolet) selected galaxies at 
$z\sim3$ (Lyman Break Galaxies [LBGs]) and at $z\sim2$ (BM/BX galaxies; 
\citealt{steidel04}; \citealt{adelberger04}) have shown that they are 
typically characterized by low extinction, modest ages, stellar masses 
$\sim10^{10}$~M$_{\sun}$, and star formation rates of 
10--100~M$_{\sun}$yr$^{-1}$ (\citealt{steidel03}; \citealt{shapley01}; 
\citealt{reddy05}). \citet{shapley01} recovered the rest-frame $V$-band 
LF of LBGs at $z\sim3$ from the rest-frame UV LF (\citealt{steidel99}; 
but see also \citealt{sawicki06}), finding that the LBG LF is characterized 
by a very steep faint end. 

LBGs dominate the UV luminosity density at $z\sim2-6$, as well as 
possibly the global star formation rate density at these redshifts 
\citep{reddy05}. However, since the Lyman break selection technique 
requires galaxies to be very bright in the rest-frame UV in order to 
be selected, it might miss galaxies that are heavily obscured by dust 
or whose light is dominated by evolved stellar populations. These objects 
can be selected in the near-infrared (NIR), which corresponds to the 
rest-frame optical out to $z\sim3$. Using the NIR selection criterion 
$J-K >2.3$ (also suggested by \citealt{saracco01}), \citet{franx03} and 
\citet{vandokkum03} discovered a new population of high-redshift galaxies 
(distant red galaxies [DRGs]) that would be largely missed by optically 
selected surveys. Follow-up studies have shown that DRGs constitute a 
heterogeneous population. They are mostly actively forming stars at 
$z\sim1.5-3.5$ (\citealt{vandokkum03}; \citealt{forster04}; 
\citealt{rubin04}; \citealt{knudsen05}; \citealt{reddy05}; 
\citealt{papovich06}). However, some show no signs of active star 
formation and appear to be passively evolving (\citealt{labbe05}; 
\citealt{kriek06}; \citealt{reddy06}), while others seem to host powerful 
active galactic nuclei (\citealt{vandokkum04}; \citealt{papovich06}). 
Compared to LBGs, DRGs have systematically older ages and larger masses 
\citep{forster04}, although some overlap between the two exists 
(\citealt{shapley05}; \citealt{reddy05}). Recently, \citet{vandokkum06} 
have demonstrated that in a mass-selected sample ($>10^{11}$~$M_{\sun}$) 
at $2<z<3$, DRGs make up 77\% in mass, compared to only 17\% from LBGs 
(see also \citealt{papovich06}), implying that the rest-frame optical LF 
determined by \citet{shapley01} is incomplete.

The global (i.e., including all galaxy types) rest-frame optical LF 
at $z>2$ can be studied by combining multiwavelength catalogs with 
photometric redshift information. \citet{giallongo05} studied the 
$B$-band LFs of red and blue galaxies. They find that the $B$-band 
number densities of red and blue galaxies have different evolution, 
with a strong decrease of the red population at $z=2-3$ compared to 
$z=0$ and a corresponding increase of the blue population, in broad 
agreement with the predictions from their hierarchical cold dark matter 
models. As all previous works at $z>2$ are based on either very deep 
photometry but small total survey area (\citealt{poli03}; 
\citealt{giallongo05}) or larger but still single field surveys 
(\citealt{gabasch04}), their results are strongly affected by 
field-to-field variations and by low number statistics, especially 
at the bright end. Moreover, \citet{gabasch04} used an I-band--selected 
data set from the FORS Deep Field. The $I$ band corresponds to the 
rest-frame UV at $z\sim2$, which means that significant extrapolation is 
required.

In this paper we take advantage of the deep NIR MUSYC survey to measure 
the rest-frame optical ($B$, $V$, and $R$ band) LFs of galaxies at 
$2 \leq z \leq 3.5$. Its unique combination of surveyed area and depth 
allows us to (1) minimize the effects of field-to-field variations, (2) 
better probe the bright end of the LF with good statistics, and (3) sample 
the LF down to luminosities $\sim0.9$~mag fainter than the characteristic 
magnitude. To constrain the faint-end slope of the LF and to increase the 
statistics, we also made use of the FIRES and the GOODS-CDFS surveys, by 
constructing a composite sample. The large number of galaxies in our 
composite sample also allows us to measure the LFs of several subsamples 
of galaxies, such as DRGs and non-DRGs (defined based on their observed 
$J-K$ color), and of intrinsically red and blue galaxies (defined based 
on their rest-frame $U-V$ color). 

This paper is structured as follows. In \S~\ref{sec-cs} we present the 
composite sample used to measure the LF of galaxies at $2 \leq z \leq 3.5$; 
in \S~\ref{sec-lf} we describe the methods applied to measure the LF and 
discuss the uncertainties in the measured LF due to field-to-field variations 
and errors in the photometric redshift estimates; the results (of all galaxies 
and of the individual subsamples considered in this work) are presented in 
\S~\ref{sec-results}, while the estimates of the number and luminosity 
densities and the contribution of DRGs (red galaxies) to the global stellar 
mass density are given in \S~\ref{sec-densities}. Our results are 
summarized in \S~\ref{sec-concl}. We assume 
$\Omega_{\rm M}=0.3$, $\Omega_{\rm \Lambda}=0.7$, and 
$H_{\rm 0}=70$~km~s$^{-1}$~Mpc$^{-1}$ throughout the paper. All magnitudes 
and colors are on the Vega system, unless identified as ``AB''. Throughout 
the paper, the $J-K$ color is in the observed frame, while the $U-V$ color 
refers to the rest frame.


\section{THE COMPOSITE SAMPLE}\label{sec-cs}

The data set we have used to estimate the LF consists of a composite 
sample of galaxies built from three deep multiwavelength surveys, all 
having high-quality optical to NIR photometry: the ``ultradeep'' Faint 
InfraRed Extragalactic Survey (FIRES; \citealt{franx03}), the Great 
Observatories Origins Deep Survey (GOODS; \citealt{giavalisco04}; Chandra 
Deep Field--South [CDF-S]), and the MUlti-wavelength Survey by Yale-Chile 
(MUSYC; \citealt{gawiser06}; \citealt{quadri06}). Photometric catalogs 
were created for all fields in the same way, following the procedures 
of \citet{labbe03}. 

FIRES consists of two fields, namely, the Hubble Deep Field--South proper 
(HDF-S) and the field around MS~1054--03, a foreground cluster at $z=0.83$. 
A complete description of the FIRES observations, reduction procedures, 
and the construction of photometric catalogs is presented in detail in 
\citet{labbe03} and \citet{forster06} for HDF-S and MS~1054--03, 
respectively. Briefly, the FIRES HDF-S and MS~1054--03 (hereafter FH and 
FMS, respectively) are $K_{\rm s}$ band--limited multicolor source catalogs 
down to $K^{\rm tot}_{\rm s}=24.14$ and $K^{\rm tot}_{\rm s}=23.14$, for a 
total of 833 and 1858 sources over fields of 
$2.5^{\prime} \times2.5^{\prime}$ and $5.5^{\prime} \times5.3^{\prime}$, 
respectively. The FH and FMS catalogs have 90\% completeness level at 
$K^{\rm tot}_{\rm s}=23.8$ and $K^{\rm tot}_{\rm s}=22.85$, respectively. 
The final FH (FMS) catalogs used in the construction of the composite 
sample has 358 (1427) objects over an effective area of 4.74 (21.2) 
arcmin$^{2}$, with $K^{\rm tot}_{\rm s}<23.14$ (22.54), which for point 
sources corresponds to a 10 (8) $\sigma$ signal-to-noise ratio ($S/N$) in 
the custom isophotal aperture.

\begin{deluxetable*}{lccccrr}[!t]
\centering
\tablewidth{420pt}
\tablecaption{The Composite Sample: Field Specifications
\label{tab-ref1}}
\tablehead{
\colhead{}                      & \colhead{}    & 
\colhead{$K^{\rm tot}_{\rm 90}$}     & \colhead{$K^{\rm tot}_{\rm lim,cs}$} & 
\colhead{Area}                       & \colhead{} & 
\colhead{} \\
   \colhead{Field}                        & \colhead{Filter Coverage} &
   \colhead{(mag)}                   & \colhead{(mag)}    & 
   \colhead{(arcmin$^{2}$)}          &  \colhead{$N$} & 
   \colhead{$N_{\rm spec}$} }
\startdata
FIRES-HDFS   & 
 U$_{\rm 300}$B$_{\rm 450}$V$_{\rm 606}$I$_{\rm 814}$J$_{\rm s}$HK$_{\rm s}$ & 
 23.80 & 23.14 & 4.74 & 358  & 68  \\
FIRES-MS1054 & 
 UBVV$_{\rm 606}$I$_{\rm 814}$J$_{\rm s}$HK$_{\rm s}$ & 
 22.85 & 22.54 & 21.2 & 1427 & 297 \\
GOODS-CDFS   & 
 B$_{\rm 435}$V$_{\rm 606}$i$_{\rm 775}$z$_{\rm 850}$JHK$_{\rm s}$ & 
 21.94 & 21.34 & 65.6 & 1588 & 215 \\
MUSYC  & UBVRIzJHK$_{\rm s}$ & 21.33 & 21.09 & 286.1 & 5507 & 116 \\
\enddata
\tablecomments{$K^{\rm tot}_{\rm 90}$ is the $K_{\rm s}$ band total magnitude 
90\% completeness limit, $K^{\rm tot}_{\rm lim,cs}$ is the $K_{\rm s}$ band 
total magnitude limit used to construct the composite sample, $N$ is 
the total number of sources down to $K^{\rm tot}_{\rm lim,cs}$, and 
$N_{\rm spec}$ is the number of objects with spectroscopic redshift.}
\end{deluxetable*}

From the GOODS/EIS observations of the CDF-S (data release version 1.0) 
a $K_{\rm s}$ band limited multicolor source catalog (hereafter CDFS) was 
constructed, described in S. Wuyts et al. (2007, in preparation). GOODS 
zero points were adopted for $J$ and $K_{\rm s}$. The $H$-band zero point 
was obtained by matching the stellar locus on a $J-K$ versus $J-H$ 
color-color diagram to the stellar locus in FIRES HDF-S and MS~1054--03. 
The difference with the official GOODS $H$-band zero point varies across 
the field, but on average our $H$-band zero points are $\sim0.1$~mag 
brighter. A total effective area of 65.6 arcmin$^{2}$ is well exposed in all 
bands. The final catalog contains 1588 objects with 
$K^{\rm tot}_{\rm s}<21.34$ in this area. At $K^{\rm tot}_{\rm s}=21.34$ 
the median $S/N$ in the $K_{\rm s}$ isophotal aperture is $\sim12$.

MUSYC consists of optical and NIR imaging of four independent 
$30^{\prime}\times30^{\prime}$ fields with extensive spectroscopic 
follow-up \citep{gawiser06}. Deeper NIR $JHK_{\rm s}$ imaging was obtained 
over four $10^{\prime}\times10^{\prime}$ subfields with the ISPI camera at 
the Cerro Tololo Inter-American Observatory (CTIO) Blanco 4~m telescope. 
A complete description of the deep NIR MUSYC observations, reduction 
procedures, and the construction of photometric catalogs will be presented 
in \citet{quadri06}. The 5~$\sigma$ point-source limiting depths are 
$J\sim22.9$, $H\sim21.8$, and $K_{\rm s}\sim21.3$. The optical $UBVRIz$ 
data are described in \citet{gawiser06}. The present work is restricted to 
three of the four deep fields: the two adjacent fields centered around 
HDF-S proper (hereafter MH1 and MH2) and the field centered around the 
quasar SDSS~1030+05 (M1030). The final MUSYC $K_{\rm s}$-selected catalog 
used in the construction of the composite sample has 5507 objects over an 
effective area of 286.1 arcmin$^{2}$, with $K^{\rm tot}_{\rm s}<21.09$, 
which for point sources corresponds to a $\sim$10~$\sigma$ $S/N$ in the 
isophotal aperture.

Table~\ref{tab-ref1} summarizes the specifications of each field, 
including wave band coverage, $K{\rm s}$ band total magnitude 90\% 
completeness limit ($K^{\rm tot}_{\rm 90}$), effective area, the 
$K{\rm s}$ band total magnitude limit used to construct the composite 
sample ($K^{\rm tot}_{\rm lim,cs}$), the number of objects, and the 
number of sources with spectroscopic redshifts.

Only a few percent of the sources in the considered catalogs have 
spectroscopic redshift measurements. Consequently, we must rely primarily 
on photometric redshift estimates. Photometric redshifts $z_{\rm phot}$ 
for all galaxies are derived using an identical code to that presented in 
\citet{rudnick01,rudnick03}, but with a slightly modified template set. 
This code models the observed spectral energy distribution (SED) using 
nonnegative linear combinations of a set of eight galaxy templates. As in 
\citet{rudnick03}, we use the E, Sbc, Scd, and Im SEDs from \citet{coleman80}, 
the two least reddened starburst templates from \citet{kinney96}, and a 
synthetic template corresponding to a 10 Myr old simple stellar population 
(SSP) with a \citet{salpeter55} stellar initial mass function (IMF). We also 
added a 1 Gyr old SSP with a Salpeter IMF, generated with the 
\citet{bruzual03} evolutionary synthesis code. The empirical templates have 
been extended into the UV and the NIR using models. Comparing the photometric 
redshifts with 696 spectroscopic redshifts (63 at $z \geqslant1.5$) collected 
from the literature and from our own observations gives a scatter in 
$\Delta z/(1+z)$ of $\sigma=0.06$. Restricting the analysis to galaxies at 
$z \geqslant1.5$ in the MUSYC fields gives $\sigma=0.12$, corresponding to 
$\Delta z \approx0.4$ at $z=2.5$. Approximately 5\% of galaxies in this 
sample are ``catastrophic'' outliers. A full discussion of the quality 
of the photometric redshifts is given elsewhere \citep{quadri06}. The 
effects of photometric redshift errors on the derived LFs are modeled in 
\S~\ref{sub-zphot}.

Rest-frame luminosities are computed from the observed SEDs and redshift 
information using the method extensively described in the Appendix of 
\citet{rudnick03}. This method does not depend directly on template fits 
to the data but rather interpolates directly between the observed fluxes, 
using the best-fit templates as a guide. We computed rest-frame luminosities 
in the $U$, $B$, $V$, and $R$ filters of \citet{bessell90}. For these filters 
we use $M_{\rm \sun, U}=+5.66$, $M_{\rm \sun, B}=+5.47$, 
$M_{\rm \sun, V}=+4.82$, and $M_{\rm \sun, R}=+4.28$. In all cases where a 
spectroscopic redshift is available we computed the rest-frame luminosities 
fixed at $z_{\rm spec}$.

Stars in all $K_{\rm s}$-selected catalogs were identified by spectroscopy, 
by fitting the object SEDs with stellar templates from \cite{hauschildt99} 
and/or inspecting their morphologies, as in \citet{rudnick03}. On average, 
approximately 10\% of all the objects were classified as stars.

\begin{figure}
\epsscale{1}
\plotone{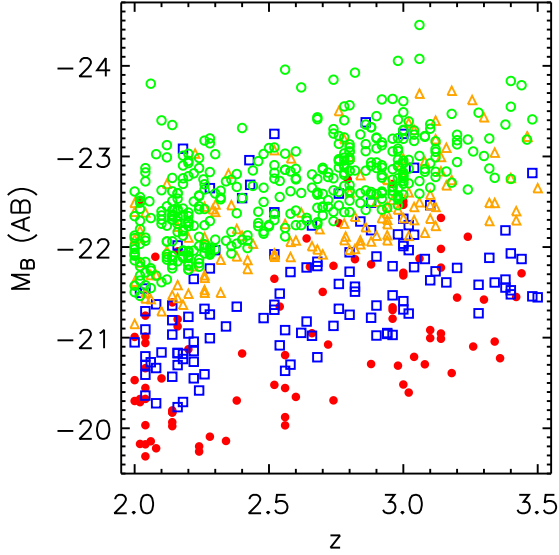}
\caption{Rest-frame $B$-band magnitude vs. redshift for the composite 
sample in the redshift range $2 \leq z \leq 3.5$; the FH, FMS, CDFS, and 
MUSYC fields are plotted as red filled circles, blue open squares, orange 
open triangles, and green open circles, respectively. 
\label{fig2_ref}}
\end{figure}

We constructed a composite sample of high-redshift ($2\leq z \leq3.5$)
galaxies to be used in the estimate of the LF in \S~\ref{sec-lf}. A large 
composite sample with a wide range of luminosities is required to sample 
both the faint and the bright end of the LF well; moreover, a large surveyed 
area is necessary to account for sample variance. The very deep FIRES allows 
us to constrain the faint end of the LF, while the large area of MUSYC 
allows us to sample the bright end of the LF very well. The CDFS catalog 
bridges the two slightly overlapping regimes and improves the number 
statistics. The final composite sample includes 442, 405, and 547 
$K_{\rm s}$-selected galaxies in the three targeted redshift intervals 
$2\leq z \leq2.5$, $2.7\leq z \leq3.3$, and $2.5< z \leq3.5$, respectively, 
for a total of 989 galaxies with $K_{\rm s}^{\rm tot}<23.14$ at 
$2\leq z \leq3.5$. Of these, $\sim$4\% have spectroscopic redshifts.
In Figure~\ref{fig2_ref} we show the rest-frame $B$-band absolute magnitude 
versus the redshift for the composite sample in the studied redshift 
range $2 \leq z \leq 3.5$.

\section{METHODOLOGY} \label{sec-lf}

\subsection{The $1/V_{\rm max}$ Method} \label{sec-1vmaxmeth}

To estimate the observed LF in the case of a composite sample, we have 
applied an extended version of the $1/V_{\rm max}$ algorithm 
\citep{schmidt68} as defined in \citet{avni80} so that several samples 
can be combined in one calculation. For a given redshift interval 
[$z_{\rm 1}$,$z_{\rm 2}$], we computed the galaxy number density $\Phi (M)$ 
in each magnitude bin $\Delta M$ in the following way: 

\begin{equation} \label{eq-1vmax}
\Phi (M) = \frac{1}{\Delta M} \sum_{i=1}^{N}{\frac{1}{V_{\rm i}}},
\end{equation}
where $N$ is the number of objects in the chosen bin and $V_{\rm i}$ 
is:

\begin{equation}
V_{\rm i} = \sum_{j=1}^{n}{\omega_{\rm j} 
\int_{z_{\rm min}}^{z_{\rm up}(i,j)}{\frac{dV}{dz} dz}},
\end{equation}
where $\omega_{\rm j}$ is the area in units of steradians corresponding 
to the $j$th field, $n$ is the number of samples combined together, 
$dV/dz$ is the comoving volume element per steradian, and $z_{\rm up}(i,j)$ 
is the minimum of $z_{\rm 2}$ and the maximum redshift at which the $i$th 
object could have been observed within the magnitude limit of the $j$th 
sample. The Poisson error in each magnitude bin was computed adopting the 
recipe of \citet{gehrels86} valid also for small numbers. 

The $1/V_{\rm max}$ estimator has the advantages of simplicity and no a 
priori assumption of a functional form for the luminosity distribution; 
it also yields a fully normalized solution. However, it can be affected by 
the presence of clustering in the sample, leading to a poor estimate of the 
faint-end slope of the LF. Although field-to-field variation represents a 
significant source of uncertainty in deep surveys (since they are 
characterized by very small areas and hence small sampled volumes), the 
majority of published cosmological number densities and related quantities 
do not properly account for sample variance in their quoted error budgets. 
Our composite sample is made of several independent fields with a large 
total effective area of $\sim380$~arcmin$^{2}$ (about a factor of 3 larger 
than the nominal area of the $K_{\rm s}$-selected CDFS-GOODS catalog used 
in \citealt{dahlen05}), which significantly reduces the uncertainties due 
to sample variance. Also, the large number of fields considered in this 
work with their large individual areas allows us to empirically measure 
the field-to-field variations from one field to the other in the estimate 
of the LF with the $1/V_{\rm max}$ method, especially at the bright end, 
and to properly account for it in the error budget.

In order to quantify the uncertainties due to field-to-field variations in 
the determination of the LF, we proceeded as follows. First, for each 
magnitude bin $\Delta M$, we measured $\Phi^{\rm j}$ for each individual 
$j$th field using equation~(\ref{eq-1vmax}). For each magnitude bin with 
$n \geqslant3$, we estimated the contribution to the error budget of 
$\Phi(M)$ from sample variance using:
\begin{equation} \label{eq-cv1}
\sigma_{\rm sv} = \frac{rms(\Phi^{\rm j})}{\sqrt{n}},
\end{equation}
with $n$ the number of individual fields used. For the magnitude bins with 
$n \leqslant2$ (usually the brightest bin and the 3-4 faintest ones), we 
adopted the mean of the $rms(\Phi^{\rm j})$ with $n \geqslant3$. The final 
1~$\sigma$ error associated to $\Phi(M)$ is then 
$\sigma=(\sigma_{\rm Poi}^{2}+\sigma_{\rm sv}^{2})^{1/2}$, with 
$\sigma_{\rm Poi}$ the Poisson error in each magnitude bin.


\subsection{The Maximum Likelihood Method} \label{sec-stymeth}

We also measured the observed LF using the STY method \citep{sandage79}, 
which is a parametric maximum likelihood estimator. The STY method has 
been shown to be unbiased with respect to density inhomogeneities (e.g., 
\citealt{efstathiou88}), it has well-defined asymptotic error properties 
(e.g. \citealt{kendall61}), and does not require binning of the data. The 
STY method assumes that $\Phi(M)$ has a universal form, i.e., the number 
density of galaxies is separable into a function of luminosity times a 
function of position: $n(M,{\bf x})=\Phi(M) \rho({\bf x})$. Therefore, the 
shape of $\Phi(M)$ is determined independently of its normalization. We 
have assumed that $\Phi(M)$ is described by a \citet{schechter76} function,

\begin{eqnarray}
\Phi (M) = (0.4 \ln{10}) \Phi^{\star} 
\big[ 10^{0.4 (M^{\star}-M)(1+\alpha)} \big] {} \nonumber\\
 \times \exp{\big[ -10^{0.4(M^{\star}-M)} \big]},
\end{eqnarray}
where $\alpha$ is the faint-end slope parameter, $M^{\star}$ is the 
characteristic absolute magnitude at which the LF exhibits a rapid 
change in the slope, and $\Phi^{\star}$ is the normalization. 

The probability of seeing a galaxy of absolute magnitude $M_{\rm i}$ at 
redshift $z_{\rm i}$ in a magnitude-limited catalog is given by

\begin{eqnarray}
p_{\rm i} \propto \frac{\Phi(M_{\rm i})}{\int_{M_{\rm faint}(z_{\rm i})}^{M_{\rm bright}(z_{\rm i})}{\Phi(M) dM}},
\end{eqnarray}
where $M_{\rm faint}(z_{\rm i})$ and $M_{\rm bright}(z_{\rm i})$ are 
the faintest and brightest absolute magnitudes observable at the redshift 
$z_{\rm i}$ in a magnitude-limited sample. The likelihood 
$\Lambda=\prod_{i=1}^{N}{p_{\rm i}}$ (where the product extends over 
all galaxies in the sample) is maximized with respect to the parameters 
$\alpha$ and $M^{\star}$ describing the LF $\Phi (M)$. The best-fit 
solution is obtained by minimizing $-2 \ln{\Lambda}$. A simple and 
accurate method of estimating errors is to determine the ellipsoid of 
parameter values defined by 
\begin{eqnarray}
\ln{\Lambda} = \ln{\Lambda_{\rm max}}-\frac{1}{2}\chi^{2}_{\rm \beta}(S),
\end{eqnarray}
where $\chi^{2}_{\rm \beta}(S)$ is the $\beta$-point of the $\chi^{2}$ 
distribution with $S$ degrees of freedom. Parameter $\chi^{2}_{\rm \beta}(S)$ 
is chosen in the standard way depending on the desired confidence level in 
the estimate (as described, e.g., by \citealt{avni76}; \citealt{lampton76}): 
$\chi^{2}_{\rm \beta}(2)=2.3$, 6.2, and 11.8 to estimate $\alpha-M^{\star}$ 
error contours with 68\%, 95\%, and 99\% confidence level (1, 2, and 
3~$\sigma$, respectively). The value of $\Phi^{\star}$ is then obtained by 
imposing a normalization on the best-fit LF such that the total number of 
observed galaxies in the composite sample is reproduced. The 1, 2, and 
3~$\sigma$ errors on $\Phi^{\star}$ are estimated from the minimum and 
maximum values of $\Phi^{\star}$ allowed by the 1, 2, and 3~$\sigma$ 
confidence contours in the $\alpha-M^{\star}$ parameter space, respectively.


\subsection{Uncertainties due to Photometric Redshift Errors} \label{sub-zphot}

Studies of high-redshift galaxies still largely rely on photometric 
redshift estimates. It is therefore important to understand how the 
photometric redshift uncertainties affect the derived LF and to 
quantify the systematic effects on the LF best-fit parameters.

\citet{chen03} have shown that at lower redshifts ($z \lesssim1$) the
measurement of the LF is strongly affected by errors associated with 
$z_{phot}$. Specifically, large redshift errors together with the
steep slope at the bright end of the galaxy LF tend to flatten the
observed LF and result in measured $M^{\star}$ systematically brighter
than the intrinsic value, since there are more intrinsically faint 
galaxies scattered into the bright end of the LF than 
intrinsically bright galaxies scattered into the faint end. Using
Monte Carlo simulations, \citet{chen03} obtained a best-fit $M^{\star}$ 
that was 0.8~mag brighter than the intrinsic value in the redshift 
range $0.5 \leq z \leq 0.8$. 

In order to quantify the systematic effect on the LF parameters
$\alpha$ and $M^{\star}$ in our redshift range of interest
($2 \leq z \leq 3.5$), we performed a series of Monte Carlo simulations. 
The details of these simulations and the results are presented in 
Appendix~\ref{app-1}. Briefly, we generated several model catalogs of 
galaxies of different brightness according to an input Schechter LF, 
extracted the redshifts of the objects from a probability distribution 
proportional to the comoving volume per unit redshift ($dV/dz$), and 
obtained the final mock catalogs after applying a limit in the observed 
apparent magnitude. To simulate the errors in the redshifts, we assumed 
a redshift error function parametrized as a Gaussian distribution function 
of 1 $\sigma$ width $\sigma^{\prime}_{\rm z}(1+z)$, with 
$\sigma^{\prime}_{\rm z}$ being the scatter in $\Delta z/(1+z_{\rm spec})$, 
and we formed the observed redshift catalog by perturbing the input galaxy 
redshift within the redshift error function. Finally, we determined the LF 
for the galaxies at $z_{1} \leq z \leq z_{2}$ using the $1/V_{\rm max}$ and 
maximum likelihood methods described in \S\S~\ref{sec-1vmaxmeth} and 
\ref{sec-stymeth}, respectively. As shown in Appendix~\ref{app-1}, the 
systematic effects on the measured $\alpha$ and $M^{\star}$ in the redshift 
interval $2 \leq z \leq 3.5$ are negligible with respect to the other 
uncertainties in the LF estimate if the errors on the photometric redshifts 
are characterized by a scatter in $\Delta z/(1+z_{\rm spec})$ of 
$\sigma^{\prime}_{\rm z} \sim 0.12$, which is the appropriate value for 
the $z \geq 2$ sample considered in this work. This is not true at $z<1$, 
where we find large systematic effects on both $M^{\star}$ and $\alpha$, 
consistent with \citet{chen03}. As explained in detail in 
Appendix~\ref{app-1}, the large systematic effects found at $z<1$ arise 
from the strong redshift dependency of both $dV/dz$ and $dM/dz$ at low-$z$; 
at $z>2$ these dependencies are much less steep, and this results in smaller 
systematic effects on the measured LF. From the Monte Carlo simulations we 
also quantified that the effects of photometric redshift errors on the 
estimated luminosity density are typically a few percent (always $<6$\%). 

We conclude that the parameters of the LF and the luminosity density 
estimates presented in this work are not significantly affected by the 
uncertainties in the photometric redshift estimates\footnote{In 
Appendix~\ref{app-1} we also investigated the effects of non-Gaussian 
redshift error probability distributions. Systematic outliers in the 
photometric redshift distribution can potentially cause systematic 
errors in the LF measurements, although these are much smaller than 
the random uncertainties in the LF estimates (if the outliers are 
randomly distributed).}. In order to include this contribution in the 
error budget, we conservatively assume a 10\% error contribution to the 
luminosity density error budget due to uncertainties in the photometric 
redshift estimates.


\section{THE OBSERVED LUMINOSITY FUNCTIONS} \label{sec-results}

\begin{figure}
\epsscale{0.95}
\plotone{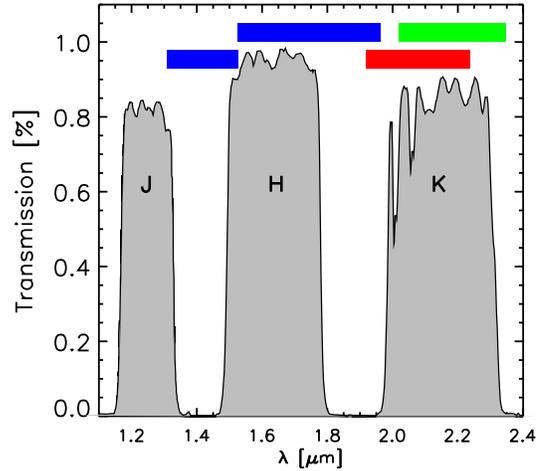}
\caption{Transmission curves of the $J$, $H$, and $K_{\rm s}$ bands 
({\it gray shaded regions}). The filled rectangles represent the observed 
wavelength ranges spanned by the central wavelengths of the rest-frame 
$R$ band at $2 \leq z \leq 2.5$ ({\it red}), of the rest-frame $V$ band at 
$2.7 \leq z \leq 3.3$ ({\it green}), and of the rest-frame $B$ band at 
$2 \leq z \leq 2.5$ and $2.5< z \leq 3.5$ ({\it blue; lower and upper 
rectangles, respectively}). For the considered redshift intervals, the 
rest-frame $R$ and $V$ bands correspond approximately to the observed 
$K_{\rm s}$ band. \label{fig-filters}}
\end{figure}

In this section we present the results of the measurement of the LF 
of galaxies at $z \geq 2$. We have measured the global LF in the rest-frame 
$R$ and $V$ band at redshift $2 \leq z \leq 2.5$ and $2.7 \leq z \leq 3.3$, 
respectively. As shown in Figure~\ref{fig-filters}, at these redshifts, 
the rest-frame $R$ and $V$ bands correspond approximately to the observed 
$K_{\rm s}$ band, which is the selection band of the composite sample. 
We also measured the global LF in the rest-frame $B$ band in the redshift 
interval $2 \leq z \leq 2.5$, to compare it with the rest-frame $R$-band LF, 
and at redshift $2.5 < z \leq 3.5$, to compare it with previous studies. 
For each redshift interval and rest-frame band we also split the 
sample based on the {\it observed} $J-K$ color ($J-K>2.3$, DRGs; $J-K\leq2.3$, 
non-DRGs) and the {\it rest-frame} $U-V$ color ($U-V \geq 0.25$, red 
galaxies; $U-V<0.25$, blue galaxies). In \S~\ref{subsec-lf} we present the 
global LF of all galaxies, and in \S~\ref{subsec-sublf} we present the LFs 
of the considered subsamples (DRGs, non-DRGs, red and blue galaxies) in the 
rest-frame $R$ band. The results for the rest-frame $V$ and $B$ bands are 
shown in Appendix~\ref{app-2}; in Appendix~\ref{app-3} we compare our results 
with those in the literature.

\subsection{Rest-Frame LF of All Galaxies} \label{subsec-lf}

\begin{figure*}[!t]
\centering
\includegraphics[width=15cm]{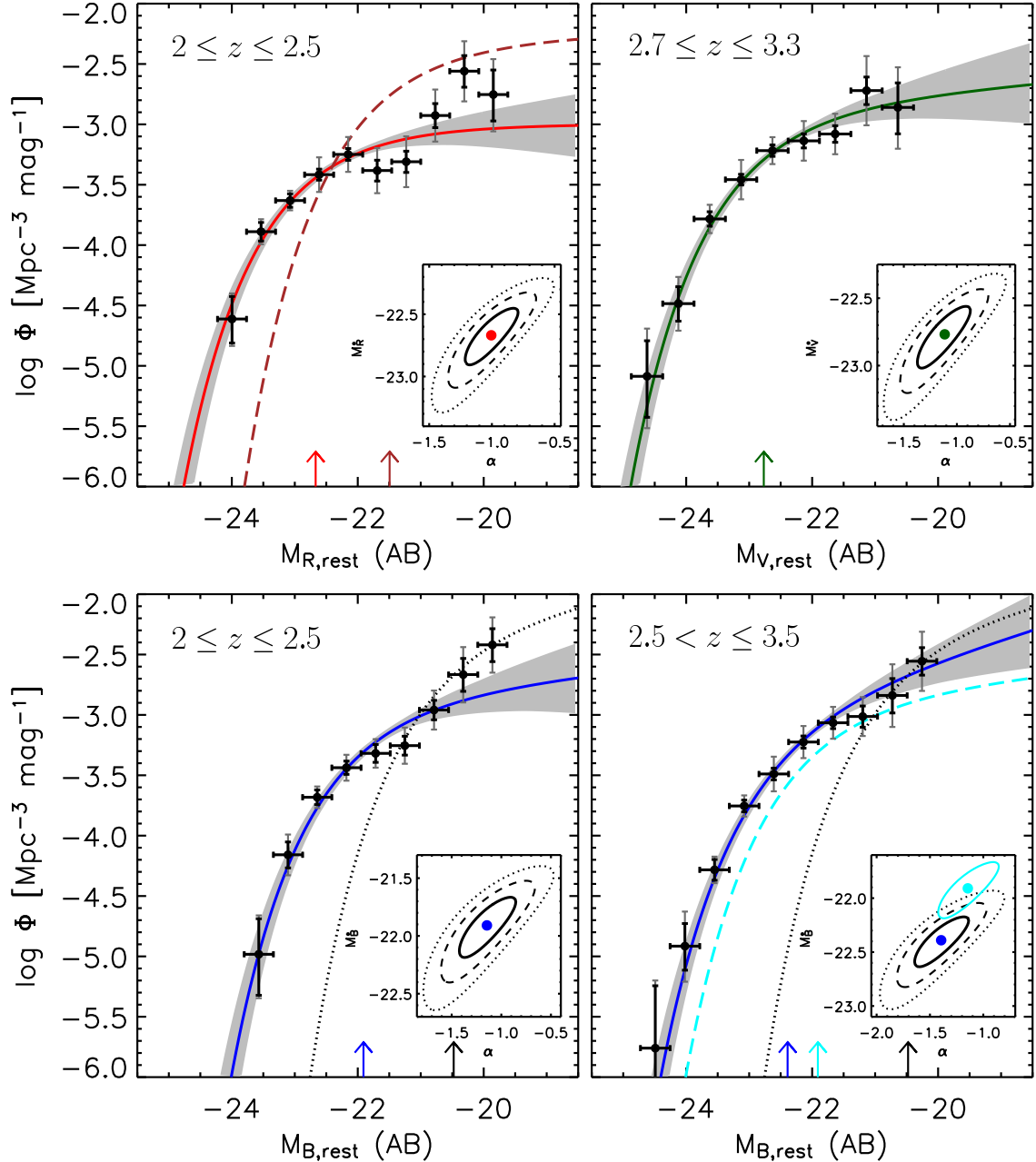}
\caption{\small {\it Top left:} Rest-frame $R$-band LF at 
$2 \leq z \leq 2.5$ for all galaxies; black filled circles 
are the $1/V_{\rm max}$ method estimates with 1~$\sigma$ 
error bars (Poisson errors only in black; field-to-field 
variation term included in gray); the red solid line is the 
LF estimated from the maximum likelihood method, and the red 
arrow represents the best-fit $M^{\star}$; the shaded region 
represents the 1~$\sigma$ uncertainties of the LF estimated 
from the maximum likelihood method; the brown dashed line 
and arrow represent the local $R$-band LF from SDSS 
(\citealt{blanton03}; $R=r-0.12$); the inset shows the best-fit 
value and the 1, 2, and 3~$\sigma$ confidence contour levels 
of the two parameters $\alpha$ and $M^{\star}$. {\it  Top right:} 
Rest-frame $V$-band global LF at $2.7 \leq z \leq 3.3$; the dark 
green solid line and the arrow represent the LF from the maximum 
likelihood method. {\it  Bottom left:} Rest-frame $B$-band global 
LF at $2 \leq z \leq 2.5$; the blue solid line and the arrow 
represent the LF from the maximum likelihood method; the black 
dotted line and arrow represent the local $B$-band LF from the 
2dFGRS (\citealt{norberg02}; $B=b_{\rm J}+0.12$). {\it  Bottom right:} 
Rest-frame $B$-band global LF at $2.5< z \leq 3.5$; the cyan dashed 
line and arrow represent the rest-frame $B$-band LF of all galaxies 
at $2 \leq z \leq 2.5$ for comparison; the black dotted line and 
arrow represent the local $B$-band LF from the 2dFGRS; in the inset, 
the 1~$\sigma$ confidence level of the estimated rest-frame $B$-band 
LF at $2 \leq z \leq 2.5$ is also plotted. 
\label{fig-LF_BVR_all}}
\end{figure*}

Figure~\ref{fig-LF_BVR_all} shows the global rest-frame $R$-  and $B$-band 
LFs for galaxies at $2 \leq z \leq 2.5$, the rest-frame $V$-band LF at
$2.7 \leq z \leq 3.3$, and the rest-frame $B$-band LF at $2.5< z \leq 3.5$. 
The large surveyed area of the composite sample allows the determination of 
the bright end of the optical LF at $z \geq 2$ with unprecedented accuracy, 
while the depth of FIRES allows us to constrain also the faint-end slope. 
This is particularly important because of the well-known correlation between 
the two parameters $\alpha$ and $M^{\star}$. The best-fit parameters with 
their 1, 2 and 3~$\sigma$ errors (from the maximum likelihood analysis) are 
listed in Table~\ref{tab-3}, together with the Schechter parameters of 
the local rest-frame $R$-band (from \citealt{blanton03}) and $B$-band 
(from \citealt{norberg02}) LFs.

\begin{deluxetable*}{lcccc}[!t]
\centering
\tablewidth{430pt}
\tablecaption{Best-fit Schechter Function Parameters for the Global LFs
\label{tab-3}}
\tablehead{
\colhead{} & \colhead{}     & 
           \colhead{$M^{\star}$} & 
           \colhead{}    & \colhead{$\Phi^{\star}$} \\
           \colhead{Redshift Range}  &   \colhead{Rest-Frame Band}         &
                    \colhead{(AB)}         &     \colhead{$\alpha$}  &  
                    \colhead{(10$^{-4}$~Mpc$^{-3}$~mag$^{-1}$)}}
\startdata
$2.0 \leq z \leq 2.5$ & $R$ & $-22.67^{+0.20,0.35,0.45}_{-0.22,0.41,0.61}$ & 
                    $-1.01^{+0.21,0.35,0.49}_{-0.20,0.34,0.47}$  & 
                    $10.65^{+2.65,4.50,5.92}_{-2.67,4.40,5.93}$ \\
$z=0$     & $R$  &  $-21.50\pm0.01$ & $-1.05\pm0.01$ & $51.1\pm1.4$ \\
$2.7 \leq z \leq 3.3$ & $V$ & $-22.77^{+0.20,0.33,0.45}_{-0.24,0.43,0.63}$ & 
                     $-1.12^{+0.23,0.40,0.57}_{-0.25,0.41,0.57}$ & 
                     $15.07^{+4.12,6.84,9.45}_{-4.41,7.03,9.28}$ \\
$2.0 \leq z \leq 2.5$ & $B$ & $-21.91^{+0.22,0.39,0.51}_{-0.26,0.49,0.73}$ & 
                     $-1.15^{+0.28,0.48,0.66}_{-0.27,0.45,0.63}$ & 
                     $14.57^{+4.60,8.07,11.07}_{-4.64,7.53,9.84}$ \\
$2.5 < z \leq 3.5$ & $B$ & $-22.39^{+0.20,0.35,0.45}_{-0.24,0.43,0.63}$ & 
                     $-1.40^{+0.25,0.42,0.58}_{-0.24,0.41,0.56}$ & 
                     $13.56^{+4.69,8.38,11.49}_{-4.51,7.08,9.17}$ \\
  $z=0$      & $B$ &  $-20.48\pm0.07$ & $-1.24\pm0.03$ & $62.9\pm3.1$ \\
\enddata
\tablecomments{The quoted errors correspond to the 1, 2, and 3~$\sigma$ 
errors estimated from the maximum likelihood analysis as described in 
\S~\ref{sec-stymeth}. The $z=0$ values are the Schechter parameters of 
the local rest-frame $R$-band (from \citealt{blanton03}) and $B$-band 
(from \citealt{norberg02}) LFs. We assumed $R_{\rm AB}=r-0.12$ and 
$B_{\rm AB}=b_{\rm J}+0.12$ to convert the local values into our 
photometric system.}
\end{deluxetable*}

At redshift $2 \leq z \leq 2.5$, the faint-end slope of the rest-frame 
$R$-band LF is slightly flatter than in the rest-frame $B$-band, although 
the difference is within the errors. In the two higher redshift bins, 
the faint-end slope of the rest-frame $V$-band LF is flatter (by $\sim0.3$) 
than in the rest-frame $B$-band, although the difference is only at the 
1~$\sigma$ level. Similarly, the faint-end slopes of the rest-frame 
$B$-band global LF in the low- and high-redshift bins are statistically 
identical. The characteristic magnitude $M^{\star}_{\rm B}$ in the low-$z$ 
interval is about 0.5~mag fainter with respect to the high-redshift one, 
although the difference is significant only at the $\sim$1.5~$\sigma$ level. 
We therefore conclude that the rest-frame $B$-band global LFs in the low- 
and high-redshift bins are consistent with no evolution within their 
errors ($<2$~$\sigma$).

In Figure~\ref{fig-LF_BVR_all} we have also plotted the local ($z \sim 0.1$) 
rest-frame $R$-band (from \citealt{blanton03}) and $B$-band (from 
\citealt{norberg02}) LFs. The faint-end slope of the $R$-band LF at 
$2 \leq z \leq 2.5$ is very similar to the faint-end slope of the local 
LF; the characteristic magnitude is instead significantly ($>3$~$\sigma$) 
brighter than the local value (by $\sim1.2$~mag), and the characteristic 
density is a factor of $\sim4.8$ smaller than the local value. The 
rest-frame $B$-band LF at $2.5< z \leq 3.5$ is characterized by a 
faint-end slope consistent with the local $B$-band LF; the characteristic 
magnitude is significantly brighter ($>3$~$\sigma$) than the local value 
by $\sim1.9$~mag, while the characteristic density is a factor of 
$\sim4.6$ smaller with respect to the local value.


\subsection{Rest-Frame LF of DRGs, Non-DRGs, and Red and Blue Galaxies} 
\label{subsec-sublf}

\begin{figure}[!b]
\epsscale{1}
\plotone{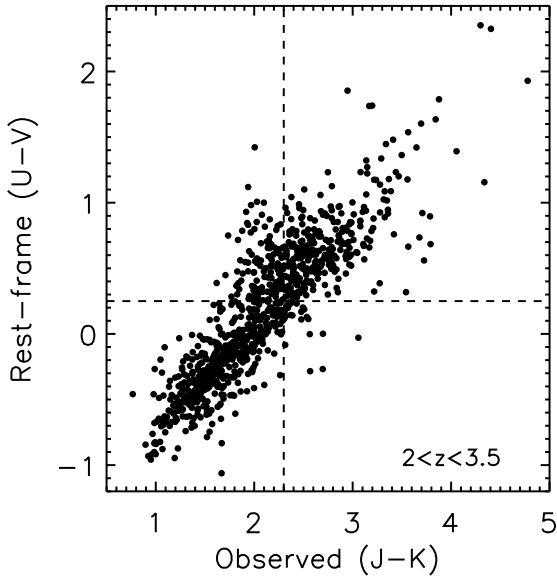}
\caption{Rest-frame $U-V$ color vs. observed $J-K$ color for the 
composite sample at $2 \leq z \leq 3.5$; the dashed vertical line 
represents the $J-K>2.3$ selection of DRGs, while the horizontal 
dashed line shows the $U-V \geq 0.25$ definition of red galaxies 
adopted in this paper.
\label{fig3_ref}}
\end{figure}

In this section we present the results of the LFs for different 
subsamples, by splitting the composite sample based on the observed 
$J-K$ color ($J-K>2.3$, DRGs; \citealt{franx03}) and on the rest-frame 
$U-V$ color (by defining the red galaxies as those having $U-V \geq 0.25$, 
which is the median value of $U-V$ of the composite sample at 
$2 \leq z \leq 2.5$). In Figure~\ref{fig3_ref}, we show the rest-frame 
$U-V$ color versus the observed $J-K$ color for the composite sample at 
$2 \leq z \leq 3.5$.

\begin{figure*}[!t]
\centering
\includegraphics[width=15cm]{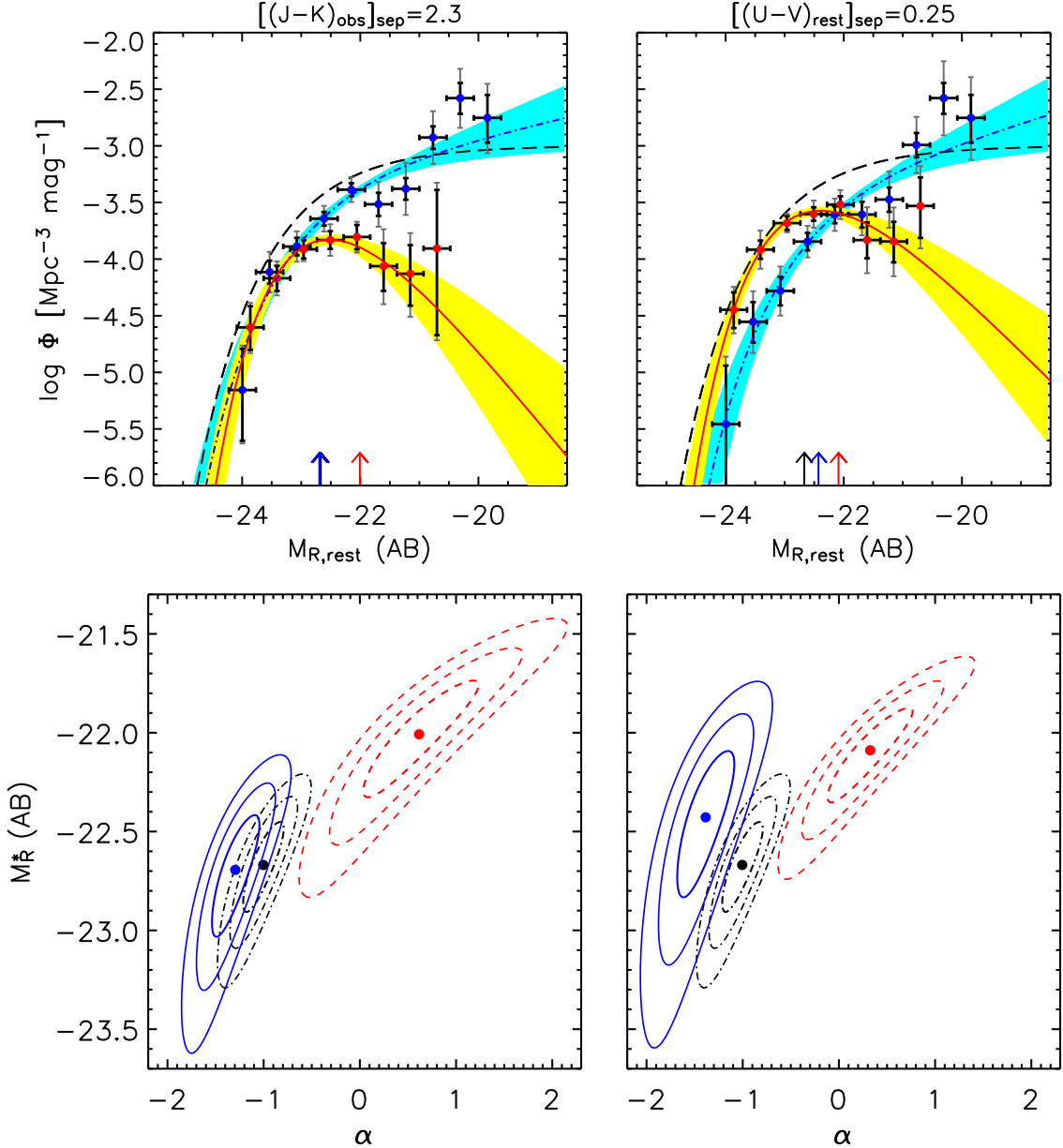}
\caption{Rest-frame R-band LF at $2 \leq z \leq 2.5$. {\it Top:} 
Rest-frame $R$-band LFs at $2 \leq z \leq 2.5$ for DRGs/non-DRGs 
({\it left}) and red/blue galaxies ({\it right}). The red solid 
line and filled circles represent the LF of DRGs and red galaxies 
estimated with the STY and $1/V_{\rm max}$ methods, respectively; 
the blue dot-dashed line and filled circles represent the LF of 
non-DRGs and blue galaxies. The rest-frame $R$-band LF of all 
galaxies is also plotted ({\it black dashed line}). The shaded 
region (yellow for DRGs/red galaxies, cyan for non-DRGs/blue 
galaxies) represents the 1~$\sigma$ uncertainties of the LF measured 
with the STY method. The arrows represent the best-fit $M^{\star}$; 
error bars as in Fig.~\ref{fig-LF_BVR_all}. {\it Bottom:} 1, 2, and 
3~$\sigma$ contour levels from the STY method for DRGs and red galaxies 
({\it red dashed lines}), for non-DRGs and blue galaxies ({\it blue 
solid lines}), and for all galaxies ({\it black dot-dashed lines}); 
the filled circles represent the best-fit Schechter values. The LF of 
non-DRGs (blue galaxies) is steeper than that of DRGs (red galaxies). 
\label{LF_R_lowz.ps}}
\end{figure*}

In Figure~\ref{LF_R_lowz.ps} we show the rest-frame $R$-band LF at 
$2 \leq z \leq 2.5$ of DRGs versus non-DRGs and red versus blue galaxies, 
together with the 1, 2, and 3~$\sigma$ contour levels in the 
$\alpha-M^{\star}_{\rm R,AB}$ parameter space from the STY analysis. The 
LFs of the different subsamples in the rest-frame $V$ band at 
$2.7 \leq z \leq 3.3$ and in the rest-frame $B$ band at $2 \leq z \leq 2.5$ 
and $2.5< z \leq 3.5$ are shown in Appendix~\ref{app-2} in 
Figures.~\ref{LF_V_highz.ps}, \ref{LF_B_lowz.ps}, and \ref{LF_B_highz.ps}, 
respectively. In Table~\ref{tab-4} the best-fit parameters and their 1, 2, 
and 3~$\sigma$ errors from the STY method are listed for all the considered 
rest-frame bands and redshift intervals.

\begin{deluxetable*}{cccccc}[!t]
\centering
\tablewidth{500pt}
\tabletypesize{\small}
\tablecaption{Best-Fit Schechter Function Parameters for Subsamples 
\label{tab-4}}
\tablehead{\colhead{} & \colhead{}     & 
           \colhead{}         & \colhead{$M^{\star}$}    & 
           \colhead{}       & \colhead{$\Phi^{\star}$} \\
           \colhead{Redshift Range}  &  \colhead{Rest-Frame Band}       &
            \colhead{Sample}      &          \colhead{(AB)}            &
          \colhead{$\alpha$} &  \colhead{(10$^{-4}$~Mpc$^{-3}$~mag$^{-1}$)} }
\startdata
$2.0 \leq z \leq 2.5$ & $R$ & $J-K>2.3$ & $-22.01^{+0.26,0.43,0.57}_{-0.30,0.55,0.81}$ &
                                  $0.62^{+0.62,1.07,1.54}_{-0.57,0.92,1.24}$ &
                                  $3.74^{+0.69,1.02,1.37}_{-1.09,1.90,2.58}$ \\
            &     & $J-K\leq2.3$ & $-22.69^{+0.27,0.43,0.57}_{-0.33,0.60,0.93}$ &
				 $-1.30^{+0.24,0.41,0.57}_{-0.24,0.40,0.55}$ &
				  $6.32^{+2.72,4.59,6.30}_{-2.41,3.76,4.78}$ \\
            &     & $U-V \geq 0.25$ & $-22.09^{+0.20,0.35,0.47}_{-0.24,0.45,0.65}$ &
				   $0.33^{+0.43,0.76,1.09}_{-0.44,0.70,0.95}$ &
				  $7.51^{+0.69,1.14,1.57}_{-1.14,2.19,3.31}$ \\
            &     & $U-V<0.25$ & $-22.43^{+0.32,0.51,0.69}_{-0.40,0.75,1.15}$ &
				  $-1.39^{+0.29,0.49,0.68}_{-0.29,0.48,0.67}$ &
				  $5.26^{+2.97,4.97,7.32}_{-2.42,3.64,4.47}$ \\

$2.7 \leq z \leq 3.3$ & $V$ & $J-K>2.3$ & $-22.63^{+0.28,0.47,0.63}_{-0.36,0.65,0.99}$ &
                                 $-0.46^{+0.52,0.89,1.27}_{-0.49,0.81,1.10}$ &
                                  $6.14^{+1.10,1.81,2.56}_{-1.71,3.09,4.34}$ \\
            &     & $J-K\leq2.3$ & $-22.63^{+0.24,0.41,0.55}_{-0.32,0.57,0.87}$ &
				  $-1.21^{+0.30,0.50,0.71}_{-0.31,0.50,0.69}$ &
				 $11.14^{+4.02,7.12,9.88}_{-4.27,6.61,8.41}$ \\
            &     & $U-V \geq 0.25$ & $-22.43^{+0.28,0.45,0.61}_{-0.32,0.59,0.87}$ &
				  $-0.11^{+0.60,1.02,1.46}_{-0.53,0.87,1.19}$ &
				  $6.25^{+0.87,1.54,2.24}_{-0.92,2.08,3.40}$ \\
            &     & $U-V<0.25$ & $-22.65^{+0.26,0.43,0.57}_{-0.30,0.57,0.85}$ &
				  $-1.26^{+0.31,0.51,0.73}_{-0.29,0.49,0.68}$ &
				$11.01^{+4.50,7.72,10.55}_{-4.16,6.52,8.36}$ \\
$2.0 \leq z \leq 2.5$ & $B$ & $J-K>2.3$  & $-21.75^{+0.42,0.67,0.87}_{-0.54,1.03,1.73}$ &
                                  $-0.54^{+0.73,1.25,1.79}_{-0.66,1.08,1.47}$ &
				  $4.12^{+1.29,2.18,3.18}_{-1.77,2.99,3.78}$ \\
            &     & $J-K\leq2.3$  & $-21.89^{+0.26,0.43,0.57}_{-0.32,0.59,0.89}$ &
				  $-1.25^{+0.31,0.53,0.73}_{-0.31,0.52,0.72}$ &
				$11.08^{+4.52,7.83,10.92}_{-4.36,6.75,8.57}$ \\
            &     & $U-V \geq 0.25$ & $-21.63^{+0.30,0.49,0.65}_{-0.38,0.69,1.05}$ &
				  $-0.40^{+0.57,0.99,1.41}_{-0.56,0.91,1.24}$ &
				  $8.63^{+1.57,2.70,3.95}_{-2.33,4.35,6.20}$ \\
            &     & $U-V<0.25$ & $-21.83^{+0.32,0.51,0.67}_{-0.42,0.77,1.19}$ &
				  $-1.33^{+0.34,0.58,0.81}_{-0.35,0.58,0.81}$ &
				 $8.19^{+4.47,7.70,10.97}_{-3.94,5.83,7.09}$ \\
$2.5< z \leq 3.5$ & $B$ & $J-K>2.3$  & $-22.35^{+0.32,0.51,0.67}_{-0.38,0.73,1.17}$ &
                                  $-1.11^{+0.51,0.86,1.21}_{-0.49,0.81,1.14}$ &
				  $4.88^{+2.09,3.35,4.55}_{-2.20,3.48,4.32}$ \\
            &     & $J-K\leq2.3$  & $-22.35^{+0.26,0.43,0.57}_{-0.30,0.57,0.87}$ &
				  $-1.45^{+0.30,0.51,0.71}_{-0.29,0.48,0.67}$ &
				 $9.50^{+4.53,7.95,11.13}_{-3.92,6.04,7.57}$ \\
            &     & $U-V \geq 0.25$ & $-22.25^{+0.32,0.51,0.67}_{-0.38,0.75,1.17}$ &
				  $-1.01^{+0.55,0.92,1.31}_{-0.52,0.87,1.21}$ &
				  $5.37^{+2.05,3.25,4.49}_{-2.34,3.77,4.73}$ \\
            &     & $U-V<0.25$ & $-22.41^{+0.26,0.43,0.57}_{-0.30,0.57,0.87}$ &
				  $-1.49^{+0.30,0.50,0.70}_{-0.27,0.47,0.65}$ &
				 $8.80^{+4.33,7.60,10.84}_{-3.64,5.62,7.00}$ \\
\enddata
\tablecomments{The quoted errors correspond to the 1, 2, and 
3~$\sigma$ errors estimated from the maximum likelihood analysis 
as described in \S~\ref{sec-stymeth}.}
\end{deluxetable*}

As shown in Figure~\ref{LF_R_lowz.ps}, the rest-frame $R$-band LF at 
$2 \leq z \leq 2.5$ of DRGs is significantly ($>3$~$\sigma$) different 
from that of non-DRGs. The faint-end slope of the non-DRG LF is much 
steeper, indicating that the contribution of DRGs to the global luminosity 
and number density at faint luminosities is very small compared to that 
of non-DRGs. The bright end of the DRG LF is instead very similar to that 
of non-DRGs, with the two subsamples contributing equally to the global LF. 
Splitting the composite sample based on the rest-frame $U-V$ color, we 
find a qualitatively similar result, with the faint-end slope of the 
blue galaxy LF being much steeper than that of red galaxies (although 
the red galaxies clearly dominate the bright end of the LF). The 
difference between the LFs of DRGs (red galaxies) and non-DRGs 
(blue galaxies) is mainly driven by the different faint-end slopes. 

A similar result holds in the rest-frame $V$ band at $2.7 \leq z \leq 3.3$, 
although it is slightly less significant (at the 2-3~$\sigma$ level): the 
non-DRG (blue galaxy) LF is very similar to that of DRGs (red galaxies) at 
the bright end, while at the faint end, the LF of non-DRGs (blue galaxies) 
is steeper than that of DRGs (red galaxies). In the rest-frame $B$ band, 
the differences between the LFs of DRGs/red galaxies and non-DRGs/blue 
galaxies become even less significant. Although DRGs/red galaxies are 
always characterized by LFs with flatter faint-end slopes, the significance 
of this result is only marginal ($<2$~$\sigma$), especially in the higher 
redshift interval. 

Within our sample, there is marginal evidence for evolution with redshift:
the rest-frame $B$-band non-DRG/blue galaxy LFs in the two targeted 
redshift bins are characterized by similar (within the errors) faint-end 
slopes, while the characteristic magnitude is brighter by 
$\sim0.5-0.6$~mag in the higher redshift bin. The LF of DRGs/red 
galaxies tends to get steeper from low to high redshifts and 
$M^{\star}$ gets brighter by $\sim0.6$~mag. However, because of the 
large uncertainties (especially for DRGs and red galaxies) on the 
measured Schechter parameters, the differences in the rest-frame 
$B$ band between the high- and the low-redshift bins are at most 
at the 2~$\sigma$ significance level.

We note that the uncertainties on the estimated Schechter parameters 
mainly arise from the small number statistics at the very faint end, 
which is probed only by FIRES. Very deep (down to the deepest FIRES) 
NIR imaging over large spatially disjoint fields is required for 
further progress in our understanding of the lowest luminosity 
galaxies at $z>2$.


\subsection{Comparison with LBGs} \label{sec-lbgs}

\citet{shapley01} computed the rest-frame optical ($V$ band) LF of 
$z\sim3$ LBGs using the distribution of optical $\cal{R}$ magnitudes 
(i.e., the rest-frame UV LF) and the distribution of $\cal{R}$-$K_{\rm s}$ 
colors as a function of $\cal{R}$ magnitude. The rest-frame UV LF of LBGs 
was taken from \citet{adelberger00} with best-fit Schechter parameters 
$\alpha=-1.57\pm0.11$, $M^{\star}_{\rm \cal{R}}=24.54\pm0.14$~mag, and 
$\Phi^{\star}=(1.5\pm0.4)\times10^{-3}$~Mpc$^{-3}$ in our adopted 
cosmology. \citet{shapley01} detected a correlation with 98\% confidence 
between $\cal{R}$-$K_{\rm s}$ color and $\cal{R}$ magnitude, such that 
fainter galaxies have redder $\cal{R}$-$K_{\rm s}$ colors. This trend was 
included in their LF analysis by using the relationship implied by the 
best-fit regression slope to the correlation, 
$d \langle {\cal{R}}-K_{\rm s} \rangle / d {\cal{R}}=0.17$ (the scatter 
around this regression is very large). The Schechter function was then 
fitted to the average LF values, obtaining best-fit Schechter parameters 
$\alpha=-1.85\pm0.15$, $M^{\star}_{\rm V,AB}=-22.99\pm0.25$~mag, and 
$\Phi^{\star}=(6.2\pm2.7)\times10^{-4}$~Mpc$^{-3}$. The overall shape of 
the rest-frame optical LF of LBGs is determined by the way in which the 
$\cal{R}$-$K_{\rm s}$ distribution as a function of $\cal{R}$ magnitude 
redistributes $\cal{R}$ magnitudes into $K_{\rm s}$ magnitudes. Therefore, 
as a result of the detected positive correlation between $\cal{R}$ and 
$\cal{R}$-$K_{\rm s}$, the faint-end slope of the LBG rest-frame optical 
LF is steeper than that of the UV LF \citep{shapley01}.

\begin{figure}[!b]
\epsscale{1.0}
\plotone{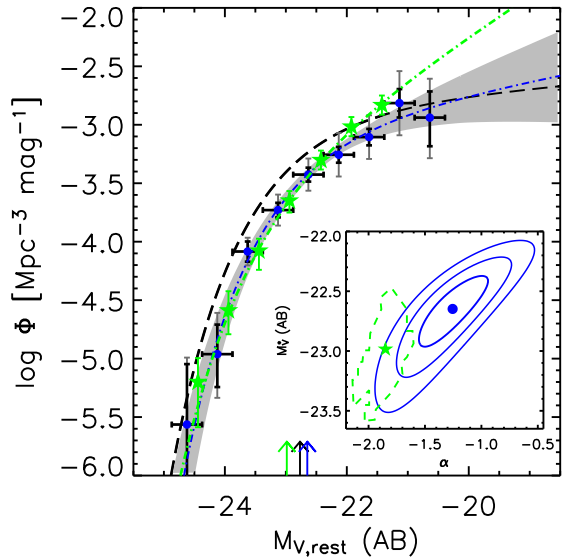}
\caption{Rest-frame $V$-band LF of blue galaxies at redshift 
$2.7 \leq z \leq 3.3$ ({\it blue dot-dashed line and filled 
circles}) and $z\sim3$ LBG LF \citep{shapley01} ({\it green 
dot-dashed and filled stars}); the shaded area represents the 
1~$\sigma$ uncertainties of the blue galaxy LF measured with 
the STY method; the global LF measured with the STY method is 
also plotted as a black long-dashed line; arrows represent the 
best-fit $M^{\star}$. The inset shows the best-fit value and the 
1, 2, and 3~$\sigma$ confidence contour levels of the two parameters 
$\alpha$ and $M^{\star}$ for blue galaxies ({\it blue filled circle 
and solid lines}) and the 68.3\% confidence intervals for the LBG 
best-fit Schechter parameters from \citep{shapley01} ({\it green 
filled star and dashed line}). \label{V_drg_lbg.ps}}
\end{figure}

In Figure~\ref{V_drg_lbg.ps} we compare the rest-frame $V$-band LF of blue 
galaxies at $2.7 \leq z \leq 3.3$ and the LBG LF from \citet{shapley01} in 
the same rest-frame band and redshift interval. The blue galaxy LF estimated 
with the $1/V_{\rm max}$ method appears consistent within the errors with 
the average LF values of LBGs (shown as stars in Figure~\ref{V_drg_lbg.ps}). 
However, the best-fit Schechter parameters from the maximum likelihood 
analysis are only marginally consistent, with the faint-end slope of the 
LBG LF being significantly steeper than the one of blue galaxies, as shown 
in the inset of Figure~\ref{V_drg_lbg.ps}. The same result is obtained if 
the rest-frame $V$-band LF of non-DRGs (rather than rest-frame blue galaxies) 
is compared to that of LBGs.

\subsection{Comparison with Previous Studies} \label{sec-liter}

In Appendix~\ref{app-3} we compare our results with previously 
published LFs. Specifically, we have compared our rest-frame 
$B$-band LF with that derived by \cite{poli03}, \cite{giallongo05}, 
and \cite{gabasch04} in the redshift intervals $2 \leq z \leq 2.5$ 
and $2.5 < z \leq 3.5$, and our rest-frame $R$-band LF with the 
rest-frame $r$-band LF derived by \cite{gabasch06} at 
$2 \leq z \leq 2.5$. We also compared our rest-frame $B$-band LFs 
of red and blue galaxies at $2.5 < z \leq 3.5$ with those measured 
by \cite{giallongo05}.


\section{Densities} \label{sec-densities}

\subsection{Number Density and Field-to-Field Variance} \label{sec-numdens}

The estimates of the number density $\eta$ obtained by integrating 
the best-fit Schechter function to the faintest observed rest-frame 
luminosity are listed in Table~\ref{tab-ndall}. For completeness, 
we also list $\eta^{\rm MUSYC}$, calculated by integrating the 
best-fit Schechter LF to the rest-frame magnitude limits of the 
NIR MUSYC, and $\eta^{+2}$, calculated by integrating the best-fit 
Schechter LF to 2~mag fainter than the faintest observed luminosities.
We find that the contribution of DRGs (red galaxies) to the total 
number density down to the faintest probed rest-frame luminosities 
is 13\%-25\% (18\%-29\%) depending on the redshift interval. By 
integrating the rest-frame $B$-band LF down to a fixed rest-frame 
magnitude limit ($M_{\rm B,AB}=-20$), we find a hint of an increase 
of the contribution of blue galaxies from the low redshift bin (62\%) 
to the higher bin (74\%), but the differences are not significant. If 
only the bright end of the LF is considered (integrating the LF down 
to the fixed NIR MUSYC limit, $M_{\rm B,AB}=-22.2$), the increase of 
the contribution of the blue galaxy population becomes significant at 
the 2~$\sigma$ level, going from 42\% in the low-$z$ bin to 66\% in 
the high-$z$ bin.

\begin{deluxetable*}{lccccc}
\centering
\tablewidth{500pt}
\tabletypesize{\small}
\tablecaption{Number Densities \label{tab-ndall}}
\tablehead{\colhead{}      & \colhead{}    &
    \colhead{} & \colhead{\tablenotemark{a}$\log{\eta}$} & 
                       \colhead{\tablenotemark{b}$\log{\eta^{\rm MUSYC}}$} &
                       \colhead{\tablenotemark{c}$\log{\eta^{+2}}$} \\
         \colhead{Redshift Range}   &     \colhead{Rest-Frame Band}     &
 \colhead{Sample} &   \colhead{(Mpc$^{-3}$)}  & \colhead{(Mpc$^{-3}$)}  & 
        \colhead{(Mpc$^{-3}$)}  }
\startdata
$2.0 \leq z \leq 2.5$ & $R$ & All        & $-2.61^{+0.07,0.11,0.16}_{-0.06,0.10,0.13}$ 
                               & $-3.18^{+0.02,0.03,0.04}_{-0.02,0.03,0.04}$ 
                               & $-2.36^{+0.17,0.29,0.42}_{-0.13,0.20,0.27}$ \\
            &     & $J-K>2.3$  & $-3.48^{+0.07,0.14,0.21}_{-0.05,0.08,0.10}$ 
                               & $-3.64^{+0.02,0.03,0.04}_{-0.02,0.03,0.04}$ 
                               & $-3.48^{+0.09,0.19,0.31}_{-0.06,0.09,0.11}$ \\
            &     & $J-K\leq2.3$  & $-2.63^{+0.08,0.14,0.20}_{-0.08,0.13,0.17}$ 
                               & $-3.37^{+0.02,0.03,0.04}_{-0.03,0.05,0.07}$ 
                               & $-2.24^{+0.25,0.42,0.59}_{-0.18,0.29,0.39}$ \\
            &     & $U-V \geq 0.25$ & $-3.19^{+0.07,0.12,0.17}_{-0.05,0.08,0.10}$ 
                               & $-3.39^{+0.02,0.03,0.04}_{-0.02,0.03,0.04}$ 
                               & $-3.18^{+0.09,0.17,0.28}_{-0.06,0.09,0.11}$ \\
            &     & $U-V<0.25$ & $-2.72^{+0.10,0.16,0.22}_{-0.09,0.14,0.20}$ 
                               & $-3.59^{+0.03,0.05,0.06}_{-0.04,0.08,0.11}$ 
                               & $-2.25^{+0.31,0.52,0.73}_{-0.21,0.34,0.46}$ \\
$2.7 \leq z \leq 3.3$ & $V$ & All        & $-2.51^{+0.07,0.11,0.16}_{-0.06,0.10,0.14}$ 
                               & $-3.27^{+0.02,0.03,0.05}_{-0.03,0.05,0.07}$ 
                               & $-2.16^{+0.19,0.33,0.47}_{-0.15,0.24,0.32}$ \\
            &     & $J-K>2.3$  & $-3.17^{+0.12,0.21,0.30}_{-0.10,0.16,0.20}$ 
                               & $-3.65^{+0.02,0.03,0.04}_{-0.03,0.06,0.10}$ 
                               & $-3.05^{+0.26,0.50,0.76}_{-0.17,0.25,0.31}$ \\
            &     & $J-K\leq2.3$  & $-2.63^{+0.08,0.13,0.18}_{-0.07,0.12,0.16}$ 
                               & $-3.49^{+0.03,0.05,0.06}_{-0.04,0.07,0.11}$ 
                               & $-2.22^{+0.24,0.41,0.59}_{-0.19,0.30,0.40}$ \\
            &     & $U-V \geq 0.25$ & $-3.25^{+0.11,0.20,0.28}_{-0.09,0.14,0.18}$ 
                               & $-3.65^{+0.02,0.03,0.04}_{-0.03,0.05,0.08}$ 
                               & $-3.19^{+0.21,0.40,0.64}_{-0.13,0.19,0.24}$ \\
            &     & $U-V<0.25$ & $-2.61^{+0.08,0.13,0.18}_{-0.07,0.12,0.16}$ 
                               & $-3.49^{+0.03,0.05,0.07}_{-0.04,0.07,0.11}$ 
                               & $-2.17^{+0.24,0.41,0.59}_{-0.20,0.32,0.41}$ \\
$2.0 \leq z \leq 2.5$ & $B$ & All        & $-2.53^{+0.08,0.13,0.19}_{-0.08,0.12,0.17}$ 
                               & $-3.38^{+0.03,0.05,0.06}_{-0.03,0.06,0.09}$ 
                               & $-2.15^{+0.25,0.42,0.60}_{-0.17,0.28,0.37}$ \\
            &     & $J-K>2.3$  & $-3.34^{+0.17,0.30,0.42}_{-0.13,0.21,0.27}$ 
                               & $-3.92^{+0.04,0.05,0.07}_{-0.05,0.11,0.17}$ 
                               & $-3.19^{+0.42,0.80,1.23}_{-0.23,0.33,0.40}$ \\
            &     & $J-K\leq2.3$  & $-2.60^{+0.09,0.15,0.21}_{-0.08,0.14,0.19}$ 
                               & $-3.53^{+0.04,0.06,0.07}_{-0.05,0.08,0.12}$ 
                               & $-2.17^{+0.30,0.51,0.73}_{-0.20,0.33,0.43}$ \\
            &     & $U-V \geq 0.25$ & $-3.07^{+0.13,0.23,0.33}_{-0.11,0.17,0.22}$ 
                               & $-3.63^{+0.03,0.04,0.06}_{-0.04,0.08,0.12}$ 
                               & $-2.95^{+0.29,0.58,0.88}_{-0.18,0.26,0.32}$ \\
            &     & $U-V<0.25$ & $-2.71^{+0.10,0.16,0.23}_{-0.09,0.15,0.21}$ 
                               & $-3.71^{+0.04,0.07,0.09}_{-0.06,0.10,0.16}$ 
                               & $-2.23^{+0.35,0.59,0.83}_{-0.22,0.37,0.49}$ \\
$2.5< z \leq 3.5$ & $B$ & All        & $-2.43^{+0.07,0.12,0.17}_{-0.07,0.12,0.16}$ 
                               & $-3.46^{+0.03,0.05,0.06}_{-0.04,0.06,0.09}$ 
                               & $-1.92^{+0.23,0.39,0.54}_{-0.18,0.30,0.40}$ \\
            &     & $J-K>2.3$  & $-3.04^{+0.15,0.26,0.36}_{-0.14,0.22,0.29}$ 
                               & $-3.89^{+0.04,0.07,0.09}_{-0.06,0.12,0.19}$ 
                               & $-2.68^{+0.42,0.76,1.10}_{-0.32,0.47,0.58}$ \\
            &     & $J-K\leq2.3$  & $-2.57^{+0.08,0.14,0.19}_{-0.08,0.13,0.18}$ 
                               & $-3.65^{+0.04,0.06,0.08}_{-0.04,0.08,0.12}$ 
                               & $-2.03^{+0.27,0.46,0.65}_{-0.22,0.35,0.47}$ \\
            &     & $U-V \geq 0.25$ & $-3.06^{+0.16,0.27,0.38}_{-0.14,0.22,0.29}$ 
                               & $-3.90^{+0.05,0.07,0.09}_{-0.07,0.12,0.20}$ 
                               & $-2.74^{+0.44,0.78,1.14}_{-0.31,0.45,0.56}$ \\
            &     & $U-V<0.25$ & $-2.56^{+0.08,0.14,0.19}_{-0.08,0.13,0.18}$ 
                               & $-3.64^{+0.04,0.06,0.08}_{-0.04,0.08,0.12}$ 
                               & $-2.00^{+0.26,0.45,0.64}_{-0.22,0.35,0.47}$ \\
\enddata
\tablenotetext{a}{The number densities are estimated by integrating 
the best-fit Schechter LF down to the faintest observed rest-frame 
luminosities: $M^{\rm last}_{\rm R,AB}=-19.6$, 
$M^{\rm last}_{\rm V,AB}=-20.3$ and $M^{\rm last}_{\rm B,AB}=-19.5$, 
$M^{\rm last}_{\rm B,AB}=-20.0$ for the low- and high-redshift intervals, 
respectively; they  correspond to observed $K_{\rm s}$ band magnitude of 
$K^{\rm tot}_{\rm s}\approx23.14$ at the lower limit of the targeted 
redshift intervals. The quoted errors correspond to the 1, 2, and 
3 $\sigma$ errors estimated from the maximum likelihood analysis as 
described in \S~\ref{sec-stymeth}.}
\tablenotetext{b}{$\eta^{\rm MUSYC}$ is the number density estimated by 
integrating the best-fit Schechter LF down to the rest-frame magnitude 
limits of the deep NIR MUSYC survey: $M^{\rm MUSYC}_{\rm R,AB}=-21.8$, 
$M^{\rm MUSYC}_{\rm V,AB}=-22.4$ and $M^{\rm MUSYC}_{\rm B,AB}=-21.7$, 
$-22.2$ for the low- and high-redshift intervals, respectively 
(corresponding to observed $K^{\rm tot}_{\rm s} \approx 21.09$).}
\tablenotetext{c}{$\eta^{+2}$ is the number density estimated by 
integrating the best-fit Schechter LF down to 2~mag fainter than the 
faintest observed luminosities.}
\end{deluxetable*}

\begin{figure}[!bh]
\epsscale{0.95}
\plotone{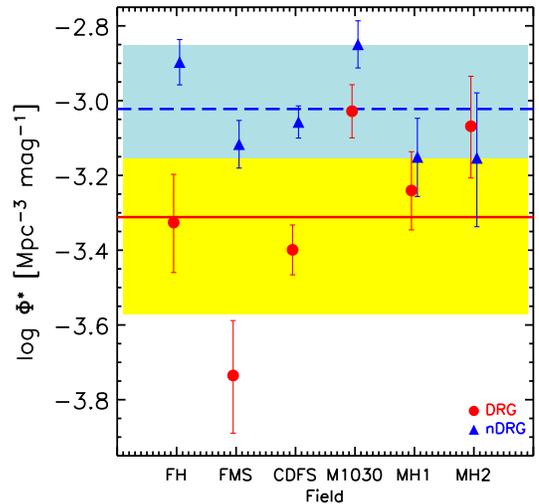}
\caption{Values of $\Phi^{\star}$ for DRGs ({\it red filled circles}) and 
non-DRGs ({\it blue filled triangles}) in the redshift range $2.5<z \leq 3.5$ 
estimated in the rest-frame $B$ band as described in \S~\ref{sec-numdens} 
for each individual field; the yellow (DRGs) and cyan (non-DRGs) shaded 
regions represent the values of $\Phi^{\star}$ from the analysis in 
\S~\ref{sec-stymeth} on the composite sample. \label{fig-phistar}}
\end{figure}

We determine the field-to-field variance in the density by fixing the 
parameters $\alpha$ and $M^{\star}$ to the best-fit values measured 
using the composite sample, and estimating $\Phi^{\star}_{\rm j}$ for 
each $j$th field separately by imposing a normalization on the LF such 
that the total number of observed galaxies in each field is reproduced.
In Table~\ref{tab-ndcomp}, the derived $\Phi^{\star}_{\rm j}$ of DRGs 
and non-DRGs in each field are listed for the three targeted redshift 
intervals and compared to $\Phi^{\star}$ measured from the composite 
sample. The results in the redshift range $2.5< z \leq 3.5$ are plotted in 
Figure~\ref{fig-phistar}. We find an overdensity of DRGs in the M1030 field 
at all redshifts, with the excess (as compared to the characteristic density 
of the composite sample) varying from a factor of $\sim1.2$ in the lowest 
redshift bin up to a factor of $\sim1.9$ in the redshift interval 
$2.5< z \leq 3.5$. We also find an underdensity of DRGs (a factor of 
0.82-0.86) in the GOODS-CDFS field, although only at $z>2.5$. The value 
of $\Phi^{\star}$ for DRGs in M1030 is a factor of $\sim1.6$--2.4 
larger than that in the GOODS-CDFS field at $z>2.5$, although they are 
similar at $2 \leq z \leq 2.5$. 

\begin{deluxetable*}{cccccccc}
\centering
\tablewidth{500pt}
\tabletypesize{\small}
\tablecaption{Characteristic Density $\Phi^{\star}$ in Each Field 
\label{tab-ndcomp}}
\tablehead{\colhead{Redshift Range}      & \colhead{Composite Sample}    &
 \colhead{MH2} & \colhead{MH1} & \colhead{M1030} & \colhead{CDFS} &   
    \colhead{FMS}  & \colhead{FH}}
\startdata
\multicolumn{8}{c}{DRGs} \\
\colrule
$2.0 \leq z \leq 2.5$   & $3.74^{+0.69}_{-1.09}$ & $2.71^{+0.93}_{-0.73}$ & 
              $3.13^{+0.76}_{-0.62}$ & $4.49^{+0.83}_{-0.71}$ & 
              $3.96^{+0.83}_{-0.70}$ & $3.78^{+1.51}_{-1.12}$ & 
	      $6.14^{+4.84}_{-2.93}$ \\
$2.7 \leq z \leq 3.3$ & $6.14^{+1.10}_{-1.71}$ & $7.41^{+2.26}_{-1.78}$ & 
              $6.45^{+1.43}_{-1.19}$ & $8.49^{+1.46}_{-1.26}$ & 
              $5.25^{+0.94}_{-0.81}$ & $2.70^{+1.23}_{-0.88}$ & 
	      $7.79^{+3.84}_{-2.69}$ \\
$2.5 < z \leq 3.5$ & $4.88^{+2.09}_{-2.20}$ & $8.54^{+3.08}_{-2.33}$ & 
              $5.75^{+1.55}_{-1.24}$ & $9.37^{+1.66}_{-1.42}$ & 
              $3.99^{+0.66}_{-0.57}$ & $1.84^{+0.74}_{-0.55}$ & 
	      $4.72^{+1.63}_{-1.25}$ \\
\cutinhead{non-DRGs}
$2.0 \leq z \leq 2.5$   &  $6.32^{+2.72}_{-2.41}$ &  $6.26^{+1.55}_{-1.27}$ & 
               $8.63^{+1.24}_{-1.09}$ &  $7.90^{+1.09}_{-0.96}$ & 
               $3.22^{+0.53}_{-0.46}$ &  $5.80^{+1.07}_{-0.91}$ & 
	      $12.65^{+2.33}_{-1.99}$ \\
$2.7 \leq z \leq 3.3$ & $11.14^{+4.02}_{-4.27}$ & $11.49^{+3.80}_{-2.93}$ & 
              $13.44^{+2.44}_{-2.09}$ & $13.52^{+2.17}_{-1.89}$ & 
               $9.65^{+1.14}_{-1.03}$ &  $6.70^{+1.46}_{-1.22}$ & 
	      $18.68^{+3.35}_{-2.87}$ \\
$2.5 < z \leq 3.5$ &  $9.50^{+4.53}_{-3.92}$ &  $7.03^{+3.46}_{-2.43}$ & 
               $7.07^{+1.90}_{-1.53}$ & $14.15^{+2.20}_{-1.92}$ & 
               $8.77^{+0.91}_{-0.83}$ &  $7.65^{+1.20}_{-1.05}$ & 
              $12.68^{+1.89}_{-1.66}$ \\
\enddata
\tablecomments{Units in 10$^{-4}$~Mpc$^{-3}$~mag$^{-1}$; 
the 1 $\sigma$ error of the characteristic density estimated 
for the individual field includes only the Poisson error 
\citep{gehrels86}. The measured rest-frame $R$-, $V$-, and $B$-band 
LFs have been used in the redshift ranges $2.0 \leq z \leq 2.5$, 
$2.7 \leq z \leq 3.3$, and $2.5 < z \leq 3.5$, respectively.}
\end{deluxetable*}

These results are qualitatively consistent with \citet{vandokkum06}, 
who showed that the GOODS-CDFS field is underdense in massive 
($M_{\star} >10^{11}$~$M_{\sun}$) galaxies at $2<z<3$, with a surface 
density that is about 60\% of the mean and a factor of 3 lower than 
that of their highest density field (M1030). However, our results seem 
to show systematically smaller underdensities for the GOODS-CDFS field 
compared to their work. In order to understand the origin of the smaller 
underdensity of DRGs found for the GOODS-CDFS field in our work compared 
to that of massive galaxies in \citet{vandokkum06}, we have estimated 
the surface density of DRGs in the redshift range $2<z<3$ down to 
$K^{\rm tot}_{\rm s}=21$. We find that the surface density of DRGs in the 
GOODS-CDFS field is $\sim70$\% of the mean and a factor of $\sim 2.2\pm0.5$ 
lower than that of the M1030 field, in good agreement with the values in 
\citet{vandokkum06}. Therefore, the smaller underdensities of DRGs found 
for the GOODS-CDFS field in our work appear to arise mainly from the 
different targeted redshift ranges. The approach adopted in this work to 
quantify field-to-field variance by comparing the $\Phi^{\star}_{\rm j}$ 
of the individual fields might also mitigate field-to-field differences, 
especially at the bright end.

We note that there are significant differences in the observed 
characteristic densities even within the MUSYC fields, although they 
have areas of $\sim100$~arcmin$^{2}$. For example, the observed 
$\Phi^{\star}$ of DRGs in the MH1 field is consistent with the one derived 
from the composite sample, but it is 0.61-0.76 times the value in the 
M1030 field. These results demonstrate that densities inferred from 
individual $\sim100$~arcmin$^{2}$ fields should be treated with caution.


\subsection{Luminosity Density} \label{sec-lumdens}

In this section we present estimates of the luminosity density. Because 
of the coupling between the two parameters $\alpha$ and $M^{\star}$, the 
luminosity density (obtained by integrating the LF over all magnitudes) 
is a robust way to characterize the contribution to the total LF from the 
different subpopulations and to characterize the evolution of the LF with 
redshift. 

The luminosity density $\rho_{\rm L}$ is calculated using:
\begin{equation}
\rho_{\rm L} = \int^{\infty}_{0}{L_{\rm \nu} \Phi(L_{\rm \nu}) dL_{\rm \nu}} = 
\Gamma(2+\alpha)\Phi^{\star}L^{\star},
\end{equation}
which assumes that the Schechter parametrization of the observed LF is 
a good approximation and valid also at luminosities fainter than probed 
by our composite sample. 
Table~\ref{tab-5} lists $\rho_{\rm L}$ with the corresponding 1, 
2, and 3 $\sigma$ errors\footnote{The 1, 2, and 3 $\sigma$ errors 
of the luminosity densities were calculated by deriving the distribution 
of all the values of $\rho_{\rm L}$ allowed within the 1, 2, and 
3 $\sigma$ solutions, respectively, of the Schechter LF parameters from 
the maximum likelihood analysis. The contribution from the uncertainties 
in the photometric redshift estimates derived in Appendix~\ref{app-1} 
was added in quadrature.} for all of the considered samples. We also 
list $\rho^{\rm last}_{\rm L}$, the luminosity density calculated to 
the faintest probed rest-frame luminosity, and $\rho^{\rm MUSYC}_{\rm L}$, 
the luminosity density calculated to the rest-frame magnitude limits of the 
deep NIR MUSYC. While the difference between $\rho_{\rm L}$ and 
$\rho^{\rm last}_{\rm L}$ is very small (negligible for DRGs and red 
galaxies, and $\sim0.1$~dex on average for non-DRGs and blue galaxies), 
the difference between $\rho_{\rm L}$ and $\rho^{\rm MUSYC}_{\rm L}$ is 
significant, especially for non-DRGs and blue galaxies ($\sim0.5$~dex on 
average).

\begin{deluxetable*}{cccccc}[!t]
\centering
\tablewidth{500pt}
\tabletypesize{\small}
\tablecaption{Luminosity Densities \label{tab-5}}
\tablehead{\colhead{Redshift Range}      & \colhead{Rest-Frame Band}     &
    \colhead{Sample} & \colhead{$\log{\rho_{\rm L}}$} &
\colhead{$\log{\rho^{\rm last\tablenotemark{a}}_{\rm L}}$} & 
\colhead{$\log{\rho^{\rm MUSYC\tablenotemark{b}}_{\rm L}}$} }
\startdata
$2.0 \leq z \leq 2.5$ & $R$ & All        & $26.74^{+0.06,0.08,0.12}_{-0.05,0.06,0.07}$ & 
                                 $26.71^{+0.05,0.06,0.07}_{-0.05,0.06,0.07}$ & 
                                 $26.53^{+0.05,0.06,0.07}_{-0.05,0.06,0.07}$ \\
            &     & $J-K>2.3$ &  $26.18^{+0.05,0.06,0.08}_{-0.05,0.06,0.07}$ & 
				 $26.18^{+0.05,0.06,0.08}_{-0.05,0.06,0.07}$ & 
				 $26.14^{+0.05,0.07,0.09}_{-0.05,0.07,0.09}$ \\
            &     & $J-K\leq2.3$  & $26.64^{+0.10,0.20,0.43}_{-0.07,0.08,0.10}$ & 
				 $26.57^{+0.05,0.06,0.07}_{-0.05,0.06,0.07}$ & 
				 $26.30^{+0.06,0.08,0.10}_{-0.06,0.09,0.11}$ \\
            &     & $U-V \geq 0.25$ & $26.43^{+0.05,0.06,0.07}_{-0.05,0.06,0.07}$ & 
				 $26.43^{+0.05,0.06,0.07}_{-0.05,0.06,0.07}$ & 
				 $26.37^{+0.05,0.06,0.07}_{-0.05,0.06,0.07}$ \\
            &     & $U-V<0.25$ & $26.50^{+0.18,0.48,2.47}_{-0.09,0.12,0.14}$ & 
				 $26.40^{+0.05,0.06,0.08}_{-0.05,0.06,0.07}$ & 
				 $26.03^{+0.07,0.10,0.13}_{-0.08,0.12,0.16}$ \\
$2.7 \leq z \leq 3.3$ & $V$ & All        & $26.97^{+0.08,0.15,0.26}_{-0.06,0.07,0.09}$ & 
                                 $26.91^{+0.05,0.06,0.07}_{-0.05,0.06,0.07}$ & 
				 $26.61^{+0.05,0.07,0.08}_{-0.06,0.08,0.10}$ \\
            &     & $J-K>2.3$ &  $26.44^{+0.07,0.12,0.24}_{-0.05,0.06,0.07}$ & 
				 $26.43^{+0.05,0.06,0.08}_{-0.05,0.06,0.07}$ & 
				 $26.26^{+0.06,0.07,0.09}_{-0.06,0.09,0.13}$ \\
            &     & $J-K\leq2.3$  & $26.82^{+0.12,0.28,0.66}_{-0.07,0.10,0.12}$ & 
				 $26.73^{+0.05,0.06,0.07}_{-0.05,0.06,0.07}$ & 
				 $26.36^{+0.06,0.08,0.11}_{-0.07,0.10,0.13}$ \\
            &     & $U-V \geq 0.25$ & $26.40^{+0.05,0.08,0.14}_{-0.05,0.06,0.07}$ & 
				 $26.39^{+0.05,0.06,0.07}_{-0.05,0.06,0.07}$ & 
				 $26.26^{+0.06,0.07,0.09}_{-0.06,0.08,0.12}$ \\
            &     & $U-V<0.25$ & $26.84^{+0.13,0.31,0.85}_{-0.08,0.11,0.13}$ & 
				 $26.74^{+0.05,0.06,0.07}_{-0.05,0.06,0.07}$ & 
				 $26.36^{+0.06,0.08,0.10}_{-0.07,0.10,0.13}$ \\
$2.0 \leq z \leq 2.5$ & $B$ & All        & $26.61^{+0.10,0.20,0.38}_{-0.07,0.09,0.11}$ & 
                                 $26.54^{+0.05,0.06,0.07}_{-0.05,0.06,0.07}$ & 
				 $26.15^{+0.06,0.08,0.11}_{-0.06,0.09,0.13}$ \\
            &     & $J-K>2.3$  & $25.90^{+0.10,0.27,2.95}_{-0.06,0.07,0.08}$ & 
				 $25.89^{+0.06,0.09,0.13}_{-0.05,0.06,0.07}$ & 
				 $25.64^{+0.08,0.11,0.15}_{-0.09,0.15,0.22}$ \\
            &     & $J-K\leq2.3$  & $26.53^{+0.15,0.35,1.15}_{-0.08,0.11,0.13}$ & 
				 $26.43^{+0.05,0.07,0.08}_{-0.05,0.06,0.07}$ & 
				 $25.99^{+0.07,0.10,0.13}_{-0.07,0.12,0.16}$ \\
            &     & $U-V \geq 0.25$ & $26.18^{+0.07,0.14,0.31}_{-0.05,0.06,0.07}$ & 
				 $26.17^{+0.05,0.07,0.10}_{-0.05,0.06,0.07}$ & 
				 $25.92^{+0.06,0.08,0.11}_{-0.07,0.11,0.16}$ \\
            &     & $U-V<0.25$ & $26.41^{+0.22,0.68,2.38}_{-0.10,0.14,0.16}$ & 
				 $26.29^{+0.05,0.07,0.09}_{-0.05,0.06,0.07}$ & 
				 $25.79^{+0.08,0.12,0.16}_{-0.09,0.14,0.21}$ \\
$2.5< z \leq 3.5$ & $B$ & All        & $26.90^{+0.16,0.37,1.00}_{-0.10,0.13,0.16}$ & 
                                 $26.75^{+0.05,0.06,0.08}_{-0.05,0.06,0.07}$ & 
				 $26.28^{+0.06,0.08,0.09}_{-0.06,0.09,0.12}$ \\
            &     & $J-K>2.3$ & $26.30^{+0.21,0.81,\infty}_{-0.10,0.13,0.14}$ &
				 $26.23^{+0.07,0.10,0.14}_{-0.06,0.07,0.09}$ & 
				 $25.86^{+0.07,0.10,0.12}_{-0.09,0.15,0.22}$ \\
            &     & $J-K\leq2.3$ & $26.76^{+0.24,0.74,\infty}_{-0.12,0.16,0.19}$ &
				 $26.59^{+0.05,0.07,0.09}_{-0.05,0.06,0.07}$ & 
				 $26.08^{+0.07,0.09,0.11}_{-0.07,0.11,0.15}$ \\
            &    & $U-V \geq 0.25$ & $26.27^{+0.20,0.67,\infty}_{-0.09,0.12,0.13}$ &
				 $26.22^{+0.07,0.10,0.14}_{-0.06,0.07,0.09}$ & 
				 $25.84^{+0.07,0.10,0.12}_{-0.09,0.15,0.23}$ \\
          &    & $U-V<0.25$  & $26.79^{+0.25,0.93,\infty}_{-0.13,0.17,0.20}$ & 
				 $26.60^{+0.05,0.07,0.09}_{-0.05,0.06,0.07}$ & 
				 $26.09^{+0.07,0.09,0.11}_{-0.07,0.10,0.14}$ \\
\enddata

\tablecomments{The luminosity densities are in units of 
erg~s$^{-1}$~Hz$^{-1}$~Mpc$^{-3}$. The quoted errors correspond 
to the 1, 2, and 3 $\sigma$ errors estimated from the maximum 
likelihood analysis as described in \S~\ref{sec-stymeth} and 
include the contribution from photometric redshift uncertainties.}
\tablenotetext{a}{$\rho^{\rm last}_{\rm L}$ is the luminosity density 
estimated by integrating the LF down to the faintest observed luminosities: 
$M^{\rm last}_{\rm R,AB}=-19.6$, $M^{\rm last}_{\rm V,AB}=-20.3$ and 
$M^{\rm last}_{\rm B,AB}=-19.5$, $M^{\rm last}_{\rm B,AB}=-20.0$ for 
the low- and high-redshift intervals, respectively; they  correspond 
to observed $K_{\rm s}$ band magnitude of 
$K^{\rm tot}_{\rm s} \approx 23.14$ at the lower limit of the targeted 
redshift intervals.}
\tablenotetext{b}{$\rho^{\rm MUSYC}_{\rm L}$ is the luminosity density 
estimated by integrating the LF down to the rest-frame magnitude limits 
of the deep NIR MUSYC survey: $M^{\rm MUSYC}_{\rm R,AB}=-21.8$, 
$M^{\rm MUSYC}_{\rm V,AB}=-22.4$ and $M^{\rm MUSYC}_{\rm B,AB}=-21.7$, 
$-22.2$ for the low- and high-redshift intervals, respectively 
(corresponding to observed $K^{\rm tot}_{\rm s} \approx 21.09$).}
\end{deluxetable*}

\begin{figure*}
\centering
\includegraphics[width=13.cm]{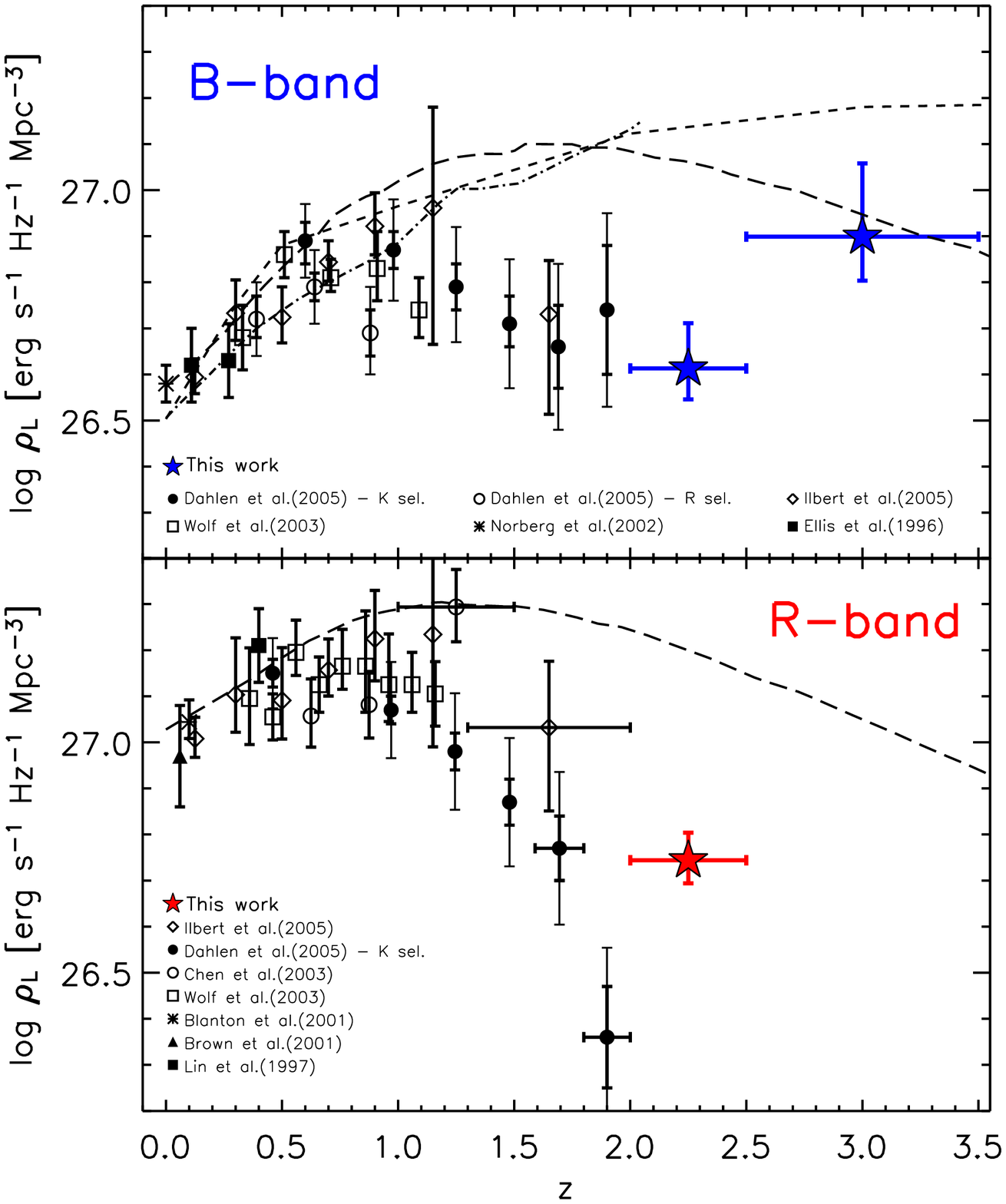}
\caption{\small {\it Top:} Total rest-frame $B$-band luminosity 
density $\rho_{\rm L}$ plotted vs. the redshift. Blue filled stars 
are the values from this work (the error bars include also the contribution 
from photometric redshift uncertainties); thick error bars represent 
statistical errors only, while thin error bars also include the sample 
variance contribution to the error budget. The $z<2$ points are taken 
from the 2dFGRS (\citealt{norberg02}; {\it asterisks}), Autofib Redshift 
Survey (\citealt{ellis96}; {\it filled squares}), COMBO-17 (\citealt{wolf03}; 
{\it open squares}), VIMOS-VLT Deep Survey (\citep{ilbert05}; {\it open 
diamonds}), and GOODS-CDFS (\citealt{dahlen05}; $R$- and $K$-selected 
catalog as open and filled circles, respectively). The long- and 
short-dashed lines correspond to the computed $B$-band rest-frame 
luminosity densities as a function of redshift using large-scale 
$\Lambda$CDM hydrodynamical simulations from \citet{nagamine00} and 
\citet{nagamine01}, respectively; the dot-dashed line is a prediction 
from a semianalytical model taken from Fig.~13 of \citet{dahlen05}.
{\it Bottom:} Total rest-frame $R$-band luminosity density $\rho_{\rm L}$ 
plotted vs. the redshift. The red filled star represents the value from 
this work. The $z<2$ points are taken from the CNOC1 Redshift Survey 
(\citealt{lin97}; {\it filled squares}), Century Survey (\citealt{brown01}; 
{\it filled triangle}), SDSS (\citealt{blanton01}; {\it asterisk}), 
COMBO-17 (\citealt{wolf03}; {\it open squares}), Las Campanas Infrared 
Survey (\citealt{chen03}; {\it open circles}), GOODS-CDFS 
(\citealt{dahlen05}; {\it filled circles}), and VIMOS-VLT Deep Survey 
(\citealt{ilbert05}; {\it open diamonds}). The long-dashed line corresponds 
to the computed $R$-band rest-frame luminosity density as a function of 
redshift using large-scale $\Lambda$CDM hydrodynamical simulations 
from \citet{nagamine00}. \label{lumdensBR_z.ps}}
\end{figure*}

In the top panel of Figure~\ref{lumdensBR_z.ps} we have plotted the total 
rest-frame $B$-band luminosity density $\rho_{\rm L}$ versus the redshift, 
including a compilation of results from the literature. Only the results 
from the literature which are not significantly affected by field-to-field 
variations, or that have taken these into account, are plotted. Our 
measurement of the total rest-frame $B$-band luminosity density is the 
only one at $z>2$ that is not significantly affected by field-to-field 
variance. From Figure~\ref{lumdensBR_z.ps}, there is an indication of a 
possible increase of the total luminosity density in the highest redshift 
bin, significant at the $\sim2.2$ $\sigma$ level. The measurement at 
$2 \leq z \leq 2.5$ is consistent with the one at $z \sim1.9$ from 
\citet{dahlen05}. In Figure\ref{lumdensBR_z.ps} we have also plotted the 
computed $B$-band rest-frame luminosity density as a function of $z$ 
predicted from large-scale $\Lambda$CDM hydrodynamical simulations from 
\citet{nagamine00,nagamine01} and from a semianalytical model taken from 
\citet{dahlen05}. While the predicted luminosity densities match the 
measurements at $z \leq1.2$ well, they clearly overpredict them at larger 
redshifts. Only the prediction at $z\sim3$ from \citet{nagamine00} is 
consistent with our measurement, although their model still overpredicts 
significantly the luminosity densities in the range $1.2\leq z \leq 2.5$.

The bottom panel of Figure~\ref{lumdensBR_z.ps} shows the total rest-frame 
$R$-band luminosity density $\rho_{\rm L}$ versus the redshift, 
including a compilation of results from the literature as in the top 
panel. As for the rest-frame $B$ band, our measurement of the rest-frame 
$R$-band $\rho_{\rm L}$ is the first one at $z>2$ for which sample variance 
does not significantly contribute to the error budget. Our point at $z>2$ 
is consistent with the trend observed at $z<2$ of decreasing luminosity 
densities with increasing redshifts, although the point at $z\sim1.9$ from 
the LF analysis of the GOODS-CDFS survey by \citet{dahlen05} is only 
$\sim43$\% of our measurement at $z\sim2.2$. From Table~\ref{tab-ndcomp} 
we see that in the redshift range $2 \leq z \leq 2.5$, the CDFS field is 
underdense in non-DRGs. The value of $\Phi^{\star}$ for all galaxies in 
the CDFS field is $6.67^{+0.83}_{-0.74} \times 10^{-4}$~Mpc$^{-4}$~mag$^{-1}$ 
(estimated as in \S~\ref{sec-numdens}), which is a factor of 
$0.63^{+0.30}_{-0.18}$ the value of $\Phi^{\star}$ from the composite sample 
(see Table~\ref{tab-3}). Therefore, the lower value of the rest-frame 
$R$-band $\rho_{\rm L}$ from \citet{dahlen05} at $1.8<z<2$ could be due to 
an underdensity of galaxies at $z\sim2$. In Figure~\ref{lumdensBR_z.ps} we 
have also plotted the computed $R$-band rest-frame luminosity density as 
function of $z$ predicted from large-scale $\Lambda$CDM hydrodynamical 
simulations from \citet{nagamine00}. As for the rest-frame $B$ band, the 
predictions match the observations well enough at $z \leq 1$, but at larger 
redshifts they significantly overpredict them. 

We can also compare our estimated luminosity densities in the rest-frame 
$B$ and $V$ bands with those from \citet{rudnick06}, who presented the 
evolution of the rest-frame optical luminosity and stellar mass densities 
at $z<3$. The luminosity density in \citet{rudnick06} was computed by 
simply adding up the luminosities of all galaxies in the targeted redshift 
bins with rest-frame $V$-band luminosities $>3\times10^{10}$~L$_{\sun}$.
If we integrate our measured LF down to the same limit adopted 
in \citet{rudnick06}, we obtain $\log{\rho_{\rm L}}=26.85\pm0.06$ and 
$26.66\pm0.06$ for the $V$ and $B$ band, respectively, at $z\sim3$, and 
$26.48\pm0.06$ for the $B$ band at $z\sim2.2$ (units in 
erg~s$^{-1}$~Hz$^{-1}$~Mpc$^{-3}$), in excellent agreement with their 
estimates ($26.79\pm0.05$, $26.66\pm0.05$, and $26.37\pm0.04$, respectively).

In \S~\ref{subsec-sublf} we showed that the contribution of DRGs and red 
galaxies to the global LF is comparable to (or larger than) that of 
non-DRGs and blue galaxies at the bright end, but it becomes significant 
smaller at the faint end, where non-DRGs and blue galaxies dominate the 
global LF. In \S~\ref{sec-numdens}, the contribution of DRGs to the 
global number density has been shown to be 13\%-25\% down to the 
faintest probed rest-frame luminosities. The contribution of DRGs to 
the global luminosity density is 19\%-29\% depending on the considered 
rest-frame band and redshift interval (see Table~\ref{tab-50}). Their 
contribution  increases up to 31\%-44\% if we cut the composite sample 
to the rest-frame absolute brightness limit of MUSYC ($\sim0.45L^{\star}$, 
$0.7L^{\star}$, $0.8L^{\star}$ of the global LF in the rest-frame 
$R$, $V$, and $B$ band, respectively), which reflects the increasing 
importance of DRGs at the bright end. A similar result holds if we 
consider the red galaxy subsample. Their contribution to the global 
luminosity density is 29\%-52\% down to the faintest observed luminosities 
and increases up to 36\%-69\% if we limit the analysis to the MUSYC 
brightness limits. From Table~\ref{tab-50} we conclude that the total 
luminosity density is dominated by non-DRGs/blue galaxies, especially 
in the bluer rest-frame optical bands, although DRGs/red galaxies 
contribute about 50\% at the bright end.

\begin{deluxetable*}{ccccc}[!t]
\centering
\tablewidth{500pt}
\tabletypesize{\small}
\tablecaption{DRGs/Red Galaxies Contribution to the Global Luminosity and 
Stellar Mass Densities \label{tab-50}}
\tablehead{\colhead{Redshift}      & \colhead{Rest-frame Band}            &
    \colhead{Sample} & \colhead{\tablenotemark{a}$f_{\rho_{\rm L}}$} &
    \colhead{\tablenotemark{b}$f_{\rho_{\rm \star}}$} }
\startdata
$2.0 \leq z \leq 2.5$ & $R$ & $J-K>2.3$  & $0.27^{+0.04}_{-0.05}$~~~$0.29^{+0.03}_{-0.03}$~~~
                                 $0.41^{+0.05}_{-0.05}$ & 
				 $0.62^{+0.22}_{-0.32}$~~~$0.66^{+0.20}_{-0.27}$~~~
				 $0.67^{+0.20}_{-0.29}$ \\
            &     & $U-V \geq 0.25$ & $0.49^{+0.06}_{-0.11}$~~~$0.52^{+0.04}_{-0.04}$~~~
                                 $0.69^{+0.04}_{-0.05}$ & 
				 $0.84^{+0.10}_{-0.25}$~~~$0.87^{+0.07}_{-0.16}$~~~
				 $0.92^{+0.05}_{-0.13}$ \\
$2.7 \leq z \leq 3.3$ & $V$ & $J-K>2.3$  & $0.29^{+0.06}_{-0.05}$~~~$0.33^{+0.04}_{-0.03}$~~~
                                 $0.44^{+0.06}_{-0.05}$ & 
				 $0.69^{+0.18}_{-0.28}$~~~$0.73^{+0.15}_{-0.23}$~~~
				 $0.79^{+0.13}_{-0.22}$ \\
            &     & $U-V \geq 0.25$ & $0.27^{+0.04}_{-0.06}$~~~$0.31^{+0.04}_{-0.04}$~~~
                                 $0.44^{+0.06}_{-0.05}$ & 
				 $0.65^{+0.20}_{-0.30}$~~~$0.69^{+0.17}_{-0.25}$~~~
				 $0.79^{+0.13}_{-0.25}$ \\
$2.0 \leq z \leq 2.5$ & $B$ & $J-K>2.3$  & $0.19^{+0.05}_{-0.05}$~~~$0.22^{+0.03}_{-0.03}$~~~
                                 $0.31^{+0.06}_{-0.05}$ & 
				 $0.62^{+0.26}_{-0.36}$~~~$0.67^{+0.21}_{-0.31}$~~~
				 $0.74^{+0.20}_{-0.39}$ \\
            &     & $U-V \geq 0.25$ & $0.37^{+0.07}_{-0.11}$~~~$0.43^{+0.04}_{-0.04}$~~~
                                 $0.58^{+0.06}_{-0.07}$ & 
				 $0.87^{+0.10}_{-0.34}$~~~$0.89^{+0.08}_{-0.21}$~~~
				 $0.90^{+0.07}_{-0.19}$ \\
$2.5< z \leq 3.5$ & $B$ & $J-K>2.3$  & $0.26^{+0.12}_{-0.10}$~~~$0.30^{+0.05}_{-0.03}$~~~
                                 $0.38^{+0.05}_{-0.06}$ & 
				 $0.78^{+0.18}_{-0.43}$~~~$0.82^{+0.13}_{-0.28}$~~~
				 $0.85^{+0.11}_{-0.31}$ \\
            &     & $U-V \geq 0.25$ & $0.23^{+0.12}_{-0.09}$~~~$0.29^{+0.05}_{-0.03}$~~~
                                 $0.36^{+0.06}_{-0.06}$ & 
				 $0.76^{+0.19}_{-0.43}$~~~$0.81^{+0.14}_{-0.28}$~~~
				 $0.84^{+0.12}_{-0.32}$ \\
\enddata
\tablecomments{The quoted errors correspond to the 1 $\sigma$ errors.}
\tablenotetext{a}{$f_{\rho_{\rm L}}$ is the contribution of DRGs/red galaxies 
to the global luminosity density; the three values correspond to the 
luminosity limits down to which the LF has been integrated: no limit, last 
observed point, and deep NIR MUSYC limit, respectively; see footnote of 
Table~\ref{tab-5} for the specific integration limits.}
\tablenotetext{b}{$f_{\rho_{\star}}$ is the contribution of DRGs/red galaxies 
to the global stellar mass density as estimated in \S~\ref{sec-massdens}; the 
three values correspond to the luminosity limits down to which the LF has 
been integrated to estimate the luminosity density.}
\end{deluxetable*}


\subsection{Stellar Mass Density Estimates} \label{sec-massdens}

Although non-DRGs and blue galaxies represent the major contribution 
to the total luminosity and number densities in the rest-frame optical 
bands, it has been shown that DRGs usually have larger mass-to-light 
ratios than non-DRGs (e.g., \citealt{forster04}; \citealt{labbe05}; 
\citealt{vandokkum04}), as is generally true for red versus blue 
galaxies (e.g., \citealt{bell01}; \citealt{kauffmann03}). It is 
therefore interesting to quantify the contribution of DRGs (red galaxies) 
to the total stellar mass density. 

Following the method described in \cite{rudnick03}, we estimated the 
stellar mass density from the measured global luminosity density 
modulo the mass-to-light ratio $M/L$. For each subsample we have 
measured the median rest-frame $U-V$ color\footnote{Rather than the 
median $U-V$ color of the sample, \cite{rudnick03} computed the global 
$U-V$ color from the relation: $U-V=-2.5 \log{j_{\rm U}} + M_{\rm \sun,U} 
+2.5 \log{j_{\rm V}}- M_{\rm \sun,V}$, where $j_{\rm U}$ 
and $j_{\rm V}$ are computed by adding the luminosities of the 
individual galaxies. The two methods return very similar values for 
the $U-V$ colors.}, estimated the corresponding $M/L$ ratio from the 
relation between $U-V$ color and $M/L$ ratio obtained from stellar 
population synthesis models, and multiplied the estimated $M/L$ ratio by 
the measured luminosity density to obtain the stellar mass density.

To convert between the measured rest-frame $U-V$ color and the 
mass-to-light ratio $M/L$, we have generated stellar population 
synthesis models with the evolutionary synthesis code developed by 
G.~Bruzual and S.~Charlot \citep{bruzual03}. We selected the ``Padova 1994'' 
evolutionary tracks, which are preferred by Bruzual \& Charlot over the more 
recent ``Padova 2000'' tracks because the latter may be less reliable and 
predict a hotter red giant branch leading to worse agreement with observed 
galaxy colors. We used the solar metallicity set of tracks. The metallicities 
of the DRGs are poorly known, with evidence for solar and supersolar 
metallicities for luminous DRGs \citep{vandokkum04}. These DRGs appear more 
metal-rich than the five LBGs at $z\sim3$ studied by \citet{pettini01} and 
similar to the seven UV-selected star-forming ``BX/MD'' objects at $z\sim2$ 
for which \citet{shapley04} inferred solar, and possibly supersolar, 
metallicities. In all cases, however, the determinations rely on limited 
samples and suffer from large uncertainties. As shown by \citet{forster04} 
adopting subsolar metallicity ($Z=0.2Z_{\sun}$), the estimated $M/L$ 
ratios are systematically lower by a factor of $\approx 2$ on average. 
Therefore, if non-DRGs (or blue galaxies) are characterized by lower 
metallicities with respect to DRGs (red galaxies), the differences in 
$M/L$ (and stellar mass densities) would be even larger than what is 
estimated assuming solar metallicities for both subsamples.
For the star formation history (SFH) we used three different prescriptions: 
a constant star formation history (CSF model), an exponentially declining in 
time SFH characterized by the parameter $\tau$ (tau-model), and an 
instantaneous burst model (SSP model). Several values of $\tau$ were used, 
from $\tau=0.1$~Gyr (the resulting model being similar to the SSP model) 
to 6~Gyr (closer to the CSF model). We adopted the \citet{chabrier03} IMF 
with lower and upper IMF mass cutoffs $m_{\rm L}=0.1$~$M_{\sun}$ and 
$m_{\rm U}=100$~$M_{\sun}$, respectively. Adopting a different IMF would 
result in different derived mass-to-light ratios, which strongly depend on 
the shape and cutoff of the low-mass IMF (for example, assuming a 
Salpeter~[1955] IMF, the estimated $M/L$ ratio would be systematically 
larger by a factor of $\sim1.7$). However, since we are interested only in the 
relative contribution of DRGs (red galaxies) to the global stellar mass 
density, the results do not depend on the adopted IMF as long as all galaxies 
are characterized by the same IMF. We assumed that the interstellar 
extinction by dust within the objects followed the attenuation law of 
\citet{calzetti00} derived empirically from observations of local 
UV-bright starburst galaxies under the formalism of a foreground 
screen of obscuring dust.

\begin{figure}
\epsscale{1.0}
\plotone{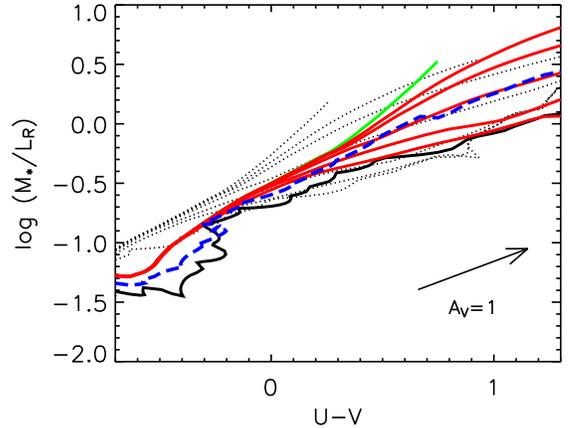}
\caption{Relation between the rest-frame $U-V$ color and the mass-to-light 
ratio in the rest-frame $R$ band, $M/L_{\rm R}$, for the model tracks 
described in \S~\ref{sec-massdens}. The dotted lines are for models 
with $A_{\rm V}=0$, while the other tracks are for $A_{\rm V}=1$. From 
top to bottom, the solid lines correspond to the CSF model ({\it green}), 
the tau-models ($\tau=6$, 3, 1, 0.3, 0.1~Gyr) ({\it red}), and the SSP model 
({\it black}). The blue dashed line, used in \S~\ref{sec-massdens} to 
estimate the contribution of DRGs and red galaxies to the total stellar 
mass density, corresponds to the median value of the tracks with 
$A_{\rm V}=1$. The arrow indicates the vector used to redden the 
$A_{\rm V}=0$ model to $A_{\rm V}=1$ model. \label{fig-mlrR}}
\end{figure}

We plot in Figure~\ref{fig-mlrR} the relation between the rest-frame $U-V$ 
color and the mass-to-light ratio in the rest-frame $R$ band, $M/L_{\rm R}$, 
for the generated model tracks in the two cases with no extinction and with 
$A_{\rm V}=1$. It is seen that dust extinction moves the tracks roughly 
parallel to the model tracks. As emphasized by \citet{bell01}, dust is a 
second-order effect for estimating stellar $M/L$ ratios. Dust extinguishes 
light from the stellar population, making it dimmer; however, dust also 
reddens the stellar population, making it appear to have a somewhat larger 
stellar $M/L$ ratio. To first order, these effects cancel out, leaving a 
dust-reddened galaxy on the same color--stellar $M/L$ ratio correlation. 

Using the relation between color and $M/L$, we convert the estimated median 
rest-frame $U-V$ color and the measurements of the luminosity densities 
$\rho_{\rm L}$ to stellar mass density estimates $\rho_{\star}$. 
Specifically, we adopted the median value of $M/L$ within the family of 
considered model tracks with $A_{\rm V}=1$ (blue dashed line plotted in 
Figure~\ref{fig-mlrR} for the rest-frame $R$ band); the error on the $M/L$ 
was chosen as half of the difference between the upper and lower envelope 
of the model tracks. The median rest-frame $U-V$ color of each subsample 
and the corresponding $M/L$ in the rest-frame $B$, $V$, and $R$ bands 
are listed in Table~\ref{tab-60}. 

\begin{deluxetable*}{cclcc}
\centering
\tablewidth{450pt}
\tabletypesize{\small}
\tablecaption{Rest-Frame $U-V$ Colors and Estimated $M/L$ Ratios 
\label{tab-60}}
\tablehead{\colhead{Redshift}      & \colhead{Rest-Frame Band}       &
    \colhead{Sample} & \colhead{\tablenotemark{a}$U-V$} &
    \colhead{\tablenotemark{b}$\log{M/L}$} }
\startdata
$2.0 \leq z \leq 2.5$ & $R$ & $J-K>2.3$  &  $0.66$ ($0.67$)  
                               &  $0.05\pm0.32$ ($0.06\pm0.32$)  \\
            &     & $J-K\leq2.3$  & $-0.06$ ($0.19$)  
                               & $-0.63\pm0.07$ ($-0.41\pm0.10$) \\
            &     & $U-V\geq0.25$ &  $0.60$ ($0.62$)  
                               & $-0.00\pm0.27$ ($0.01\pm0.28$)  \\
            &     & $U-V<0.25$ & $-0.23$ ($-0.13$) 
                               & $-0.78\pm0.04$ ($-0.71\pm0.09$) \\
$2.7 \leq z \leq 3.3$ & $V$ & $J-K>2.3$  &  $0.57$ ($0.57$)  
                               & $-0.08\pm0.28$ ($-0.08\pm0.28$) \\
            &     & $J-K\leq2.3$  & $-0.19$ ($-0.14$) 
                               & $-0.80\pm0.05$ ($-0.75\pm0.05$) \\
            &     & $U-V\geq0.25$ &  $0.59$ ($0.57$)  
                               & $-0.05\pm0.30$ ($-0.08\pm0.28$) \\
            &     & $U-V<0.25$ & $-0.18$ ($-0.14$) 
                               & $-0.75\pm0.05$ ($-0.75\pm0.05$) \\
$2.0 \leq z \leq 2.5$ & $B$ & $J-K>2.3$  &  $0.66$ ($0.67$)  
                               &  $0.00\pm0.40$ ($0.00\pm0.40$)  \\
            &     & $J-K\leq2.3$  & $-0.06$ ($0.19$)  
                               & $-0.85\pm0.05$ ($-0.60\pm0.15$) \\
            &     & $U-V\geq0.25$ &  $0.60$ ($0.62$)  
                               & $-0.10\pm0.35$ ($-0.05\pm0.35$) \\
            &     & $U-V<0.25$ & $-0.23$ ($-0.13$) 
                               & $-1.15\pm0.15$ ($-0.85\pm0.05$) \\
$2.5< z \leq 3.5$ & $B$ & $J-K>2.3$  &  $0.56$ ($0.56$)  
                               & $-0.13\pm0.33$ ($-0.13\pm0.33$) \\
            &     & $J-K\leq2.3$  & $-0.26$ ($-0.20$) 
                               & $-1.15\pm0.15$ ($-1.10\pm0.20$) \\
            &     & $U-V\geq0.25$ &  $0.57$ ($0.57$)  
                               & $-0.13\pm0.33$ ($-0.13\pm0.33$) \\
            &     & $U-V<0.25$ & $-0.25$ ($-0.20$) 
                               & $-1.15\pm0.15$ ($-1.10\pm0.20$) \\
\enddata

\tablenotetext{a}{Median rest-frame $U-V$ color of the considered 
subsample; the numbers in parentheses correspond to the median 
$U-V$ colors of the subsamples brighter than the MUSYC magnitude limit.}
\tablenotetext{b}{$\log{M/L}$ is the logarithm of the mass-to-light 
ratio in solar units assuming $A_{\rm V}=1$~mag, calculated as 
described in \S~\ref{sec-massdens}; the numbers in parentheses 
correspond to the mass-to-light ratios of the samples brighter 
than the MUSYC magnitude limit.}
\end{deluxetable*}

Although the values of the stellar mass densities of the individual 
subsamples might be affected by very large uncertainties, the 
relative contribution to the global stellar mass density of DRGs 
(red galaxies) and non-DRGs (blue galaxies) should be more robust. 
The contribution $f_{\rm \rho_{\star}}$ of the DRGs (red galaxies) 
to the global stellar mass density is listed in Table~\ref{tab-50}. 

Adopting the same assumptions for the stellar population synthesis 
models of DRGs (red galaxies) and non-DRGs (blue galaxies) (i.e., the 
median value of the considered model track), we see 
from Table~\ref{tab-60} that DRGs (red galaxies) have $M/L$ ratios 
systematically higher than non-DRGs (blue galaxies) by a factor of 
$\sim5$-11 depending on the rest-frame band (higher in the bluer 
bands). The differences in $M/L$ are smaller when the brighter samples 
(down to the MUSYC limit) are considered, with the $M/L$ of DRGs (red 
galaxies) being a factor of $\sim3$-9 (4-9) larger with respect to non-DRGs 
(blue galaxies). For comparison, from the analysis of {\it Spitzer}-IRAC 
imaging on HDF-S, \citet{labbe05} found  that the average mass-to-light 
ratio of DRGs in the rest-frame $K$ band is about a factor of $\sim3$ larger 
than LBGs, finding a correlation between $M/L_{\rm K}$ and rest-frame 
$U-V$ color. Consistently, from SED modeling of the $2 \leq z \leq 3.5$ 
FIRES galaxies, \citet{forster04} found that the median rest-frame $V$-band 
$M/L$ of DRGs is $\sim1.2$-2.3~M$_{\sun}$~L$^{-1}_{\rm V, \sun}$ 
($\sim0.4$ for LBGs). The higher values of $M/L$ for DRGs agree very well 
also with the results from SED fitting of individual galaxies in 
\citet{vandokkum04}, who found for DRGs a median value 
$M/L_{\rm V} \approx 0.8$~$(M/L)_{\sun}$. Finally, our estimated 
$M/L_{\rm V}$ are in excellent agreement with those estimated in 
\citet{rudnick06}.

Because of the systematically larger mass-to-light ratios, DRGs (red 
galaxies) dominate the global stellar mass density, with contributions 
in the range 66\%-82\% (69\%-89\%) down to the faintest probed rest-frame 
luminosities. The contribution of DRGs (red galaxies) increases up to 
67\%-85\% (79\%-92\%) if the brightest sample is considered (down to 
the rest-frame magnitude limits of the deep NIR MUSYC).
These numbers are consistent with the results of \citet{vandokkum06}. 
From a complete mass-selected sample ($M_{\star}>10^{11}$~$M_{\sun}$) 
constructed with MUSYC, FIRES, and GOODS-CDFS, they estimated 
that DRGs in the redshift interval $2<z<3$ make up 77\% of the total 
stellar mass, in very good agreement with our results.
Our results are in qualitative good agreement also with the results from 
\citet{rudnick06}, who found that the DRGs contribute 64\% of the 
stellar mass density at $z\sim2.8$ and 30\%-50\% at $z\sim2$.

We stress that the estimated contributions of DRGs and red 
galaxies to the global stellar mass density are very uncertain 
and need confirmation from detailed SED analysis of mass-limited 
(rather than luminosity-limited) samples.


\section{SUMMARY AND CONCLUSIONS} \label{sec-concl}

In this paper we have measured the rest-frame optical ($R$, $V$, and 
$B$ band) luminosity functions of galaxies at redshifts $2 \leq z \leq 3.5$ 
from a composite sample constructed with the deep NIR MUSYC, the ultra-deep 
FIRES, and the GOODS-CDFS. The large surveyed area ($\sim378$~arcmin$^{2}$, 
76\% of which comes from the deep NIR MUSYC) of the composite sample and 
the large range of luminosities spanned allows us to measure the bright end 
of the LF and to constrain the faint-end slope. Moreover, the several 
independent fields and their large area enabled us to largely reduce 
uncertainties due to sample variance, especially at the bright end. We have 
used Monte Carlo simulations to show that the uncertainties in the photometric 
redshift estimates do not significantly affect the measured parameters of 
the LF in the studied redshift regimes. There is a hint for a steepening 
in the faint-end slope of the LF from the rest-frame $R$ band to the 
$B$ band, although the differences are not significant. 
The measured LF faint-end slopes at $z>2$ are consistent, within the 
errors, with those in the local LFs. The characteristic magnitudes are 
significantly brighter than the local ones (e.g., $\sim1.2$~mag in the 
rest-frame $R$ band), while the measured values for $\Phi^{\star}$ are 
typically a factor of $\sim5$ smaller with respect to the local values.

The large number of objects in the composite sample allowed the first 
measurement of the LF of DRGs (defined based on their observed $J-K$ 
color), which we compared to that of non-DRGs in the same redshift range. 
The DRG population is characterized by a very different LF than that of 
non-DRGs, especially at the faint end. While at 
the bright end the LF of DRGs is similar to that non-DRGs, at the faint end 
the latter one has a significantly steeper faint-end slope, especially 
in the rest-frame $R$ band. The significance of the difference between the 
LFs of DRGs and non-DRGs decreases going to bluer rest-frame bands and to 
higher redshifts, although this is mainly caused by decreasing constraints 
on the faint end of the LF of DRGs. Qualitatively similar results are found 
if we compare the LFs of red (rest-frame $U-V \geq 0.25$) and blue 
(rest-frame $U-V < 0.25$) galaxies in the same redshift intervals, with 
the former equally contributing (or even dominating) at the bright end and 
the latter dominating the faint end.

In the rest-frame $V$ band we have also compared the LFs of blue galaxies 
(non-DRGs) with those of LBGs in the same redshift range. Although 
the two LFs agree very well at the bright end, the faint-end slope 
estimated by \citet{shapley01} is much steeper than the one measured in 
this paper. As the rest-frame optical LF of LBGs was estimated in 
\citet{shapley01} from the rest-frame UV LF and the observed distribution 
of $\cal{R}$-$K_{\rm s}$ colors as a function of $\cal{R}$ magnitude, their 
steeper slope could be a result of an overestimate of the regression 
slope of the correlation between $\cal{R}$-$K_{\rm s}$ and $\cal{R}$ and/or 
of the faint-end slope of the adopted rest-frame UV LF of LBGs. Supporting 
the former possibility is the work of \citet{labbe06}, who do not find any 
positive correlation between $R$ magnitudes and $R-K_{\rm s}$ colors in their 
deeper sample. Support for the latter comes from the very recent work of 
\citet{sawicki06}, who measured the rest-frame UV LF for $z \sim3$ LBGs 
from the Keck Deep Fields (KDF; \citealt{sawicki05}) and find a faint-end 
slope $\alpha=-1.43^{+0.17}_{-0.09}$, significantly shallower than the one 
adopted in \citet{shapley01}. Alternatively, our blue $K_{\rm s}$-selected 
(i.e., rest-frame optical selected) galaxies might simply constitute a 
different population than the $z\sim3$ LBGs (rest-frame UV selected), with 
different characterizations of the LF at the faint end. We also caution that 
our measurements of the faint-end slopes still have significant uncertainties 
due to small number statistics.

We generally find good agreement between our measured rest-frame 
$B$-band LFs at $2.5 < z \leq 3.5$ and those previously published 
by \cite{poli03}, \cite{giallongo05}, and \cite{gabasch04}. In the 
redshift range $2 \leq z \leq 2.5$, the agreement between our 
rest-frame $B$- and $R$-band LFs and those measured by 
\cite{gabasch04,gabasch06} from the FDF survey is less good, 
especially for the $R$ band. Their Schechter parameters $\alpha$ 
and $M^{\star}$ are consistent with ours only at the 2 $\sigma$ 
level in the rest-frame $R$ band, while their estimated $\Phi^{\star}$ 
is larger than ours by a factor of $\sim$1.3-1.6 in the rest-frame 
$B$ and $R$ band, respectively. We have shown that this disagreement 
may be due to the spectroscopically confirmed existence of 
a (proto)cluster at $z=2.35$ in the single field FDF survey.

From the measured LFs we have estimated the number and luminosity densities 
of the global population of high-$z$ galaxies and of various subsamples. 
The contribution of DRGs (red galaxies) to the global number 
density is only $\sim13$\%-25\% (18\%-29\%) down to the faintest probed 
rest-frame luminosities. However, we have shown that field-to-field 
variations can be very significant (up to a factor of $\sim3$), especially 
for relatively bright samples, in accord with the highly clustered nature 
of high luminosity and red galaxies (\citealt{adelberger05}; 
\citealt{daddi03}; \citealt{quadri06b}). The contribution of DRGs 
(red galaxies) to the global luminosity density is $\sim20$\%-30\% 
(30\%-50\%), higher in the redder rest-frame bands (which are less affected 
by extinction and better tracers of the underlying stellar mass) and at 
lower redshifts. With respect to the lower $z$ luminosity density estimates 
from the literature, we confirm the trend of slowly decreasing rest-frame 
$R$-band luminosity densities beyond $z\sim1$, with $\rho_{\rm L, R}$ at 
$z\sim2.3$ being a factor of $\sim2$ smaller than the local one. In the 
rest-frame $B$ band, the measured global luminosity density at $z\sim2.3$ 
is similar to the local value. At $z \sim 3$, the estimated global luminosity 
density may be a factor of $\sim2$ higher, similar to the values around 
$z\sim1$.

Finally, using stellar population synthesis models, we have derived the 
mass-to-light ratios of the considered subsamples by converting the 
estimated median rest-frame $U-V$ color into $M/L$. In the rest-frame 
$R$ and $V$ bands, the mass-to-light ratios of DRGs (red galaxies) are 
a factor of $\sim5$ larger than non-DRGs (blue galaxies), consistent with 
previous works. In the rest-frame $B$ band the difference in $M/L$ is 
higher, up to a factor of $\sim11$. Using the estimated $M/L$, we have 
quantified the contribution of DRGs and red galaxies to the global 
stellar mass density, finding that the total stellar mass budget is 
dominated by DRGs (red galaxies), whose contribution is of order 
$\sim60$\%-80\% of the global value. We caution that our $M/L$ ratios 
estimates are very rough and characterized by very large uncertainties 
and need confirmation from detailed SED analysis.

The main limitation of this work is the small number statistics at the 
very faint end of the LF, which is probed only by the ultra-deep FIRES. 
The faint-end slopes of the DRG and red galaxy subsamples are 
especially uncertain. To make further progress in the determination of the 
LF of different galaxy populations at $z>2$ and to better constrain the 
global LFs, it is crucial to better probe the faint end of the LFs. This 
can only be achieved with ultradeep NIR imaging with high-quality 
optical data over many spatially disjoint fields, in order to improve the 
statistics at the faint end and to mitigate the effect of field-to-field 
variations. 
Although we have shown that well-behaved photometric redshift 
errors do not affect significantly the measurement of the LF, the heavy 
reliance on photometric redshifts is another limitation of this work, since 
``catastrophic'' failures and systematic errors could potentially affect 
the LF measurements. Obtaining large numbers of spectroscopic redshifts 
for $K$-selected high-$z$ sources has proven difficult and extremely time 
consuming. Even though the success rate for measuring spectroscopic redshift 
for bright galaxies is high with NIR spectroscopy \citep{kriek06}, only 
the use of multiobject NIR spectrographs will make it 
possible to construct a large sample of high-$z$ $K$-selected galaxies with 
spectroscopic redshift measurements.
Further advances can be expected from further extension of the wavelength 
range into the red. Scheduled {\it Spitzer} IRAC observations on the deep 
NIR MUSYC fields will allow us to (1) separate old and passively evolving 
galaxies from heavily obscured and strongly active star-forming galaxies 
(see \citealt{labbe05}), making it possible to study the LF of physically 
different types of galaxies; (2) extend the robust measurement of the LF at 
redshift $2<z<3$ into the rest-frame NIR, which is much closer to a 
selection by stellar mass; (3) convert the measured rest-frame NIR 
luminosity function into a mass function and study the evolution of the 
stellar mass density; and (4) extend the study of the rest-frame optical 
LFs to even higher redshifts.


\acknowledgments
 
We thank all the members of the MUSYC collaboration for their contribution 
to this research. MUSYC has greatly benefited from the support of 
Fundaci\'on Andes and the Yale Astronomy Department. D.M. is supported by 
NASA LTSA NNG04GE12G. The authors acknowledge support from NSF CARRER 
AST-0449678. E.G. is supported by NSF Fellowship AST-0201667. P.L. is 
supported by Fondecyt Grant \#1040719. We thank the anonymous referee 
for comments and suggestions which helped improve the paper.


\appendix

\section{Appendix A \\ ~~~~  \\ Effects of Uncertainties in $z_{\rm phot}$: Monte Carlo Simulations}
\label{app-1}

In order to quantify the systematic effects on the LF parameters $\alpha$ 
and $M^{\star}$ due to the uncertainties in the photometric redshift 
estimates, we performed a series of Monte Carlo simulations. First, we 
generated several model catalogs of 25,000 galaxies with redshifts 
between $z_{\rm 1,MC}$ and $z_{\rm 2,MC}$ and with luminosities 
drawn from an input Schechter LF. While in the Monte Carlo simulations 
of \citet{chen03} the redshifts of the objects in the mock catalogs 
were extracted from a random uniform distribution, we took into 
account the fact that, under the assumption of no evolution in the 
number density, the probability of a galaxy existing at the redshift 
$z$ is proportional to the volume:
\begin{equation} \label{eq-pz}
p(z) \propto \frac{dV}{dz} \propto \frac{d_{L}^{2}(z)}{(1+z)^{2}} 
\frac{1}{\sqrt{\Omega_{\Lambda} + \Omega_{M}(1+z)^{3}}},
\end{equation}
where $d_{L}$ is the luminosity distance. 
Since in a real survey galaxies are selected down to a limiting 
apparent magnitude, the final mock catalogs are obtained after applying a 
limit in the observed apparent magnitude. The effect of a limiting apparent 
magnitude is that, at a fixed observed magnitude, intrinsically fainter 
sources are systematically excluded from the catalog at higher redshift. 
Next, as done by \citet{chen03}, we assumed a redshift error function 
parameterized as a Gaussian distribution function of 1 $\sigma$ width
$\sigma^{\prime}_{\rm z}(1+z)$, with $\sigma^{\prime}_{\rm z}$ 
the scatter in $\Delta z/(1+z_{\rm spec})$, and we formed an observed 
redshift catalog by perturbing the input galaxy redshift within the 
redshift error function. Finally, we determined the LF for the galaxies 
at $z_{1} \leq z \leq z_{2}$ using the $1/V_{\rm max}$ and maximum likelihood 
methods described in \S~\ref{sec-lf}. Note that we ignore $K$-correction 
in our Monte Carlo simulations.

\begin{figure*}
\centering \includegraphics[width=17cm]{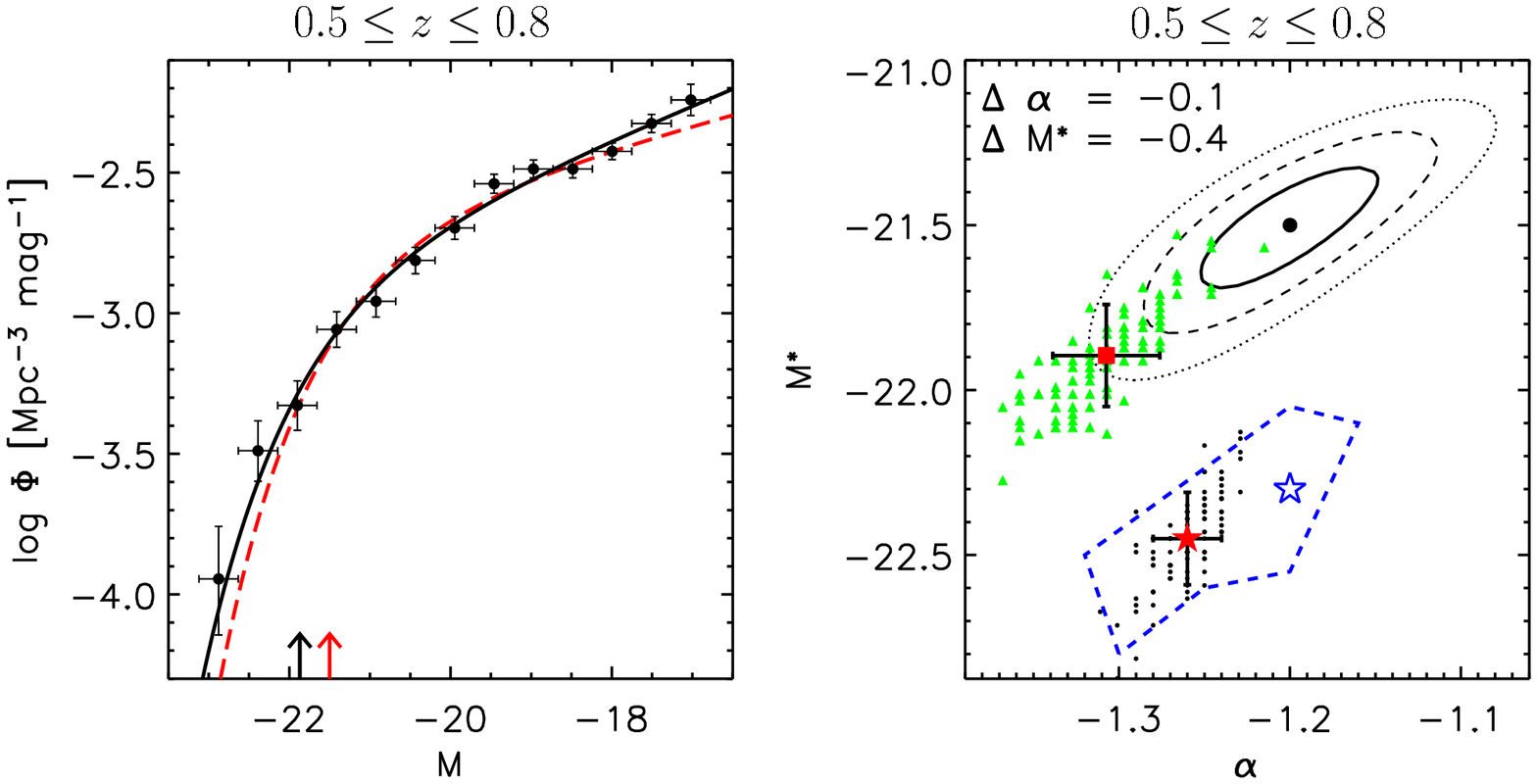}
\caption{{\it Left:} Intrinsic Schechter LF ({\it red dashed line}) 
compared to the observed LF at $0.5 \leq z \leq 0.8$ corresponding 
to the median  Monte Carlo realization (black solid line and filled 
circles correspond to the maximum likelihood and $1/V_{\rm max}$ 
method, respectively); the best-fit $M^{\star}$ are also plotted as
arrows. {\it Right:} Measured Schechter parameters $M^{\star}$
plotted vs. $\alpha$; the black filled circle represents the input
parameters of the ``observed'' model catalog, together with the 1, 2, 
and 3 $\sigma$ contour levels; the green filled triangles represent 
the 100 Monte Carlo realizations; the red filled square is the median
realization, and the plotted error bars represent the rms in $\alpha$
and $M^{\star}$. The amount of the systematic effects in $\alpha$ and
$M^{\star}$ are also specified. The red filled star represents the 
median realization when the redshifts of the objects are extracted from 
a random uniform distribution, as done in \citet{chen03}; the small 
black filled circles represent the corresponding 100 Monte Carlo 
realizations. The result from \citet{chen03} is also shown ({\it blue 
open star}) together with its 99\% uncertainties ({\it blue dashed contour}).
\label{fig-chen4V}}
\end{figure*}

We first studied the effects of the photometric redshift uncertainties 
at $z<1$, by using $z_{\rm 1,MC}=0.15$, $z_{\rm 2,MC}=1.6$, and assuming 
an input Schechter LF with parameters $\alpha=-1.2$ and $M^{\star}=-21.5$ 
(as in \citealt{chen03}). In order to compare our results with those from 
the Monte Carlo simulations in \cite{chen03}, we used 
$\sigma^{\prime}_{\rm z}=0.15$ and we measured the LF in the same redshift 
range $0.5 \leq z \leq 0.8$. We find that the median measured $M^{\star}$ 
is brighter 
than the intrinsic value by $\sim0.4$~mag and that the measured $\alpha$ is 
steeper on average than the intrinsic value by $\sim0.1$. This result is 
shown in Figure~\ref{fig-chen4V}: in the left panel, the input Schechter 
LF is compared to the median Monte Carlo realization; in the right 
panel, the 100 Monte Carlo realizations are plotted in the $\alpha-M^{\star}$ 
plane and compared to the best-fit values and the corresponding 1, 2, and 
3 $\sigma$ contour levels of the LF measured on the redshift-unperturbed mock 
catalog.

\begin{figure*}
\centering \includegraphics[width=14cm]{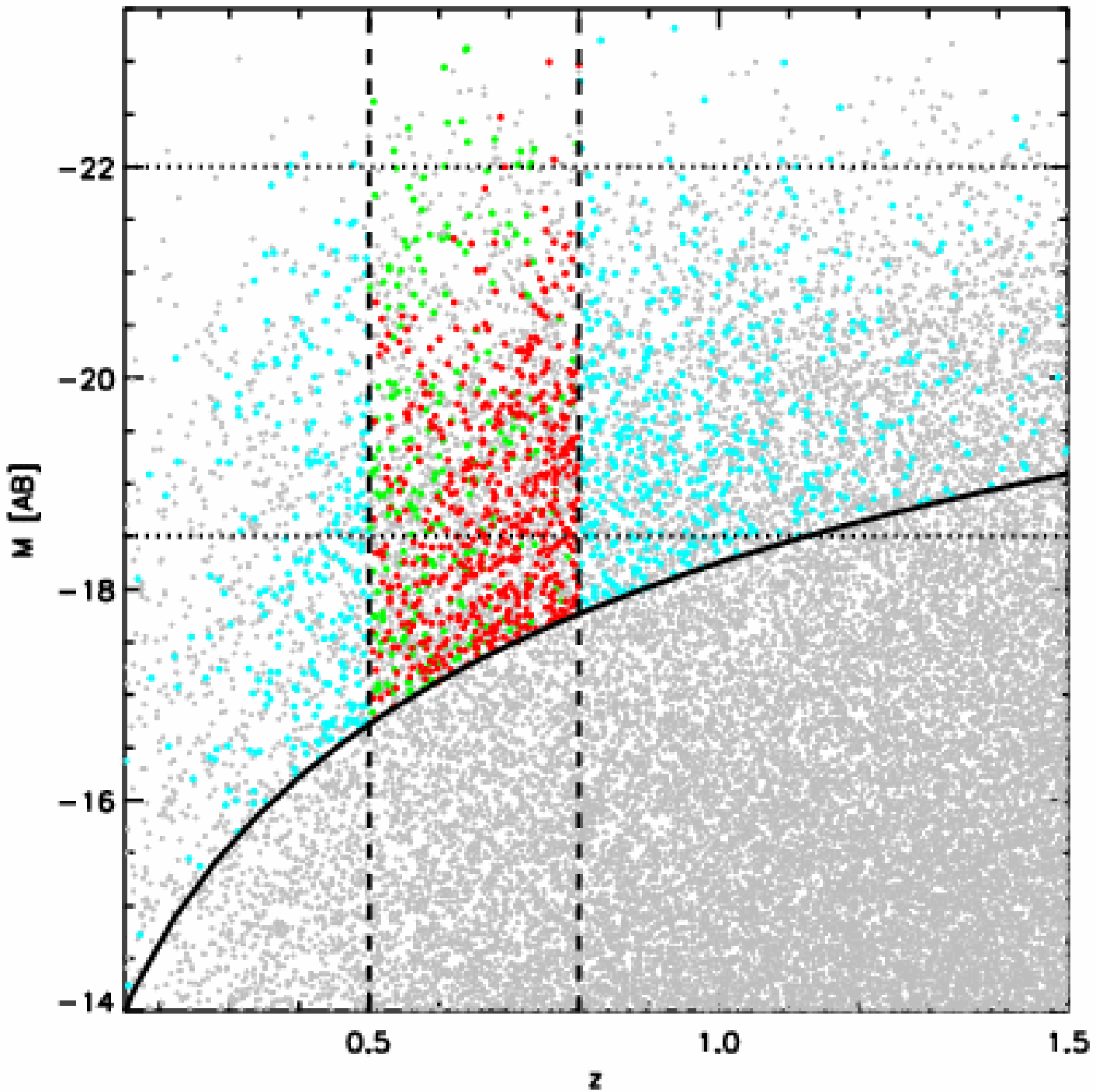}
\caption{Rest-frame absolute magnitudes $M$ plotted vs. the redshifts 
$z$ of the redshift-unperturbed mock catalog ({\it gray points}). The 
solid line represents the limiting magnitude of the sample as a function 
of redshift; the two vertical dashed lines represent the redshift 
interval within which the LF is measured. Cyan filled circles at $z>0.8$ 
are the sources that in the redshift-perturbed mock catalog scatter from 
high $z$ into the considered redshift bin ({\it red filled circles}); 
the cyan filled circles at $z<0.5$ are the sources that scatter from low 
$z$ into the considered redshift bin ({\it green filled circles}). The 
loci at magnitudes fainter than the lower dotted line and brighter than 
the upper dotted line are the regions where we observe the excess of 
sources at the faint and bright ends (evident in the left panel of 
Fig.~\ref{fig-chen4V}), respectively. \label{fig-mc_Mvsz}}
\end{figure*}

As shown in Figure~\ref{fig-chen4V}, the measured systematic effects on 
$\alpha$ and $M^{\star}$ caused by the redshift uncertainties arise from 
an excess of sources at both the faint and the bright end of the LF. A 
careful analysis of the Monte Carlo simulations reveals the origin of 
these excesses. Because of the uncertainties in the redshifts, sources 
can scatter from high to low redshifts, and vice versa. 
As the probability of a galaxy existing at redshift $z$ is proportional 
to $dV/dz$ (which peaks at $z\sim2.5$), the number of sources that scatter 
from higher redshifts to lower ones is much larger than vice versa. This 
is evident in Figure~\ref{fig-mc_Mvsz}, where the rest-frame absolute 
magnitudes of the mock catalog are plotted versus redshift: at a fixed 
magnitude, the number of sources is larger at higher redshifts.

Since an object in the mock catalog is characterized by a fixed apparent 
magnitude, a new redshift estimate translates into a new rest-frame absolute 
magnitude; e.g., when a source ``scatters'' from high to low redshift, the 
estimated absolute brightness is fainter than the intrinsic value, and the 
object moves, in Figure~\ref{fig-mc_Mvsz}, from right to left along a line 
parallel to the black solid line. Moreover, at any redshift, there are more 
sources at faint magnitudes than at bright ones because of the shape of the 
LF. Therefore, the sources scattering from high redshifts into the 
considered redshift bin (plotted in Fig.~\ref{fig-mc_Mvsz} in cyan) 
preferentially end up at fainter magnitudes (plotted in red), producing the 
excess at the faint end of the LF with respect to the input LF.
The excess at the bright end is instead mainly produced by those sources that 
scatter into the considered redshift range coming from lower redshifts 
(represented in Fig.~\ref{fig-mc_Mvsz} with cyan and green filled circles); 
since at the bright end the number of objects is very small, even a handful of 
new sources can significantly increase the measured density with respect 
to the intrinsic one. Because at low redshifts the dependency of $dM/dz$ 
with redshift is strong, even small uncertainties in the redshift estimate 
have large effects on the rest-frame absolute magnitude. For example, at 
$z\sim0.7$, a $\Delta z=0.25$ results in a $\Delta M \sim0.7-1$~mag 
(only 0.1-0.2~mag at $z=2.25$).

The systematic effect on $M^{\star}$ that we derive 
($\Delta M^{\star} \sim-0.4$~mag) is about half the effect found by 
\cite{chen03}. Also almost no systematic effect in $\alpha$ was found in 
their work. It seems very likely that these differences might be due to 
the fact that the redshifts are drawn from a random uniform distribution 
in \cite{chen03} while in our Monte Carlo simulations they are extracted 
from the probability function specified in Eq.~\ref{eq-pz}. In the former 
case, there would be a much larger number of low-$z$ sources that can 
scatter into the considered redshift bin, resulting in a larger excess 
of sources at the bright end and therefore a larger systematic effect 
in $M^{\star}$. If we repeat our Monte Carlo simulations extracting 
the redshifts of the sources from a random uniform distribution, we 
obtain $\Delta M^{\star} \sim -0.9$~mag and $\Delta \alpha \sim -0.06$, 
consistent with the result of \cite{chen03} as shown in the right 
panel of Figure~\ref{fig-chen4V}.

\begin{figure*}
\centering \includegraphics[width=17cm]{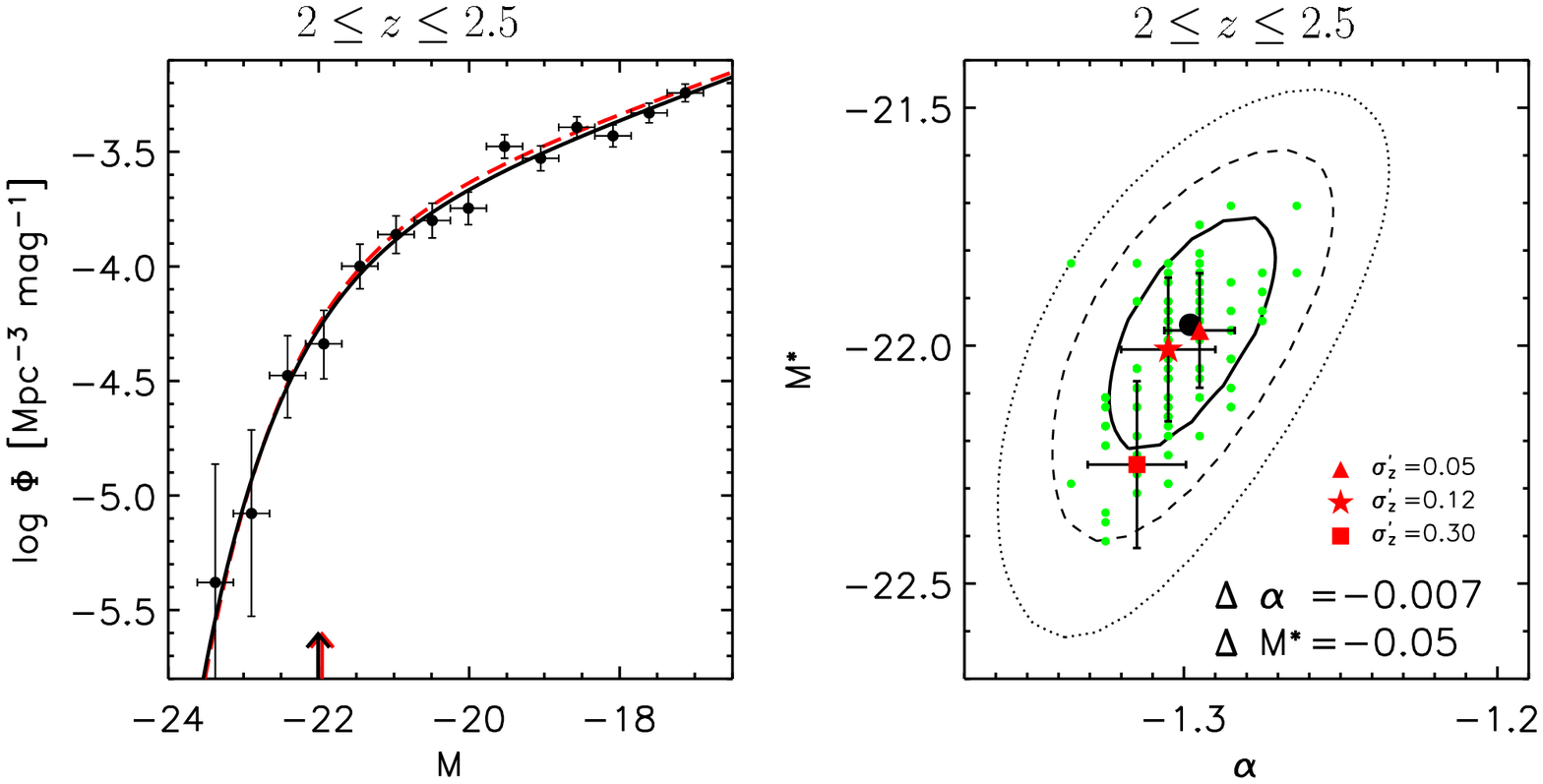}
\caption{{\it Left:} Input Schechter LF ({\it red dashed line}) compared 
to the observed LF at $2 \leq z \leq 2.5$ of the median Monte Carlo 
realization (black solid line and filled circles correspond to the 
maximum likelihood and $1/V_{\rm max}$ methods, respectively); the 
best-fit $M^{\star}$ are also plotted as arrows. {\it Right:} 
Measured Schechter parameters $M^{\star}$ plotted vs. $\alpha$; the filled 
black circle represents the input parameters of the ``observed'' model 
catalog, together with the 1, 2, and 3 $\sigma$ contour levels; small 
green filled circles represent the 100 Monte Carlo realizations using 
$\sigma^{\prime}_{\rm z}=0.12$; the red filled star is the corresponding 
median realization; the red filled triangle and square represent the median 
realization of the Monte Carlo simulations with $\sigma^{\prime}_{\rm z}=0.05$ 
and $\sigma^{\prime}_{\rm z}=0.3$, respectively. The amount of the 
systematic effect in $\alpha$ and $M^{\star}$ is also specified for 
$\sigma^{\prime}_{\rm z}=0.12$. \label{fig-mock7wV}}
\end{figure*}

Next, we repeated our Monte Carlo simulations at higher redshift by 
generating model catalogs with galaxies at redshifts between 
$z_{\rm 1,MC}=1$ and $z_{\rm 2,MC}=6$, since the goal of this paper 
is to measure the LF of galaxies in the redshift intervals 
$2 \leq z \leq 2.5$ and $2.5 < z \leq 3.5$. In Figure~\ref{fig-mock7wV} 
we plot the results of our Monte Carlo 
simulations at $2 \leq z \leq 2.5$, assuming an input Schechter LF with 
$\alpha=-1.3$ and $M^{\star}=-22$ and for $\sigma^{\prime}_{\rm z}=0.12$ 
(which corresponds to the photometric redshift errors in the deep NIR MUSYC 
for $z>1.5$ objects).

The systematic effects on the measured $\alpha$ and $M^{\star}$ are now 
very small, $\Delta M^{\star}=-0.05$~mag and $\Delta \alpha=-0.007$, 
and negligible with respect to the other uncertainties on the estimated 
best-fit parameters. Similar results are obtained in the redshift bin 
$2.5 < z \leq 3.5$. At $2 \leq z \leq 3.5$, the effect due to 
$p(z) \propto dV/dz$ 
is much smaller than at $z<1$, since $p(z)$ peaks at $z \sim 2.5$ and then 
decreases very slowly, so that the number of high- and low-$z$ sources 
scattering into the considered redshift bin is similar to the number of 
sources scattering out. Also, at $z>2$, the error on the rest-frame 
absolute magnitude corresponding to a redshift error is significantly smaller 
than at $z<1$; e.g., $\Delta M=0.3-0.4$~mag for $\Delta z \sim0.4$ at 
$z=2.25$. Therefore, the measured LF is similar to the input one and the 
systematic effects on $\alpha$ and $M^{\star}$ are negligible compared to 
the uncertainties in the LF estimates for reasonable values of 
$\sigma^{\prime}_{\rm z}$ ($\lesssim0.12$).

We repeated the above Monte Carlo simulations assuming different
input $\alpha$ ($-1.5$, $-1$, and $-0.5$) to study the behavior of the 
systematic effects as function of the faint-end slope. No significant 
differences are found: $-0.04<\Delta M^{\star}<0.01$~mag for $\alpha=-1$ and 
$-0.5$ (depending on the considered redshift bin); for $\alpha=-1.5$, the 
systematic effect is slightly larger ($\Delta M^{\star} \sim-0.07/+0.08$~mag 
for $2 \leq z \leq 2.5$ and $2.5< z \leq 3.5$, respectively), but 
also the uncertainties on 
the best-fit $M^{\star}$ increase with steeper faint-end slopes (since the
observed LF appears more like a power law) so that the systematic 
effects on the measured best-fit Schechter parameters remain very small  
with respect to the uncertainties on the best-fit values.

Finally, we investigated the effects of non-Gaussian redshift error 
probability distributions. First, using a model catalog with galaxies 
at redshifts between $z_{\rm 1,MC}=0.1$ and $z_{\rm 2,MC}=6$, we 
simulated the effect of a 5\% ``catastrophic'' outliers by assigning 
random redshifts to 5\% of the mock catalog. Adopting the input LF 
with $\alpha=-1.3$ and $M^{\star}=-22.0$ and assuming 
$\sigma^{\prime}_{\rm z}=0.12$, we find larger systematic effect in 
both $\alpha$ and $M^{\star}$ by a factor of almost 2. 
Next, we built the mean redshift probability 
distribution of the deep NIR MUSYC by averaging the individual 
redshift probability distribution for each galaxy calculated by the 
used photometric redshift code (for details see \citealt{rudnick03}). 
The average  MUSYC redshift probability distribution is well modeled by a 
Lorentzian function, rather than a Gaussian function. We find a larger 
systematic effect in $\alpha$, twice as much as the corresponding 
effect assuming a Gaussian parametrization for the redshift probability 
distribution, but similar systematic effect in $M^{\star}$, although in 
the opposite direction. To summarize, although the systematic effects in 
$\alpha$ and $M^{\star}$ expectedly get larger when we simulate 
``catastrophic'' outliers or we adopt a redshift error function with 
broader wings compared to the Gaussian model, they remain much smaller 
than the random uncertainties in the LF estimates. 

We also quantified the systematic effect on the luminosity density 
estimates. We find that the effect is of the order of a few percent 
(always $<$6\%) depending on the considered redshift interval and on the 
input Schechter LF. 

In order to include this contribution in the error budget, we 
conservatively assume a 10\% error contribution to the luminosity 
density error budget due to uncertainties in the photometric 
redshift estimates.


\section{Appendix B \\ ~~~~  \\ Rest-Frame $V$- and $B$-band Luminosity Functions for DRGs/non-DRGs and \\ Red/Blue Galaxies}
\label{app-2}

\begin{figure*}
\centering
\includegraphics[width=14.5cm]{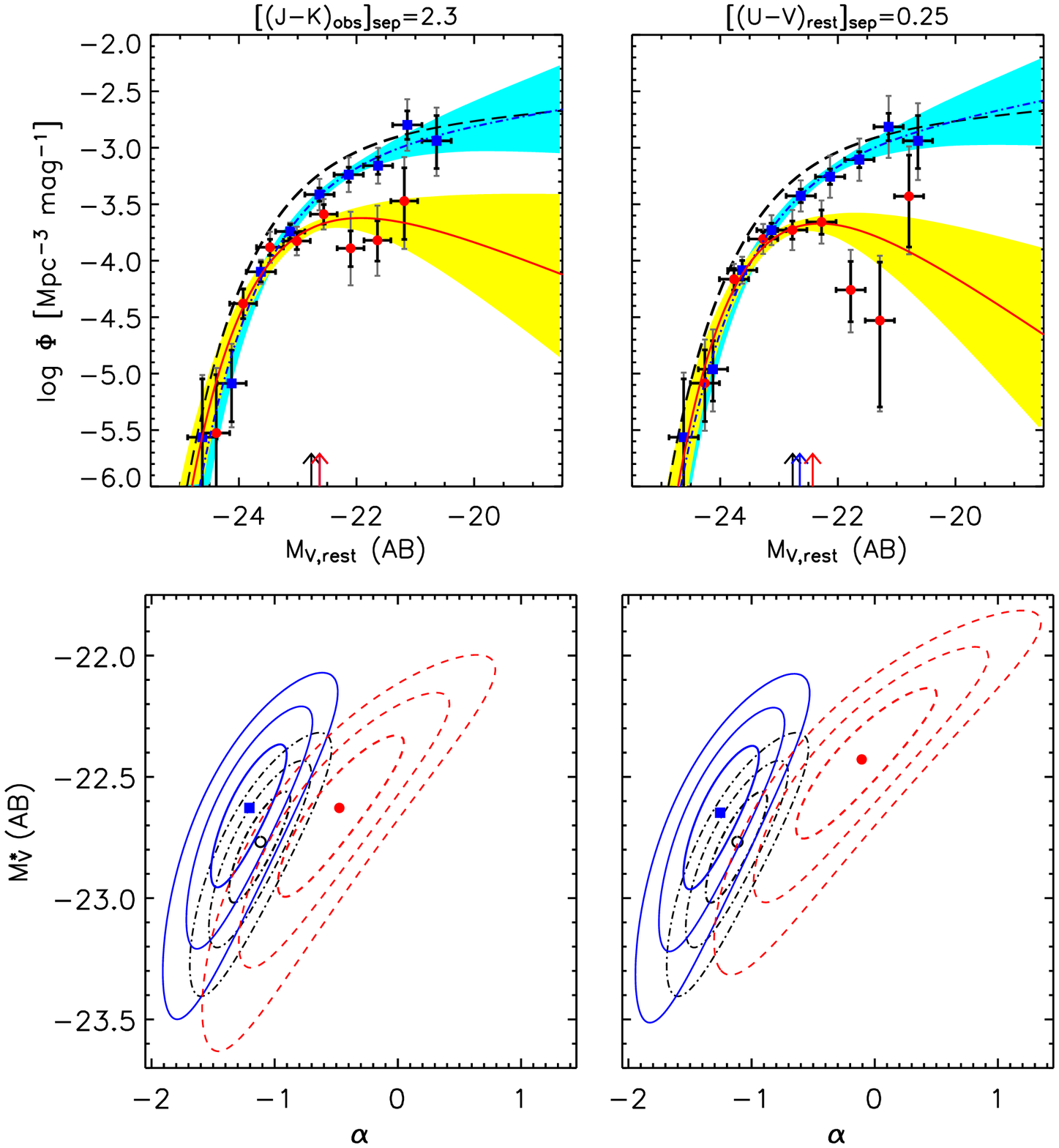}
\caption{Rest-frame V-band LF at $2.7\leq z \leq3.3$. {\it Top:} 
Rest-frame $V$-band LFs at $2.7 \leq z \leq 3.3$ for DRGs/non-DRGs 
({\it left}) and red/blue galaxies ({\it right}). The red solid line 
and filled circles represent the LF of DRGs and red galaxies estimated 
with the STY and $1/V_{\rm max}$ methods, respectively; the blue 
dot-dashed line and filled squares represent the LF of non-DRGs and 
blue galaxies. The rest-frame $V$ band LF of all galaxies is also plotted 
({\it black dashed line}). The shaded regions represent the 1 $\sigma$ 
uncertainties of the LFs measured with the STY method. The arrows represent 
the best-fit $M^{\star}$; error bars as in Fig.~\ref{LF_R_lowz.ps}. 
{\it Bottom:} 1, 2, and 3 $\sigma$ contour levels from the STY method for 
DRGs and red galaxies ({\it red dashed lines}), for non-DRGs and blue 
galaxies ({\it blue solid lines}), and for all galaxies ({\it black 
dot-dashed lines}); the best-fit Schechter values are also plotted 
(red filled circle for DRGs and red galaxies, blue filled square for 
non-DRGs and blue galaxies, and open circle for all galaxies). 
\label{LF_V_highz.ps}}
\end{figure*}

\begin{figure*}
\centering
\includegraphics[width=14.5cm]{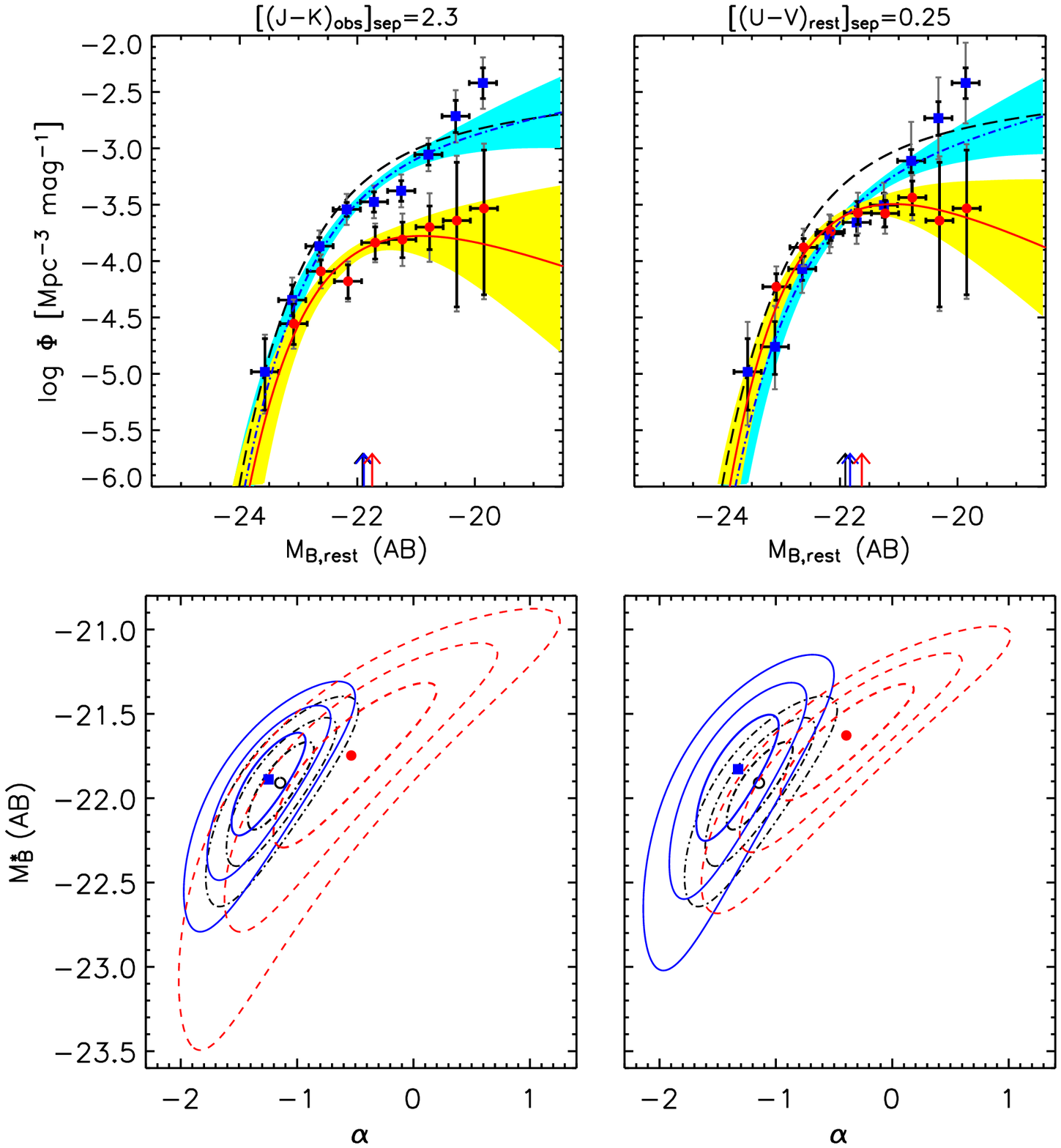}
\caption{Rest-frame B-band LF at $2\leq z \leq2.5$; symbols as in
Fig.~\ref{LF_V_highz.ps}. \label{LF_B_lowz.ps}}
\end{figure*}

\begin{figure*}
\centering
\includegraphics[width=14.5cm]{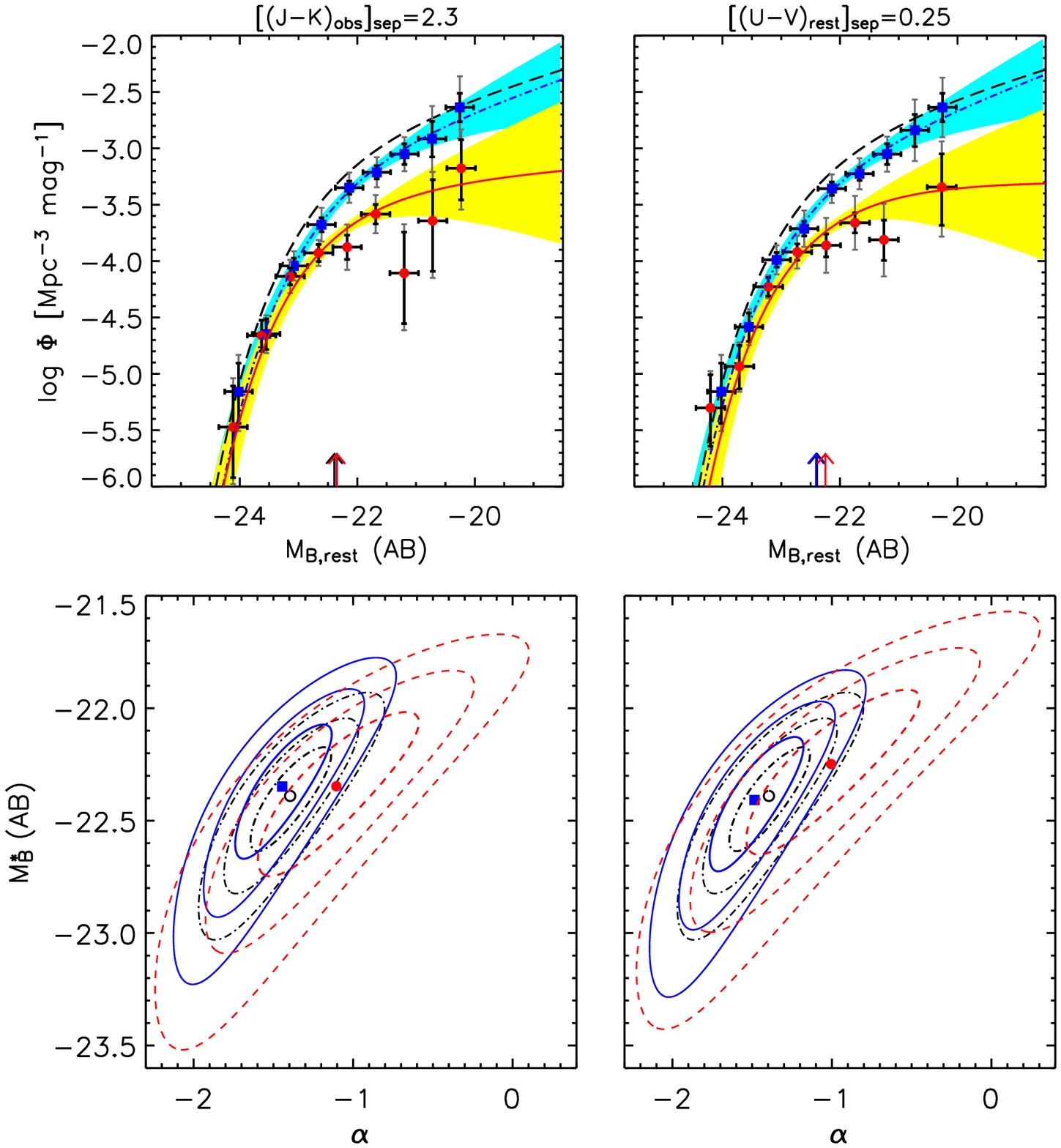}
\caption{Rest-frame B-band LF at $2.5< z \leq3.5$; symbols as in
Fig.~\ref{LF_V_highz.ps}. \label{LF_B_highz.ps}}
\end{figure*}

In the body of the paper we showed the rest-frame $R$ band LFs of 
DRGs, non-DRGs, red and blue galaxies (see Fig.~\ref{LF_R_lowz.ps}). For 
completeness, we show here the comparison of the LFs of DRGs (red 
galaxies) and non-DRGs (blue galaxies) discussed in \S~\ref{subsec-sublf} 
in the rest-frame $V$ band at $2.7\leq z \leq3.3$ (Fig.~\ref{LF_V_highz.ps}) 
and in the rest-frame $B$ band at $2\leq z \leq2.5$ 
(Fig.~\ref{LF_B_lowz.ps}) and at $2.5< z \leq3.5$ 
(Fig.~\ref{LF_B_highz.ps}). The corresponding best-fit 
Schechter parameters are listed in Table~\ref{tab-4}.

\section{Appendix C \\ ~~~~ \\ Comparison with Previously Published Luminosity Functions}
\label{app-3}

Here we compare our results to previous rest-frame optical LF studies 
at $z>2$, which were based on smaller samples. We note that these 
studies are affected by significant uncertainties due to field-to-field 
variance (as they are based on a single field or on a very small total 
surveyed area) and by small number statistics at the bright end. 

\subsubsection{C1. POLI ET AL. (2003) AND GIALLONGO ET AL. (2005)}

\begin{figure}
\epsscale{1}
\plotone{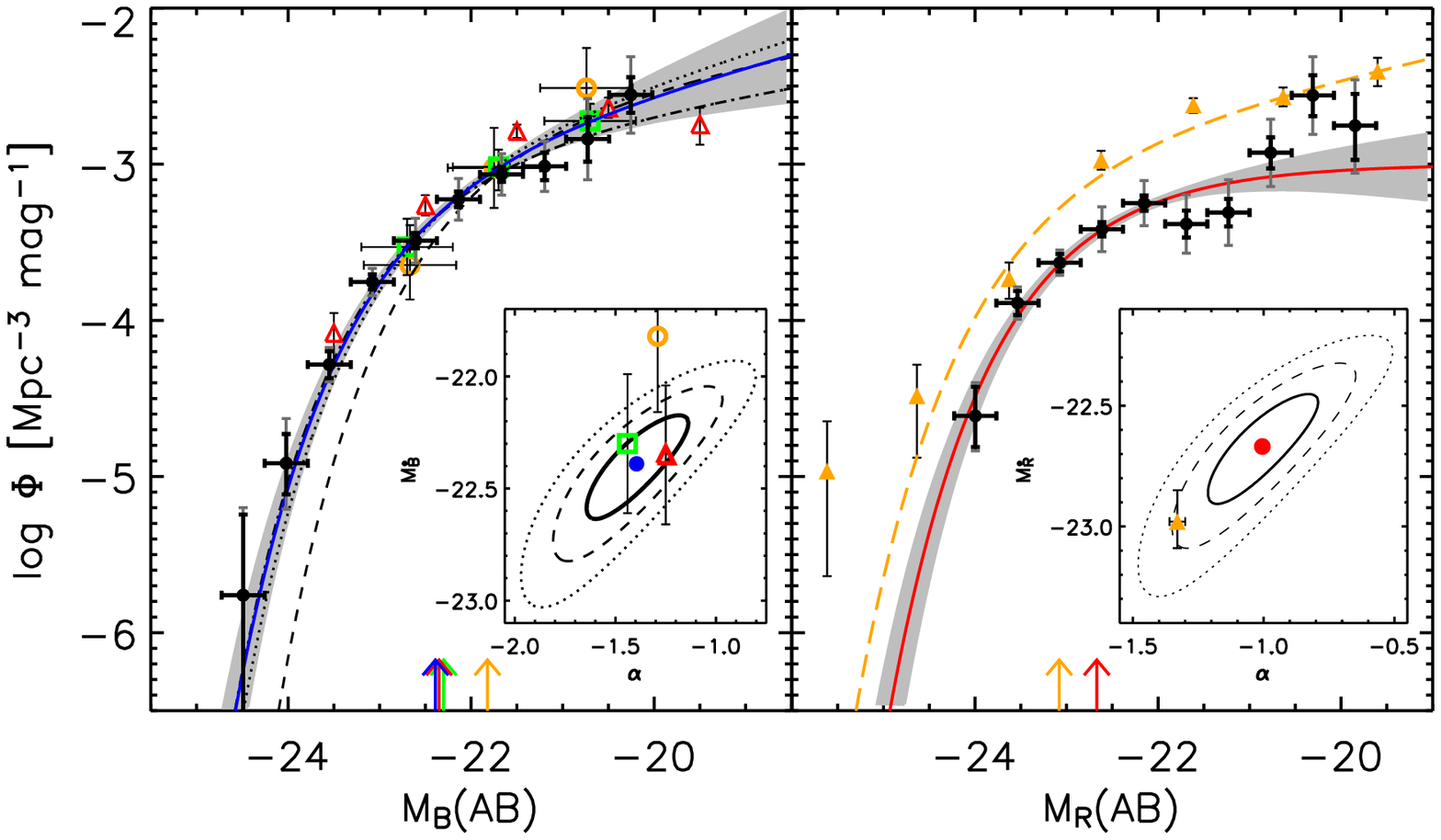}
\caption{{\it Left:} Rest-frame $B$-band LF at $2.5< z \leq3.5$ 
from this work ({\it blue solid line and black filled circles}) compared 
to the LF from Poli et al. (2003; {\it dotted line and green open 
squares}), from Giallongo et al. (2005; {\it dashed line and orange 
open circles}), and from Gabasch et al. (2004; {\it dot-dashed line 
and red open triangles}). In the inset, the 1, 2, and 3 $\sigma$ 
contour levels of the best-fitSchechter parameters $\alpha$ and 
$M^{\star}$ are compared to the best-fit values estimated by 
\cite{poli03}, \cite{giallongo05}, and \cite{gabasch04}. 
{\it Right:} Rest-frame $R$-band LF at $2\leq z \leq2.5$ from this work 
({\it red solid line and black filled circles}) compared to the LF 
from Gabasch et al. (2006; {\it orange dashed line and filled triangles}; 
$R=r-0.12$). In the inset, the 1, 2, and 3 $\sigma$ contour levels of 
the best-fit Schechter parameters $\alpha$ and $M^{\star}$ are compared 
to the best-fit values estimated by \cite{gabasch06}. \label{fig-poli}}
\end{figure}

\cite{poli03} analyzed a sample of 138 $K$-selected galaxies down to
$K_{\rm AB}=25$ to construct the rest-frame $B$-band LF in the
redshift range $1.3<z<3.5$. The total area of their composite sample is
$\sim68$~arcmin$^{2}$, a factor of $\sim5.6$ smaller than the area
sampled in this work. \cite{giallongo05} repeated the analysis in
\cite{poli03} with an improved composite sample (although with the
same area) and allowing the Schechter parameters $\Phi^{\star}$ and
$M^{\star}$ to vary with the redshift, while $\alpha$ is kept constant
at the low-redshift value ($z<1$). A direct comparison between the LFs
of \cite{poli03} and \cite{giallongo05} and the LF measured in this
work is shown in Figure~\ref{fig-poli} ({\it left panel}) for the redshift 
range $2.5<z \leq3.5$.

First, we note in Figure~\ref{fig-poli} how much better the bright end of 
the LF is constrained from our work: the large area of the composite 
sample ($\sim76$\% of which comes from the deep NIR MUSYC alone) 
allows us to sample the LF up to $B$-band magnitudes $\sim1.5$~mag brighter 
than done in \cite{poli03} and \cite{giallongo05}. Our measurements of 
the LF using the $1/V_{\rm max}$ method are consistent within the errors 
with those in both \cite{poli03} and \cite{giallongo05}. The best-fit LF 
estimated with the maximum likelihood analysis in \cite{poli03} is 
consistent, within the errors, with our best-fit solution. However, the 
best-fit LF estimated with the maximum likelihood analysis in 
\cite{giallongo05} is significantly different from ours, as clearly shown 
in the inset of Figure~\ref{fig-poli} ({\it left panel}). While their 
faint-end slope is consistent with our best-fit $\alpha$, their lack of 
constraints on the bright end of the LF results in a much fainter 
$M^{\star}_{\rm B}$ (by $\sim0.6$~mag).

\subsubsection{C2. GABASCH ET AL. (2004, 2006)}

\cite{gabasch04,gabasch06} analyzed a sample of 5558 
$I$-selected galaxies down to $I_{\rm AB}=26.8$ (50\% compleness 
limit) from the FDF survey \citep{heidt03} to 
study the evolution of the rest-frame $B$- and $r$-band LF over 
the redshift range $0.5<z<5$. The total area of their sample is
$\sim40$~arcmin$^{2}$ over a single field, a factor of $\sim9.5$ 
smaller than the total area sampled in this work. 

A direct comparison between the $B$-band LF from \cite{gabasch04} 
and ours is possible in the redshift range $2.5< z \leq3.5$, and it 
is shown in the left panel of Figure~\ref{fig-poli} ({\it red symbols}). 
The best-fit Schechter parameters $\alpha$, $M^{\star}$, and $\Phi^{\star}$ 
are consistent with ours within the errors. 
In the redshift range $2\leq z \leq2.5$ we compared our estimated 
$B$-band LF with the one defined by their best-fit Schechter parameters 
($\alpha$, $M^{\star}$, and $\Phi^{\star}$) estimated at $z \sim2.25$. 
While their best-fit Schechter parameters $\alpha$ and $M^{\star}$ are 
consistent with ours, their best-fit $\Phi^{\star}$ is a factor of 
$\sim1.3$ larger.
This difference can be entirely accounted for by field-to-field variations 
(see \S~\ref{sec-numdens}). In fact, the FDF survey consists of a single 
pointing of only $\sim40$~arcmin$^{2}$ and thus it is potentially 
strongly affected by sample variance. Moreover, \cite{gabasch04} 
spectroscopically identified an overdensity of galaxies at $z=2.35$ 
(possibly a [proto]cluster, with more than 10 identical redshifts), which 
can potentially strongly bias the estimate of $\Phi^{\star}$ in this 
redshift bin.

In the right panel of Figure~\ref{fig-poli} we compare our measured $R$-band 
LF with that measured by \cite{gabasch06} in the redshift range 
$1.91< z <2.61$. The agreement between the two LFs is much worse than for 
the rest-frame $B$-band. Their best-fit Schechter parameters $\alpha$ and 
$M^{\star}$ are now consistent with ours only at the $\sim2$ $\sigma$ level, 
with their $M^{\star}$ about 0.3~mag brighter than ours; their best-fit 
$\Phi^{\star}$ is a factor of $\sim1.6$ larger than our best-fit value. 
As mentioned 
above, an overdensity of galaxies was spectroscopically found in this 
redshift interval in the FDF. Although it is hard to quantify the effect 
of the presence of a (proto)cluster at this redshift on the measured LF, 
it is interesting to make a connection with the work of \cite{steidel05}, 
who spectroscopically identified a protocluster at $z=2.3$ in the 
HS~1700+643 field. In the spectroscopically identified galaxy sample, 
\cite{steidel05} found 19 (out of 55) objects at $z=2.300\pm0.015$ in 
the redshift range $2< z <2.5$. Within the 
$\sim 7^{\prime}\times7^{\prime}$ field of view (similar to the FDF 
field of view) over which the protocluster objects are distributed, the 
protocluster and ``field'' galaxy sky distributions are the same. Assuming 
that the distribution of the spectroscopically identified galaxies is 
representative of the whole sample, the estimated density would be a factor 
of $\sim$1.3-1.6 that of ``field'' galaxies only, consistent with the 
difference found in $\Phi^{\star}$ between our composite sample and the 
FDF in the redshift range $2< z <2.5$.


\subsubsection{C3. COMPARISON WITH GIALLONGO ET AL. (2005) FOR RED/BLUE GALAXIES}

\begin{figure}
\epsscale{1}
\plotone{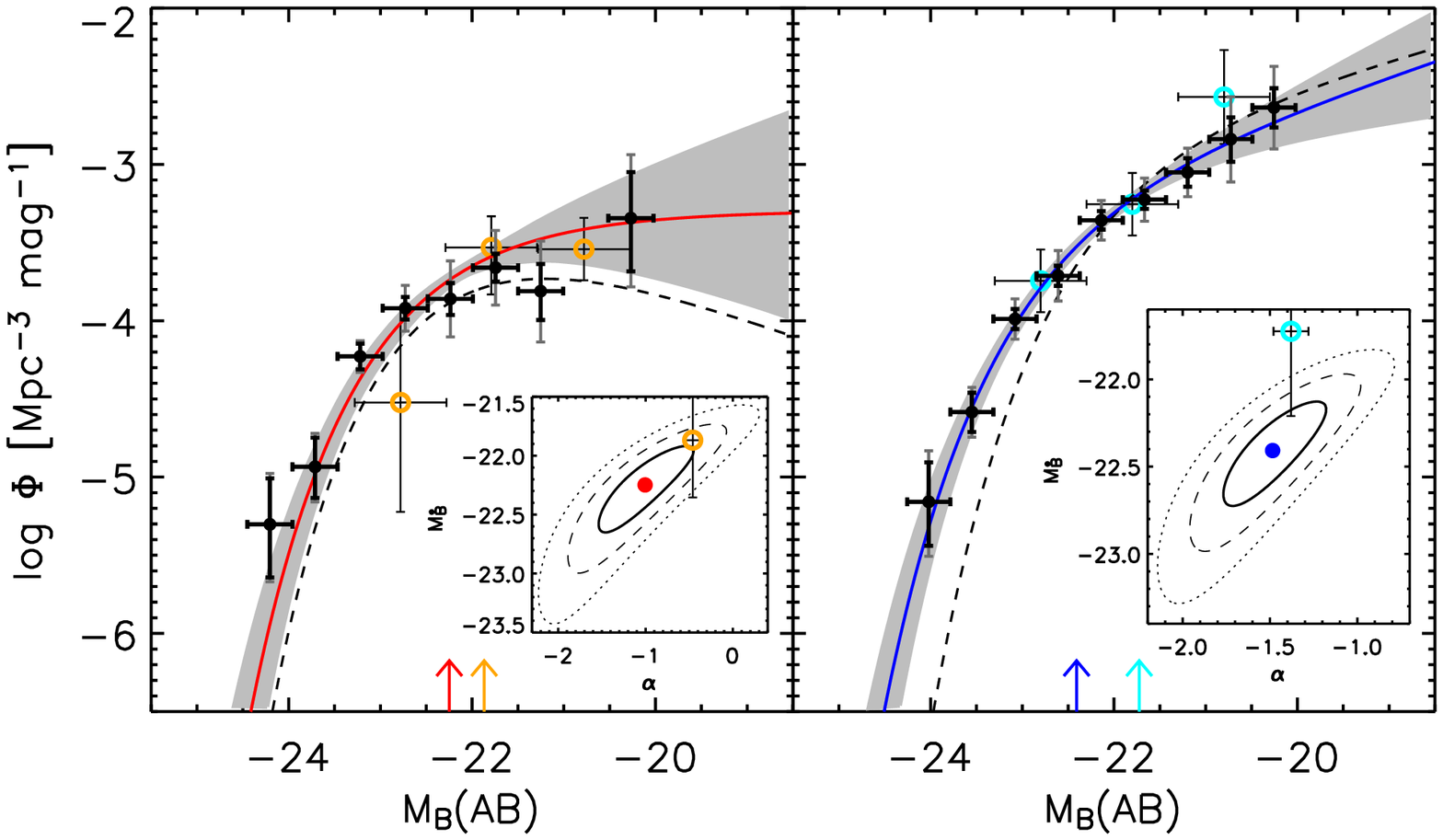}
\caption{{\it Left:} Rest-frame $B$-band LF at $2.5< z \leq3.5$ from
this work for red galaxies ($U-V \geq 0.25$; {\it red solid line and
black filled circles}) compared to the one from Giallongo et al. 
(2005; {\it dashed line and orange open circles}); in the inset, the 1, 
2, and 3 $\sigma$ contour levels of the best-fit Schechter parameters
$\alpha$ and $M^{\star}$ are compared to the best-fit values estimated
by Giallongo et al. (2005; {\it orange open circle}), which assumed a 
constant $\alpha=-0.46$. {\it Right:} Rest-frame $B$-band LF at
$2.5< z \leq3.5$ from this work for blue galaxies ($U-V < 0.25$; 
{\it blue solid line and black filled circles}) compared to the one from
Giallongo et al. (2005; {\it dashed line and cyan open circles}); in the 
inset, the 1, 2, and 3 $\sigma$ contour levels of the best-fit 
Schechter parameters $\alpha$ and $M^{\star}$ are compared to the 
best-fit values estimated by Giallongo et al. (2005; {\it cyan open circle}), 
which assumed a constant $\alpha=-1.38$. \label{fig-Bgiallongo}}
\end{figure}

\cite{giallongo05} measured the rest-frame $B$-band LF of red and blue
galaxies at redshift $2.5<z<3.5$ using a sample of 138 $K$-selected
galaxies in the redshift range $1.3<z<3.5$ down to $K_{\rm AB}=25$. 
Their red and blue populations were defined on the basis of
an ``S0 color track''. From their Figure~1, the average rest-frame $U-V$
color of their model at redshift $2.5<z<3.5$ is $\sim0.22$, very
similar to our definition of blue and red galaxies ($U-V<0.25$ and 
$U-V \geq 0.25$, respectively). In Figure~\ref{fig-Bgiallongo} we have
compared the LFs of red and blue galaxies from this work to those
presented in \cite{giallongo05}.

As for the LF of all galaxies, we are able to constrain the bright end of 
the LF much better by sampling the LF to luminosities $\sim1.5$~mag brighter. 
For the red galaxy population, their measurements of the LF with the 
$1/V_{\rm max}$ method are consistent within the errors with ours; their 
best-fit faint-end slope is also consistent, within the errors, with our 
estimate, although $\alpha$ was fixed at the local value in 
\cite{giallongo05}; our $M^{\star}$ is $\sim0.4$~mag brighter 
than their best-fit value, but still consistent within the errors at the 
1 $\sigma$ level. The best-fit normalization of \cite{giallongo05} is a 
factor of $\sim1.5$ smaller than ours, only marginally consistent at the 
2 $\sigma$ level.

For the blue population, the $1/V_{\rm max}$ measurements of the LF from 
\cite{giallongo05} are consistent with ours. However, the agreement gets 
worse when we compare our LF estimated using the STY method with theirs, 
especially at the bright end, where they lack information. While their 
assumed faint-end slope is consistent with our best-fit solution, the 
characteristic magnitude from \cite{giallongo05} is $\sim0.7$~mag fainter 
than ours. Their best-fit solution is, however, very uncertain, characterized 
by very large error bars and still consistent with ours at the 
$\sim1$ $\sigma$ level.



\begin{thebibliography}{}
\bibitem[Adelberger et al.(2005)]{adelberger05} Adelberger, K.~L., Erb, D.~K., Steidel, C.~C., Reddy, N.~A., Pettini, M., \& Shapley, A.~E. 2005, \apj, 620, L75
\bibitem[Adelberger \& Steidel (2000)]{adelberger00} Adelberger, K.~L., \& Steidel, C.~C. 2000, \apj, 544, 218
\bibitem[Adelberger et al.(2004)]{adelberger04} Adelberger, K.~L., Steidel, C.~C., Shapley, A.~E., Hunt, M.~P., Erb, D.~K., Reddy, N.~A., \& Pettini, M. 2004, \apj, 607, 226
\bibitem[Avni (1976)]{avni76} Avni, Y. 1976, \apj, 210, 642
\bibitem[Avni \& Bahcall (1980)]{avni80} Avni, Y., \& Bahcall, J. N. 1980, 
  \apj, 235, 694
\bibitem[Bell \& de Jong (2001)]{bell01} Bell, E.~F., de Jong, R.~S. 2001, \apj, 550, 212
\bibitem[Bessell (1990)]{bessell90} Bessell, M.~S. 1990, \pasp, 102, 1181
\bibitem[Blanton et al.(2001)]{blanton01} Blanton, M.~R., et al. 2001, \aj, 121, 2358
\bibitem[Blanton et al.(2003)]{blanton03} Blanton, M.~R., et al. 2003, \apj, 592, 819
\bibitem[Brown et al.(2001)]{brown01} Brown, W.~R., Geller, M.~J., Fabricant, D.~G., Kurtz, M.~J. 2001, \aj, 2001, 122, 714
\bibitem[Bruzual \& Charlot (2003)]{bruzual03} Bruzual, G., \& Charlot, S. 2003, \mnras, 344, 1000
\bibitem[Calzetti et al.(2000)]{calzetti00} Calzetti, D., Armus, L., Bohlin, R.~C., Kinney, A.~L., Koornneef, J., \& Storchi-Bergmann, T. 2000, \apj, 533, 682
\bibitem[Chabrier (2003)]{chabrier03} Chabrier, G. 2003, \pasp, 115, 763
\bibitem[Chen et al.(2003)]{chen03} Chen, H.-W, et al. 2003, \apj, 586, 745
\bibitem[Cole et al.(2001)]{cole01} Cole, S., et al. 2001, \mnras, 326, 255
\bibitem[Coleman et al.(1980)]{coleman80} Coleman, G.~D., Wu, C.-C., Weedman, D.~W. 1980, \apjs, 43, 393
\bibitem[Daddi et al.(2003)]{daddi03} Daddi, E., et al. 2003, \apj, 588, 50
\bibitem[Dahlen et al.(2005)]{dahlen05} Dahlen, T., Mobasher, B., Somerville, R.~S., Moustakas, L.~A., Dickinson, M., Ferguson, H.~C., Giavalisco, M. 2005, \apj, 631, 126
\bibitem[Davis et al.(2003)]{davis03} Davis, M., et al. 2003, Proc. SPIE, 4834, 161
\bibitem[Efstathiou et al.(1988)]{efstathiou88} Efstathiou, G., Ellis R.~S., \& Peterson, B.~A. 1988, \mnras, 232, 431
\bibitem[Ellis et al.(1996)]{ellis96} Ellis R.~S., Colless, M., Broadhurst, T., Heyl, J., Glazebrook, K. 1996, \mnras, 280, 235
\bibitem[F{\"o}rster Schreiber et al.(2004)]{forster04} F{\"o}rster Schreiber, N.~M., et al. 2004, \apj, 616, 40
\bibitem[F{\"o}rster Schreiber et al.(2006)]{forster06} F{\"o}rster Schreiber, N.~M., et al. 2006, \aj, 131, 1891
\bibitem[Franx et al.(2003)]{franx03} Franx, M., et al. 2003, \apj, 587, L79
\bibitem[Gabasch et al.(2004)]{gabasch04} Gabasch, A., et al. 2004, \aap, 421, 41
\bibitem[Gabasch et al.(2006)]{gabasch06} Gabasch, A., et al. 2006, \aap, 448, 101
\bibitem[Gawiser et al.(2006)]{gawiser06} Gawiser, E., et al. 2006, \apjs, 162, 1
\bibitem[Gehrels (1986)]{gehrels86} Gehrels, N. 1986, \apj, 303, 336
\bibitem[Giallongo et al.(2005)]{giallongo05} Giallongo, E., Salimbeni, S., Menci, N., Zamorani, G., Fontana, A., Dickinson, M., Cristiani, S., Pozzetti, L. 2005, \apj, 622, 116
\bibitem[Giavalisco et al.(2004)]{giavalisco04} Giavalisco, M., et al. 2004, \apj, 600, L93
\bibitem[Hauschildt et al.(1999)]{hauschildt99} Hauschildt, P.~H., Allard, F., \& Baron, E. 1999, \apj, 512, 377
\bibitem[Heidt et al.(2003)]{heidt03} Heidt, J., et al. 2003, \aap, 398, 49
\bibitem[Ilbert et al.(2005)]{ilbert05} Ilbert, O., et al. 2005, \aap, 439, 86
\bibitem[Kauffmann et al.(2003)]{kauffmann03} Kauffmann, G., et al. 2003, \mnras, 341, 33
\bibitem[Kendall \& Stuart (1961)]{kendall61} Kendall, M.~G., \& Stuart, A. 1961, The Advanced Theory of Statistics, Vol.~2 (London: Griffin \& Griffin)
\bibitem[Kinney et al.(1996)]{kinney96} Kinney, A.~L., Calzetti, D., Bohlin, R.~C., McQuade, K., Storchi-Bergmann, T., \& Schmitt, H.~R. 1996, \apj, 467, 38
\bibitem[Knudsen et al.(2005)]{knudsen05} Knudsen, K.~K., et al. 2005, \apj, 632, L9
\bibitem[Kochanek et al.(2001)]{kochanek01} Kochanek, C.~S., et al. 2001, \apj, 560, 566
\bibitem[Kriek et al.(2006)]{kriek06} Kriek, M. et al. 2006, \apj, 649, L71
\bibitem[Labb\'e et al.(2003)]{labbe03} Labb\'e, I., et al. 2003, \aj, 125, 1107
\bibitem[Labb\'e et al.(2005)]{labbe05} Labb\'e, I., et al. 2005, \apj, 624, L81
\bibitem[Labb\'e et al.(2006)]{labbe06} Labb\'e, I., et al. 2006, \apj, submitted
\bibitem[Lampton et al.(1976)]{lampton76} Lampton, M., Margon, B., 
\& Bowyer, S. 1976, \apj, 208, 177
\bibitem[Le F\'evre et al.(2004)]{lefevre04} Le F\'evre, et al. 2004, \aap, 417, 839
\bibitem[Lilly et al.(1995)]{lilly95} Lilly, S.J., Tresse, L., Hammer, F., Crampton, D., Le F\`evre, O. 1995, \apj, 455, 108
\bibitem[Lin et al.(1997)]{lin97} Lin, H., Yee, H.~K.~C., Carlberg, R.~G., Ellingson, E. 1997, \apj, 475, 494
\bibitem[Nagamine et al.(2000)]{nagamine00} Nagamine, K., Cen, R., \& Ostriker, J.~P. 2000, \apj, 541, 25
\bibitem[Nagamine et al.(2001)]{nagamine01} Nagamine, K., Fukugita, M., Cen, R., \& Ostriker, J.~P. 2001, \mnras, 327, L10
\bibitem[Norberg et al.(2002)]{norberg02} Norberg, P., et al. 2002, \mnras, 336, 907
\bibitem[Papovich et al.(2006)]{papovich06} Papovich, C., et al. 2006, \apj, 640, 92
\bibitem[Pettini et al.(2001, 0.1-0.5~Z$_{\sun}$)]{pettini01} Pettini, M., Shapley, A.~E., Steidel, C.~C., Cuby, J.-G., Dickinson, M., Moorwood, A.~F.~M., Adelberger, K.~L., \& Giavalisco, M. 2001, \apj, 554, 981
\bibitem[Poli et al.(2003)]{poli03} Poli, F., et al. 2003, \apj, 593, L1
\bibitem[Quadri et al.(2007a)]{quadri06b} Quadri, R., et al. 2007a, \apj, 654, 138
\bibitem[Quadri et al.(2007b)]{quadri06} Quadri, R., et al. 2007b, \aj, submitted (astro-ph/0612612)
\bibitem[Reddy et al.(2005)]{reddy05} Reddy, N.~A., Erb, D.~K., Steidel, C.~C., Shapley, A.~E., Adelberger, K.~L., Pettini, M. 2005, \apj, 633, 748
\bibitem[Reddy et al.(2006)]{reddy06} Reddy, N.~A., Steidel, C.~C., Fadda, D., Yan, L., Pettini, M., Shapley, A.~E., Erb, D.~K., Adelberger, K.~L. 2006, \apj, 644, 792
\bibitem[Rubin et al.(2004)]{rubin04} Rubin, K.~H.~R., van Dokkum, P.~G., Coppi, P., Johnson, O., F{\"o}rster Schreiber, N.~M., Franx, M., \& van der Werf, P. 2004, \apj, 613, L5
\bibitem[Rudnick et al.(2001)]{rudnick01} Rudnick, G., et al. 2001, \aj, 122, 2205
\bibitem[Rudnick et al.(2003)]{rudnick03} Rudnick, G., et al. 2003, \apj, 599, 847
\bibitem[Rudnick et al.(2006)]{rudnick06} Rudnick, G., et al. 2006, \apj, 650, 624
\bibitem[Salpeter (1955)]{salpeter55} Salpeter, E.~E. 1955, \apj, 121, 161
\bibitem[Sandage et al.(1979)]{sandage79} Sandage, A., Tammann, G. A., \& Yahil, A. 1979, \apj, 232, 352
\bibitem[Saracco et al.(2001)]{saracco01} Saracco, P., Giallongo, E., Cristiani, S., D'Odorico, S., Fontana, A., Iovino, A., Poli, F., \& Vanzella, E. 2001, \aap, 375, 1
\bibitem[Sawicki \& Thompson (2005)]{sawicki05} Sawicki, M., \& Thompson, D. 2005, \apj, 635, 100
\bibitem[Sawicki \& Thompson (2006)]{sawicki06} Sawicki, M., \& Thompson, D. 2006, \apj, 642, 653
\bibitem[Schechter (1976)]{schechter76} Schechter, P. 1976, \apj, 203, 297
\bibitem[Schmidt (1968)]{schmidt68} Schmidt, M. 1968, \apj, 151, 393
\bibitem[Shapley et al.(2004)]{shapley04} Shapley, A.~E., Erb, D.~K., Pettini, M., Steidel, C.~C., \& Adelberger, K.~L. 2004, \apj, 612, 108
\bibitem[Shapley et al.(2001)]{shapley01} Shapley, A.~E., Steidel, C.~C., Adelberger, K.~L., Dickinson, M., Giavalisco, M., \& Pettini, M. 2001, \apj, 562, 95
\bibitem[Shapley et al.(2005)]{shapley05} Shapley, A.~E., Steidel, C.~C., Erb, D.~K., Reddy, N.~A., Adelberger, K.~L., Pettini, M., Barmby, P., \& Huang, J. 2005, \apj, 626, 698
\bibitem[Steidel et al.(1999)]{steidel99} Steidel, C.~C., Adelberger, K.~L., Giavalisco, M., Dickinson, M., \& Pettini, M. 1999, \apj, 519, 1
\bibitem[Steidel et al.(2005)]{steidel05} Steidel, C.~C., Adelberger, K.~L., Shapley, A.~E., Erb, D.~K., Reddy, N.~A., \& Pettini, M. 2005, \apj, 626, L44
\bibitem[Steidel et al.(2003)]{steidel03} Steidel, C.~C., Adelberger, K.~L., Shapley, A.~E., Pettini, M., Dickinson, M., \& Giavalisco, M. 2003, \apj, 592, 728
\bibitem[Steidel et al.(1996)]{steidel96} Steidel, C.~C., Giavalisco, M., Pettini, M., Dickinson, M., \& Adelberger, K.~L. 1996, \apj, 462, L17
\bibitem[Steidel et al.(2004)]{steidel04} Steidel, C.~C., Shapley, A.~E., Pettini, M., Adelberger, K.~L., Erb, D.~K., Reddy, N.~A., \& Hunt, M.~P. 2004, \apj, 604, 534
\bibitem[van Dokkum et al.(2003)]{vandokkum03} van Dokkum, P.~G., et al. 2003, \apj, 587, L83
\bibitem[van Dokkum et al.(2004)]{vandokkum04} van Dokkum, P.~G., et al. 2004, \apj, 611, 703
\bibitem[van Dokkum et al.(2006)]{vandokkum06} van Dokkum, P.~G., et al. 2006, \apj, 638, 59
\bibitem[Wolf et al.(2003)]{wolf03} Wolf, C., Meisenheimer, K., Rix, H.-W., Borch, A., Dye, S., Kleinheinrich, M. 2003, \aap, 401, 73
\bibitem[Zucca et al.(2006)]{zucca06} Zucca, E., et al. 2006, \aap, 445, 879
\end{thebibliography}
\end{document}